\documentclass[onecolumn,secnumarabic,nobibnotes,aps,superscriptaddress]{revtex4-2}
\usepackage{color, xcolor, colortbl}
\usepackage{graphicx}
\usepackage{geometry}
\usepackage{amsthm,amsmath,amssymb,amsfonts}
\usepackage{dcolumn}
\usepackage{epstopdf}
\usepackage{bm}
\usepackage{appendix}
\usepackage{multirow}
\usepackage{braket}
\usepackage[english]{babel}
\usepackage{hyperref}
\usepackage[capitalize]{cleveref}
\usepackage{xpatch}
\usepackage{tikz}
\usepackage{adjustbox}
\usepackage{xspace}
\usepackage{enumitem} 
\usepackage{comment}
\usepackage{pifont}

\newcommand{\bvec}[1]{\mathbf{#1}}

\newcommand{\vk}{\bvec{k}}

\newcommand{\vq}{\bvec{q}}
\renewcommand{\vr}{\bvec{r}}

\newcommand{\vt}{\bvec{t}}

\newcommand{\vF}{\bvec{F}}
\newcommand{\vG}{\bvec{G}}

\newcommand{\vK}{\bvec{K}}

\newcommand{\vR}{\bvec{R}}

\newcommand{\Tr}{\operatorname{Tr}}

\newcommand{\I}{\mathrm{i}}

\newcommand{\mc}[1]{\mathcal{#1}}

\newcommand{\wt}[1]{\widetilde{#1}}

\newcommand{\abs}[1]{\left\lvert#1\right\rvert}
\newcommand{\norm}[1]{\left\lVert#1\right\rVert}

\newcommand{\ud}{\,\mathrm{d}}

\newcommand{\CC}{\mathbb{C}}

\providecommand{\definitionname}{Definition}
\providecommand{\assumptionname}{Assumption}
\providecommand{\corollaryname}{Corollary}
\providecommand{\lemmaname}{Lemma}
\providecommand{\propositionname}{Proposition}
\providecommand{\remarkname}{Remark}
\providecommand{\theoremname}{Theorem}

\definecolor{purp}{RGB}{160, 32, 240}

\definecolor{forestgreen}{rgb}{0.13, 0.55, 0.13}

\definecolor{airforceblue}{rgb}{0.36, 0.54, 0.66}

\usetikzlibrary{quantikz}

\usetikzlibrary{fit}
\tikzset{%
  highlight/.style={rectangle,rounded corners,fill=blue!15,draw,fill opacity=0.3,thick,inner sep=0pt}
}

\usepackage[ruled,vlined]{algorithm2e}
\usepackage[normalem]{ulem}

\newcommand{\DeptMath}{Department of Mathematics, University of California, Berkeley, California 94720 USA}
\newcommand{\DeptPhys}{Department of Physics, University of California, Berkeley, California 94720 USA}
\newcommand{\Caltech}{Division of Chemistry and Chemical Engineering, California Institute of Technology, Pasadena, California 91125, United States}
\newcommand{\LBLMath}{Applied Mathematics and Computational Research Division, Lawrence Berkeley National Laboratory, Berkeley, CA 94720, USA}

\usepackage{listings}
\usepackage{xcolor}

\definecolor{codegreen}{rgb}{0,0.6,0}
\definecolor{codegray}{rgb}{0.5,0.5,0.5}
\definecolor{codepurple}{rgb}{0.58,0,0.82}
\definecolor{backcolour}{rgb}{0.95,0.95,0.92}

\lstdefinestyle{mystyle}{
    backgroundcolor=\color{backcolour},   
    commentstyle=\color{codegreen},
    keywordstyle=\color{magenta},
    numberstyle=\tiny\color{codegray},
    stringstyle=\color{codepurple},
    basicstyle=\ttfamily\footnotesize,
    breakatwhitespace=false,         
    breaklines=true,                 
    captionpos=b,                    
    keepspaces=true,                 
    numbers=left,                    
    numbersep=5pt,                  
    showspaces=false,                
    showstringspaces=false,
    showtabs=false,                  
    tabsize=2
}

\usepackage{svg}
\usepackage{caption}
\usepackage{subcaption}

\begin{document}

\title{Interacting models for twisted bilayer graphene: a quantum chemistry approach}
\author{Fabian M. Faulstich}
\thanks{These three authors contributed equally.}
\affiliation{\DeptMath}
\author{Kevin D. Stubbs}
\thanks{These three authors contributed equally.}
\affiliation{\DeptMath}
\author{Qinyi Zhu}
\thanks{These three authors contributed equally.}
\affiliation{\DeptMath}
\author{Tomohiro Soejima}
\affiliation{\DeptPhys}
\author{Rohit Dilip}
\affiliation{\Caltech}
\author{Huanchen Zhai}
\affiliation{\Caltech}
\author{Raehyun Kim}
\affiliation{\DeptMath}
\author{Michael P. Zaletel}
\affiliation{\DeptPhys}
\author{Garnet Kin-Lic Chan}
\affiliation{\Caltech}
\author{Lin Lin}
\email{linlin@math.berkeley.edu}
\affiliation{\DeptMath}
\affiliation{\LBLMath}
\date{\today}

\begin{abstract}
The nature of correlated states in twisted bilayer graphene (TBG) at the magic angle has received intense attention in recent years. 
We present a numerical study of an interacting Bistritzer-MacDonald (IBM) model of TBG using a suite of methods in quantum chemistry, including Hartree-Fock, coupled cluster singles, doubles (CCSD), and perturbative triples (CCSD(T)), as well as a  quantum chemistry formulation of the density matrix renormalization group method (DMRG). Our treatment of TBG is agnostic to gauge choices, and hence we present a new gauge-invariant formulation to detect the spontaneous symmetry breaking in interacting models.
To benchmark our approach, we focus on a simplified spinless, valleyless IBM model.
At integer filling ($\nu=0$), all numerical methods agree in terms of energy and $C_{2z} \mc{T}$ symmetry breaking.
Additionally, as part of our benchmarking, we explore the impact of different schemes for removing ``double-counting'' in the IBM model.
Our results at integer filling suggest that cross-validation of different IBM models may be needed for future studies of the TBG system.
After benchmarking our approach at integer filling, we perform the first systematic study of the IBM model near integer filling (for $\abs{\nu}< 0.2$). 
In this regime, we find that the ground state can be in a metallic and $C_{2z} \mc{T}$ symmetry breaking phase. The ground state appears to have low entropy, and therefore can be relatively well approximated by a single Slater determinant. Furthermore, we observe many low entropy states with energies very close to the ground state energy in the near integer filling regime.

\end{abstract}

\maketitle

\section{Introduction}
\label{sec:Introduction}

The correlated insulating and superconducting phases of magic angle twisted bilayer graphene (TBG) have received intense research attention in the past few years~\cite{CaoFatemiDemir2018,CaoFatemiFang2018,LuStepanovYang2019,YankowitzChenPolshyn2019,BultinckKhalafLiuEtAl2020,ChatterjeeBultinckZaletel2020,SoejimaParkerBultinckEtAl2020,XieMacDonald2020,WuSarma2020,DasLuHerzog-Arbeitman2021,BernevigSongRegnaultEtAl2021,LiuKhalafLee2021,SaitoGeRademaker2021,JiangLaiWatanabe2019,PotaszXieMacDonald2021}.
Since each moir\'e unit cell of magic angle TBG contains around ten thousand carbon atoms, to take into account electron correlations among different moir\'e unit cells, a faithful atomistic model of TBG would involve hundreds of thousands of carbon atoms. This is extremely challenging for numerical studies of TBG even at the level of tight-binding models. 
As a result, the Bistritzer-MacDonald (BM) model~\cite{BistritzerMacDonald2010}, which is a continuum tight-binding model, has been widely adopted as the starting point for further numerical studies. However, a 
non-interacting tight-binding model cannot support either the correlated insulating or the superconducting phase, and electron-electron correlations must be properly taken into account.
The BM model reveals that the flat bands of interest are energetically separated from the other bands. Therefore, as a reasonable starting point, one can study the interacting moir\'e physics ``downfolded'' to the flat bands. This gives rise to the ``interacting Bistritzer-MacDonald'' (IBM) model, which takes the form of an extended Hubbard model with pairwise long-range interactions.
Though the IBM model is not uniquely defined, and a unifying physical description of the correlated phases has yet to emerge, such a downfolding procedure has been used by a number of recent works for studying phase diagrams of TBG beyond the tight-binding approximation~\cite{BultinckKhalafLiuEtAl2020,ChatterjeeBultinckZaletel2020,SoejimaParkerBultinckEtAl2020,XieMacDonald2020,WuSarma2020,DasLuHerzog-Arbeitman2021,BernevigSongRegnaultEtAl2021,LiuKhalafLee2021,SaitoGeRademaker2021,JiangLaiWatanabe2019,PotaszXieMacDonald2021}. 

To solve the IBM model numerically, the simplest approximation is Hartree-Fock (HF) theory. The IBM model at the HF level can host a diverse range of phases due to spontaneous symmetry breaking \cite{BultinckKhalafLiuEtAl2020,KwanWagnerBultinck2021,XieMacDonald2020}.
In certain parameter regimes, the Coulomb energy scale ($10\sim 20$ meV) of the IBM model is larger than the dispersion of the flat bands ($\sim 5$ meV). Hence electron correlation effects may become significant, and post-Hartree-Fock calculations are needed in order to validate and/or correct the physical picture provided by HF theory. 
Recent studies using exact diagonalization (ED)~\cite{PotaszXieMacDonald2021, xie2021TBGVI} and the density matrix renormalization group (DMRG)~\cite{KangVafek2020, SoejimaParkerBultinckEtAl2020, parker2021strain, WangParkerSoejimaEtAl2022} suggest that HF theory provides a good approximation to the description of ground state properties of TBG at least at integer filling ($\nu=0$) (the filling parameter $\nu$ refers to the number of electrons per $\vk$-point relative to the charge neutrality point).

As compared to many other models of interacting physics, such as the Hubbard model, the IBM model includes a long-range Coulomb interaction which complicates the numerical description. 
The techniques to treat such long-range interactions are well studied in the ab initio quantum chemistry community. 
Using mature quantum chemistry software packages, it is thus possible to study a wide range of ground state and excited state properties of the IBM model at the correlated electron level.  
In addition, away from integer filling or the chiral limit, we expect in some scenarios that the HF solution can still be a good starting point for post-Hartree-Fock calculations.
This is the regime where many common quantum chemistry approximations, such as the coupled cluster approaches (see \cite{SzaboOstlund1989,ShavittBartlett2009}), often excel in terms of a balance between efficiency and accuracy. 
In particular, these approaches can be less expensive than ED or DMRG, and therefore can be applied to systems of larger sizes.
This work provides a description of the IBM model compatible with quantum chemistry language and implementations, and an initial study of the performance of various quantum chemistry methods for the ground state properties of the IBM model. 

\subsection{Overview of Implementation}
The BM Hamiltonian is defined by taking two copies of graphene, rotating them relative to each other by an angle $\theta$, and adding inter-layer coupling terms. The relative strength of this inter-layer coupling is controlled by two parameters $w_0$ and $w_1$, which control the strength of AA hopping and AB hopping, respectively. Following Ref.~\cite{SoejimaParkerBultinckEtAl2020}, we fix $\theta = 1.05$ and $w_1 = 109$ meV and vary the ratio $w_0 / w_1$ between $0$ and $0.95$. The value used in Ref.~\cite{SoejimaParkerBultinckEtAl2020} is $w_0=80$ meV, $w_1=109$ meV, which corresponds to a ratio $w_0/w_1\approx 0.73$. However, in-plane lattice relaxation, which expands AB regions and contracts AA regions~\cite{CarrFangZhuEtAl2019}, as well as out-of-plane relaxation, which increases the interlayer separation in AA regions relative to AB regions~\cite{NamKoshino2017}, could change the value of $w_0/w_1$. The limit $w_0/w_1=0$ is referred to as the chiral limit~\cite{TarnopolskyKruchkovVishwanath2019}. The (non-interacting) BM model at the chiral limit exhibits additional symmetries, which have been used extensively in the theoretical studies of the BM model (e.g., the existence of flat bands at certain magic angles)~\cite{WatsonLuskin2021,BeckerEmbreeWittsten2022,LedwithTarnpolskyKhalaf2020,VafekKang2021,BeckerHummerZworski_12022,BeckerHumbertZworski2022,BeckerEmbreeWittstenEtAl2021,TarnopolskyKruchkovVishwanath2019}.

In this work, we follow Ref.~\cite{SoejimaParkerBultinckEtAl2020} and assume that the IBM model contains only valley $K$ and spin $\uparrow$; in other words, the model is spinless and valleyless. 
This model neglects certain electron-electron interactions (even at the mean-field level) and limits the exploration of certain phases, such as the Kramers intervalley-coherent (K-IVC) state~\cite{BultinckKhalafLiuEtAl2020} in the full model. 
On the other hand, ED calculations for the full model suggest that the TBG system is often spin and valley polarized~\cite{PotaszXieMacDonald2021}. Post-Hartree-Fock calculations of the phase diagram for the IBM model with valley and spin degrees of freedom will be studied in the future. 
Our implementation is based on the Python-based Simulations of Chemistry Framework (PySCF)~\cite{SunBerkelbachBluntEtAl2018,SunZhangBanerjeeEtAl2020}. After constructing the quantum many-body Hamiltonian by means of the form factors from the BM model~\cite{BultinckKhalafLiuEtAl2020,SoejimaParkerBultinckEtAl2020}, HF and post-HF calculations, as well as calculations with integer and non-integer fillings, can be carried out on the same footing. 
Our post-HF calculations are performed using the coupled cluster singles and doubles (CCSD) method, the perturbative non-iterative energy correction to CCSD (called CCSD(T))~\cite{RaghavachariTrucksPopleEtAt1989}, and the density matrix renormalization group (DMRG)~\cite{White1992} method, in particular its quantum chemical formulation (with specific algorithmic choices designed for quartic Hamiltonians, sometimes called QC-DMRG~\cite{white1999ab,chan2002highly,chan2016matrix}) as implemented in Block2~\cite{zhai2021low}.

 
\subsection{Symmetry Breaking Order Parameters}
In order to study the phase diagram of TBG, we need to define order parameters to quantify the spontaneous symmetry breaking in the density matrix.
The order parameters are often basis dependent, and hence basis changes (sometimes called gauge fixing) tailored for each symmetry may be needed. 
For instance, one of the most important symmetries of TBG is the $C_{2z}\mc{T}$ symmetry, which characterizes the quantum anomalous Hall (QAH) state. 
The $C_{2z}\mc{T}$ order parameter is defined in the Chern band basis~\cite{BultinckKhalafLiuEtAl2020,SoejimaParkerBultinckEtAl2020,BernevigSongRegnaultEtAl2021,KangVafek2020}, which needs to be carefully constructed due to the topological obstruction in constructing the Wannier states. 
We present a new set of gauge-invariant order parameters defined using the sewing matrices~\cite{FangGilbertBernevig2012,BernevigSongRegnaultEtAl2021}, which can be applied to both unitary and antiunitary symmetries without the need for basis change. 
These gauge-invariant order parameters can therefore be computed conveniently in the band basis of the BM model, and can be used to quantify the symmetry breaking in the density matrix.
Our numerical results verify that the phase diagrams obtained from the gauge-invariant and gauge-dependent order parameters previously reported in the literature are consistent.

\subsection{Subtraction Schemes}
The construction of the BM model already implicitly takes electron interactions into consideration via the single-electron dispersion. Hence, adding an additional Coulomb interaction term to the BM model leads to double-counting errors. In the literature, there are a number of different proposals for removing the double-counting effects.
These different choices lead to model discrepancies which can be an important source of uncertainty in TBG modeling. 
We compare the average scheme (AVG)~\cite{KangVafek2020,BernevigSongRegnaultEtAl2021}, and the decoupled scheme (DEC)~\cite{BultinckKhalafLiuEtAl2020,SoejimaParkerBultinckEtAl2020,XieMacDonald2020} for removing such double-counting effects. The former defines a model that is particle-hole symmetric, and the latter uses a more physical reference density matrix. 
While the results obtained from the two IBM models qualitatively agree, important differences remain even when all other simulation parameters are reasonably converged. For instance, we find that the $C_{2z}\mc{T}$ order parameter in the average scheme is very close to an integer $1$ near the chiral limit, indicating that the system is fully polarized in the Chern basis, and the order parameter undergoes a sharp transition to $0$ around $w_0/w_1 \approx 0.8$.  In the decoupled scheme, the $C_{2z}\mc{T}$ order parameter is around $0.8$ at the chiral limit, and the order parameter changes non-monotonically as the ratio increases, the transition region becomes much wider for the same system size. 
In the average scheme, the interaction Hamiltonian at the chiral limit is positive semidefinite~\cite{BernevigSongRegnaultEtAl2021} and exhibits an enlarged $U(4)\times U(4)$ symmetry~\cite{BultinckKhalafLiuEtAl2020,BernevigSongRegnaultEtAl2021}.
The ground state energy at integer filling is zero, which can be exactly achieved using a single Slater determinant given by the HF solution. 
With the decoupled scheme, the correlation energy is nonzero even at the chiral limit and $\nu=0$, and we find that the correlation energy is generally larger than that in the average scheme. 
The differences due to model discrepancies can be even larger than post-HF electron correlation effects. 
As a result, in the absence of an interacting model for the TBG system that is fully based on first principles, we may need to investigate an ensemble of interacting models to cross-validate the results.

\subsection{Integer Versus Non-Integer Fillings}
At integer filling $(\nu=0$), i.e., the charge neutral point, we find that total energies from HF, CCSD, CCSD(T), and DMRG largely agree with each other, and correlation energies (defined as the difference between the HF energy and the post-HF energy) are generally less than $0.5$ meV per moir\'e site.
Using the gauge-invariant order parameters, our results confirm that at integer filling, the system is either in a $C_{2z}\mc{T}$ symmetry breaking and insulating state, or in a $C_{2z}\mc{T}$ trivial and metallic state~\cite{SoejimaParkerBultinckEtAl2020}.
We also perform the first systematic study of the IBM model near integer filling (for $\abs{\nu}<0.2$). In this regime, we find states that are $C_{2z}\mc{T}$ symmetry breaking and metallic.
Furthermore, the IBM model can host many states that are energetically close to the ground state, and it can be difficult to converge to the ``true'' global minima for all levels of theories. To highlight this difficulty, we explored two different initialization schemes: (1) Fixed initialization, which uses the one-particle reduced density matrix that follows that in Ref.~\citenum{SoejimaParkerBultinckEtAl2020}. (2) Random initialization, which uses a random one-particle reduced density matrix satisfying the electron number constraint.

We observe that the energy corrections provided by post-HF methods can be larger than that in the integer filling case. Although this trend agrees with the exact diagonalization calculations of the full IBM model in~\cite{PotaszXieMacDonald2021}, the quantitative magnitude of the corrections in our simulations can depend on the local minima attained at the HF level. The various local minima are not simply an artifact of the HF approximation. For example, we find also that the results of the DMRG calculations can also strongly depend on the HF orbitals, and all DMRG calculations yield solutions with relatively low Fermi-Dirac entropy, which suggests that these minima are all close to single Slater determinants, i.e. solutions that can be described relatively well by the HF approximation.

\section{Preliminaries}
\label{sec:Preliminaries}
The tight-binding models for monolayer graphene, bilayer graphene, and the BM model for twisted bilayer graphene have been extensively studied in the literature.
We therefore only provide a minimal introduction to the BM model and the wavefunctions involved; we refer the reader to e.g. Refs.~\cite{BultinckKhalafLiuEtAl2020,BernevigSongRegnaultEtAl2021} and the references therein for a more detailed discussion.
Throughout this paper, we adopt atomic units, except for energies which are reported in millielectron volts.

Recall that the BM model depends on two parameters $w_0$ and $w_1$, which control the strength of AA hopping and AB hopping respectively. Through this article we fix $\theta = 1.05$, $w_1 = 109$ meV and vary the ratio $w_0 / w_1$ between $0$ and $0.95$. 
Subsequently, we denote the moir\'e unit cell by $\Omega$, its area by $|\Omega|$, and the moir\'e Bravais lattice by $\mathbb{L}$. 
Correspondingly, we denote the moir\'e Brillouin zone (mBZ) by $\Omega^*$ and the moir\'e reciprocal lattice $\mathbb{L}^*$. 
The mBZ is discretized using a Monkhorst-Pack (MP) grid~\cite{MonkhorstPack1976} of size $N_{\vk}=n_{k_x}\cdot n_{k_y}$.
When the MP grid includes the $\Gamma$-point of the mBZ, the computation can be identified with a moir\'e supercell consisting of $N_{\vk}$ unit cells with a sample area $N_{\vk}\abs{\Omega}$. 
A given BM wavefunction, i.e., a BM band, can be labeled by a tuple $(n,\vk,s,\tau)$, where $n$ is the band index, $\vk\in \mathrm{mBZ}$ is the $\vk$-point index, $s\in\{\uparrow,\downarrow\}$ is the spin index, and $\tau\in\{\vK,\vK'\}$ is the valley index. 
Since the spin and valley indices often do not appear explicitly in the Hamiltonian, they are also referred to as \textit{flavor indices}. 
For simplicity, we follow the assumption in Ref.~\cite{SoejimaParkerBultinckEtAl2020}, and drop the flavor indices $s,$ $\tau$, i.e., the system is spinless and valleyless.

Let $\vr$ be the real space index in the moir\'e supercell, by Bloch's theorem, we can express a BM orbital in real space as 
\begin{equation}
\psi_{n\vk}(\vr,\sigma,l)
=\frac{1}{\sqrt{N_{\vk}}}e^{\I \vk\cdot\vr}u_{n\vk}(\vr,\sigma,l)
=\frac{1}{\sqrt{N_{\vk}}\abs{\Omega}}\sum_{\vG} e^{\I(\vk+\vG)\cdot \vr}u_{n\vk}(\vG,\sigma,l).
\end{equation}
Here  $\vG\in \mathbb{L}^*$ is the plane-wave index, $\sigma\in\{A=1,B=-1\}$ is the sublattice index, $l\in\{1,-1\}$ is the layer index. We  also refer to $(\vG,\sigma,l)$ or $(\vr,\sigma,l)$ as \textit{internal indices}. 
Note that  $u_{n\vk}(\vr,\sigma,l)$ is periodic with respect to $\mathbb{L}$, i.e., $u_{n\vk}(\vr,\sigma,l)=u_{n\vk}(\vr+\vR,\sigma,l), \forall \vR\in\mathbb{L}$.
The normalization condition is chosen such that $u_{n\vk}$ is normalized within the moir\'e unit cell.
Moreover, the factor $\frac{1}{\sqrt{N_{\vk}}}$ ensures that $\psi_{n\vk}$ is normalized within the moir\'e supercell. 
With some abuse of notation, we use $u_{n\vk}(\vr,\sigma,l)$ and $u_{n\vk}(\vG,\sigma,l)$ to denote the coefficients of a BM wavefunction in real space and reciprocal space, respectively. 
In practical calculations, the number of plane-wave indices $\vG$ needs to be truncated to a finite size. 
Throughout this article, we omit the range of summation unless otherwise specified.
Subsequently, we refer to the set of all plane waves indexed by $\vG$ with sublattice index $\sigma$ and layer index $l$ as the primitive basis of the BM model and denote the corresponding Fock space by $\mc{F}$.
Let $\hat{c}_{\vk}^{\dagger}(\vG,\sigma,l),\hat{c}_{\vk}(\vG,\sigma,l)$ be the creation and annihilation operators acting on $\mc{F}$, respectively. 
Then the creation and annihilation operators corresponding to the band $n\vk$ are
\begin{equation}
\label{eqn:creation-op-def}
\begin{split}
\hat{f}^{\dag}_{n\vk} 
=&\sum_{\vG,\sigma,l}\hat{c}_{\vk}^{\dagger}(\vG,\sigma,l)u_{n\vk}(\vG,\sigma,l)
,\\
\hat{f}_{n\vk} 
=&\sum_{\vG,\sigma,l}\hat{c}_{\vk}(\vG,\sigma,l)u^*_{n\vk}(\vG,\sigma,l).
\end{split}
\end{equation}
Here $u_{n\vk}^*$ denotes the complex conjugation of $u_{n\vk}$.
The band creation and annihilation operators satisfy the canonical anticommutation relation, i.e.,  $\{\hat{f}^{\dag}_{n\vk},\hat{f}_{n'\vk'}\}
= \delta_{nn'} \delta_{\vk\vk'}$, and define the band basis of the BM model.
Note that the definition of the band creation and annihilation operators can be periodically extended outside the mBZ according to 
\begin{equation}
\hat f^{\dag}_{n(\vk+\vG)}=\hat f^{\dag}_{n\vk}, \quad \hat f_{n(\vk+\vG)}=\hat f_{n\vk}, \quad \vG\in\mathbb{L}^*.
\label{eqn:period_f}
\end{equation}

\section{Interacting Bistritzer-MacDonald model} 
\label{sec:ibm_model}

\begin{figure*}
\begin{center}
\begin{subfigure}[c]{.3\textwidth}
    \includegraphics[width=\textwidth]{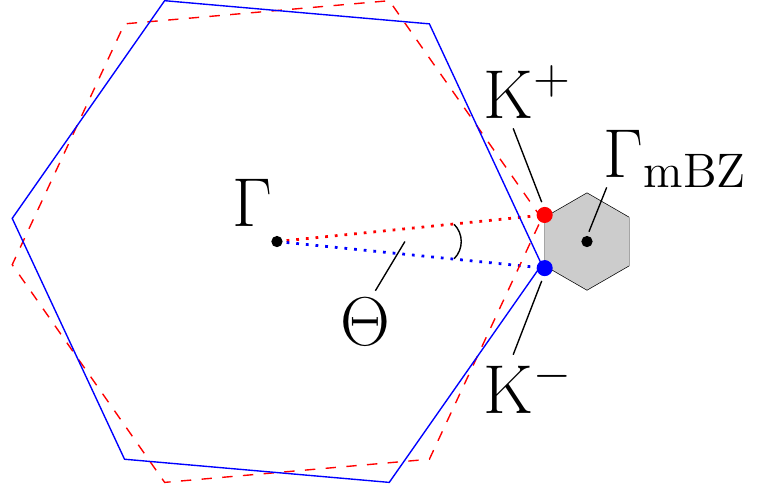} 
\end{subfigure}
\hfill
\begin{subfigure}[c]{.3\textwidth}
    \includegraphics[width=\textwidth]{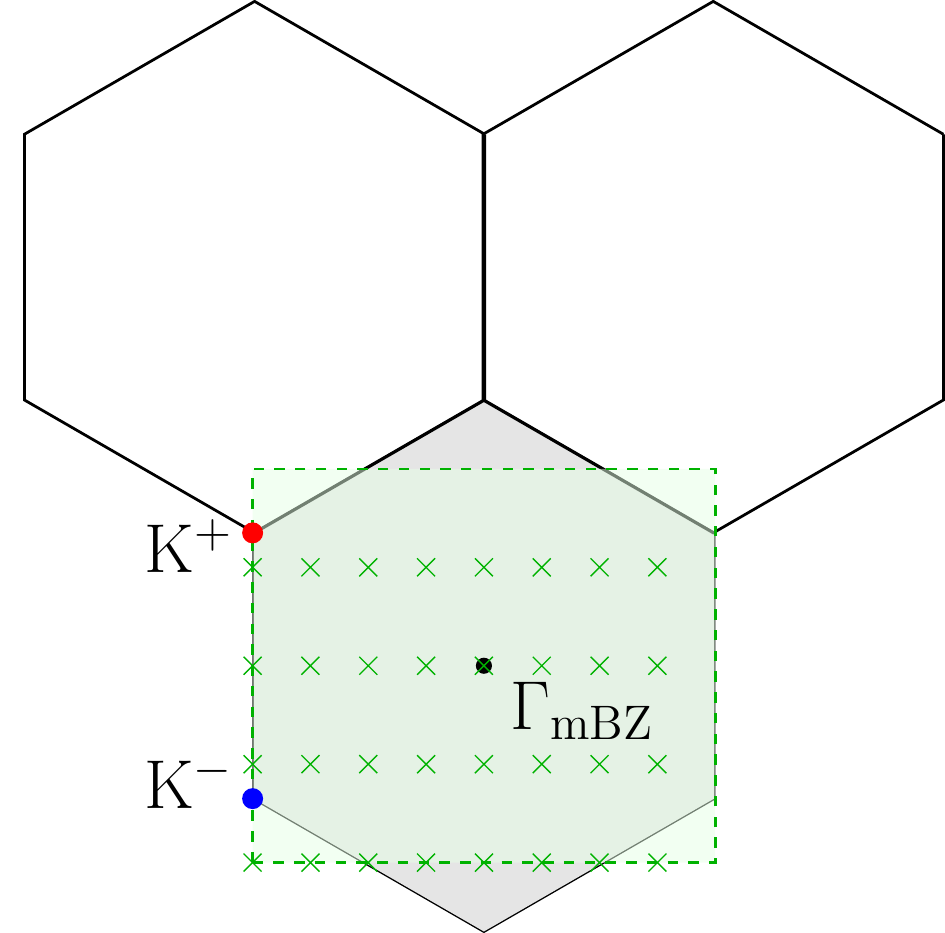} 
\end{subfigure}
\hfill
\begin{subfigure}[c]{.3\textwidth}
    \includegraphics[width=\textwidth]{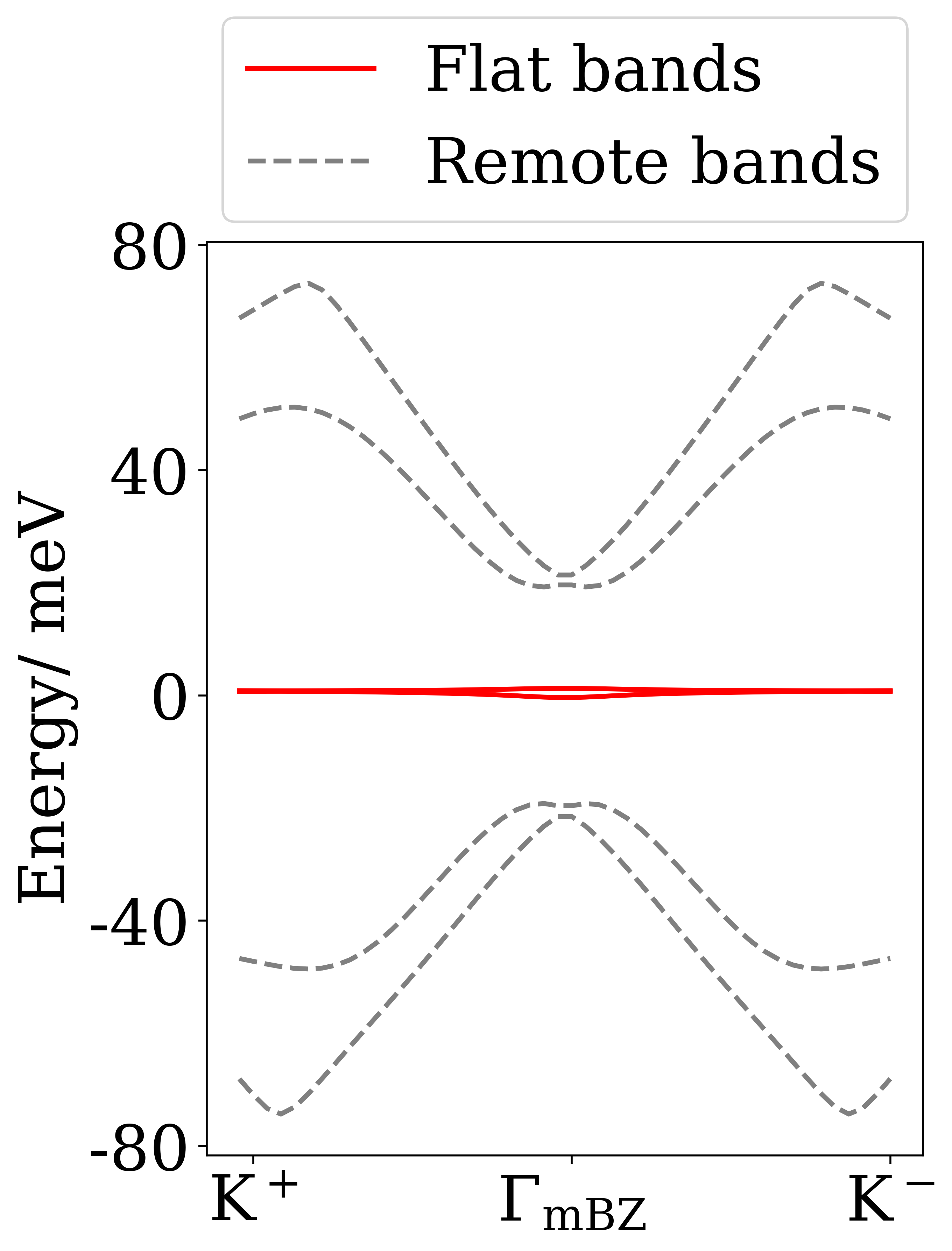}
\end{subfigure}
\begin{subfigure}[c]{.3\textwidth}
    \caption{}
    \label{fig:tbg_layers}
\end{subfigure}
\hfill
\begin{subfigure}[c]{.3\textwidth}
    \caption{}
    \label{fig:mbz_num_lattice}
\end{subfigure}
\hfill
\begin{subfigure}[c]{.3\textwidth}
    \caption{}
    \label{fig:flat_bands}
\end{subfigure}
\end{center}
\caption{
\raggedright(\subref{fig:tbg_layers}) Two monolayer graphene Brillouin zones (BZ) depicted by a dashed red line and solid blue line, respectively, aligned by their $\Gamma$-point and twisted by an angle $\Theta$, with the corresponding Dirac points $K^+$ and $K^-$. The moir\'e Brillouin zone (mBZ) is indicated by the grey shaded region centered at $\Gamma_{\rm mBZ}$. 
(\subref{fig:mbz_num_lattice}) Choice of the rectangular unit cell in reciprocal space (green shaded area encircled by green dashed line, see Ref.~\cite{SoejimaParkerBultinckEtAl2020}) relative to the mBZ. The green crosses show a mBZ discretization grid of $8\times 4$ $\vk$-points. 
The Dirac points of the monolayers $K^+$ and $K^-$ are included as reference points. 
(\subref{fig:flat_bands}) Band structure of the BM Hamiltonian over the mBZ, with the corresponding flat bands (solid red lines) and remote bands (dashed grey lines). The interacting BM Hamiltonian is then projected onto the subspace spanned by the two flat bands; the system's parameters are $\theta = 1.05^{\circ}$, $w_1 = 109$ meV and $w_0/w_1=0.7$. 
}

\end{figure*}

For values of the ratio $w_0 / w_1 \in [0, 0.95]$, the spinless, valleyless BM Hamiltonian has a direct gap between two bands with roughly zero energy and the remainder of the spectrum (see e.g., \cref{fig:flat_bands} for $w_0/w_1=0.7$).
We refer to these two bands as the flat bands of the BM model and label them by the index $n\in\{-1,1\}$.
The Hamiltonian of the IBM model restricted to these flat bands takes the form
\begin{equation}
\begin{split}
\hat{H}_{\mathrm{IBM}}
=& \hat{H}_0+\hat{H}_I\\
=&\sum_{\vk\in \Omega^*} \sum_{mn}\hat f_{m\vk}^{\dagger} [h(\vk)]_{mn} \hat f_{n\vk}\\
&+\frac{1}{2}
\sum_{\substack{\vk,\vk',\vk''\in \Omega^*\\\vk''' = \vk+\vk'-\vk''}} \sum_{mm'nn'}\braket{m\vk, m'\vk'|n\vk'', n' \vk'''}\hat f_{m\vk}^{\dagger} \hat f_{m'\vk'}^{\dagger} \hat f_{n'\vk'''} \hat f_{n\vk''},
\end{split}
\label{eqn:IBM_Ham}
\end{equation}
where $\hat{H}_0$ and $\hat{H}_I$ are the quadratic term and the quartic term, respectively. 
The ground state of the IBM model is then defined as the solution to the minimization problem
\begin{equation}
E_0=\min_{\substack{\ket{\Psi}\in\mc{F}, ~\braket{\Psi|\Psi}=1\\\braket{\Psi|\hat{N}|\Psi}=N_e}} \quad \braket{\Psi|\hat{H}|\Psi},
\end{equation}
where $N_e=(\nu+1) N_{\vk}$ is the total number of electrons, and $\hat{N}=\sum_{\vk}\sum_{n} \hat f_{n\vk}^{\dagger} \hat f_{n\vk}$ is the total number operator.
The number of electrons per $\vk$-point is given by $\nu+1$ and we subsequently refer to $\nu$ as the \textit{filling factor}. 
Note that in this convention, the particle filling is reported with respect to the charge neutral point.
Since there are only two bands per $\vk$-point, the only non-trivial integer value for the filling factor is $\nu=0$, which is also called the integer filling case (or the particle-hole symmetric case) of the IBM model in the spinless, valleyless regime.

The main object of interest in this work is the one-particle reduced density matrix (1-RDM) corresponding to the ground state $\ket{\Psi}$ defined as
\begin{equation}
\label{eqn:1rdm}
[P(\vk)]_{nm}=\braket{\Psi|\hat{f}_{m\vk}^{\dagger}\hat{f}_{n\vk}|\Psi}.
\end{equation}
We emphasize that the 1-RDM is well-defined in the entire moir\'e reciprocal space due to the periodic extension in \cref{eqn:period_f}.
Using the 1-RDM, we find that for any $\vk$, $\vk'$ in the moir\'e reciprocal space,
\begin{equation}
\braket{\Psi|\hat{f}_{m\vk}^{\dagger}\hat{f}_{n\vk'}|\Psi}
=P(\vk)_{nm} \sum_{\vG\in\mathbb{L}^*}\delta_{\vk,\vk'+\vG}.
\label{eqn:average_f}
\end{equation}

The quartic term $\hat{H}_I$ describes the (screened) Coulomb interaction via the two-electron repulsion integrals (ERI) denoted by $\braket{m\vk, m'\vk'|n\vk'', n' \vk'''}$.
The coefficients of the quadratic term can be written as
\begin{equation}
h(\vk)=h_{\mathrm{BM}}(\vk)-h_{\mathrm{sub}}(\vk),
\label{eqn:ibm_quadratic}
\end{equation}
where $[h_{\mathrm{BM}}(\vk)]_{mn}=\varepsilon^{\mathrm{BM}}_n(\vk) \delta_{mn}$ is given by the BM band energy. The second term $h_{\mathrm{sub}}(\vk)$ is called the subtraction Hamiltonian, which removes the double-counting of the Coulomb interaction within the flat bands, and is defined in terms of the Hartree-Fock potential (see \cref{sec:subtraction}). 
The derivation of the Coulomb interaction term is presented in \cref{sec:ham_i}. 

\section{Computational methods and implementation}

\subsection{Hartree-Fock Theory}

The Hartree-Fock approximation is the starting point for various correlated 
electronic-structure methods~\cite{Helgaker2014}. 
The underlying assumption is that the many-body wavefunction takes the form of a Slater determinant, i.e., 
\begin{equation}
\ket{\Psi_S}=\prod_{\vk}\prod_{i\in\mathrm{occ}} \hat b_{i\vk}^{\dag}\ket{\mathrm{vac}},
\end{equation}
where $\ket{\mathrm{vac}}$ is the vacuum state, and 
\begin{equation}
\hat b_{i\vk}^{\dag}=\sum_{n}\hat f_{n\vk}^{\dag} \Xi_{ni}(\vk)
\end{equation}
defines the creation operator for the Hartree-Fock orbitals for each $\vk\in\Omega^*$. 

For integer filling, the number of occupied orbitals per $\vk$-point is $\nu+1$ (indexed by $\mathrm{occ}$). 
The 1-RDM associated with a given Slater determinant $\ket{\Psi_S}$ is then
\begin{equation}
[P(\vk)]_{nm}=\braket{\Psi_S|\hat{f}_{m\vk}^{\dagger}\hat{f}_{n\vk}|\Psi_S}=
\sum_{i\in\mathrm{occ}} \Xi_{ni}(\vk)\Xi_{mi}^*(\vk).
\end{equation}
Following the standard derivation of Hartree-Fock theory (see e.g.,~\cite{SzaboOstlund1989,MartinReiningCeperley2016}), we begin with the characterization of the Hartree-Fock energy, i.e., 
\begin{equation}
\label{eqn:e_hf}
E_{\rm HF}
=
\min_{P\in \mathcal{M}}{\rm Tr}[PH]
=
\min_{P\in \mathcal{M}} \mathcal{E}^{\rm (HF)}(P),
\end{equation}
where $\mathcal{M}$ is the set of 1-RDMs associated with the possible single Slater determinants of the system.
A common way to seek the solution to~\cref{eqn:e_hf} is by finding a stationary point of $\mathcal{E}^{\rm (HF)}$, which is equivalent to diagonalizing the Fock operator~\cite{SzaboOstlund1989} $\hat{F}[P]= \hat{H}_0+ \hat{V}_{HF}[P]$, where $\hat{V}_{HF}[P]$ is the Hartree-Fock potential.

The Hartree-Fock potential can be written in terms of the so-called \textit{form factor} matrix, $\Lambda_{\vk}(\vq+\vG)$. Simply speaking, the form factor is given by the Fourier coefficients of the pair product of the periodic Bloch functions of the BM model $\{ u_{n\vk} \}$. 
This matrix is calculated via the following formula (see \cref{eqn:form_factor}): 
\begin{equation}
[\Lambda_{\vk}(\vq+\vG)]_{mn} =\frac{1}{\abs{\Omega}}
\sum_{\vG'\in\mathbb{L}^*} \sum_{\sigma,l}u^*_{m\vk}(\vG',\sigma,l) u_{n(\vk+\vq+\vG)}(\vG',\sigma,l).
\end{equation}
With this definition, the Hartree-Fock potential takes the compact form
\begin{equation}
\hat{V}_{\mathrm{HF}}[P]
=\hat{J}[P]- \hat{K}[P]
=\sum_{\vk\in \Omega^*} \hat f_{m\vk}^{\dagger}[v_{\mathrm{hf}}[P](\vk)]_{mn}
\hat f_{n\vk},
\end{equation}
where the matrix elements are given by
\begin{equation}
\label{eqn:hf-rho}
\begin{split}
[v_{\mathrm{hf}}[P](\vk)]_{mn}=&\frac{1}{|\Omega|} 
\sum_{\vG\in\mathbb{L}^*} V(\vG) \left(\frac{1}{N_{\vk}}\sum_{\vk'\in \Omega^*} \Tr[ \Lambda_{\vk'}(-\vG) P(\vk')]\right)[\Lambda_{\vk}(\vG)]_{mn}\\
&-\frac{1}{ |\Omega| N_{\vk}} \sum_{\vq'} \sum_{m'n'}
 V(\vq') [\Lambda_{\vk}(\vq')]_{mn'} [P(\vk+\vq')]_{n'm'}[\Lambda_{\vk+\vq'}(-\vq')]_{m'n}.
\end{split}
\end{equation}
We here employ the quantum chemistry notation where $H_0$ is the core Hamiltonian, $\hat{J}[\cdot]$ and $\hat{K}[\cdot]$ are the Coulomb and exchange operators, respectively. 
For completeness, the derivation of the expressions of $\hat{J}[\cdot]$ and $\hat{K}[\cdot]$ are given in \cref{sec:hf_deriv}.
This non-linear eigenvalue problem is then determined by self-consistently evaluating the 1-RDM~\cite{Lehtola2020}.

In quantum chemistry discussions of Hartree-Fock theory, it is also common to require that $\hat{f}_{nk}^\dag$ commutes with the electronic spin operator $\hat{S}_z$. When no such restriction is used, the theory is termed \textit{generalized} Hartree-Fock theory (GHF). In the current treatment, the electronic spin is fully polarized. However, there is a pseudospin variable, namely the sublattice index $\sigma$. We will have no restriction that $\hat{f}_{nk}^\dag$ commutes with the sublattice pseudospin operator. Thus we will refer later to carrying out GHF calculations, in the sense of no restriction on the pseudospin.

%

\subsection{Subtraction Hamiltonian}\label{sec:subtraction}
 

Since the BM band energies already take the electron-electron interaction between the two layers of graphene into account, the screened Coulomb potential in the IBM model would double count such interactions.
As a remedy, one can introduce a subtraction Hamiltonian, see~\cref{eqn:ibm_quadratic}.
At the level of Hartree-Fock theory~\cite{BultinckKhalafLiuEtAl2020}, this subtraction Hamiltonian can be evaluated by means of the Hartree-Fock potential~\cref{eqn:hf-rho} with respect to an {\it a priori} chosen reference density $P^0$, i.e., $h_{\mathrm{sub}}(\vk)=v_{\mathrm{hf}}[P^0](\vk)$.
Then, since the mapping $P \mapsto v_{\mathrm{hf}}[P]$ is linear, the Fock operator including the subtraction part, denoted $\hat{F}_{\rm sub}$, can be written as

\begin{equation}
\label{eqn:subtracted_HF}
\hat{F}_{\rm sub}[P(\vk)]
=
\hat{F}[P(\vk)]-\hat{V}_{\rm HF}[P^0(\vk)]
=
\sum_{\vk\in\rm{mBZ}}\sum_{mn}\hat f^{\dagger}_{m\vk}[h_{\mathrm{BM}}(\vk)]_{mn}
\hat f_{n\vk} + \hat{V}_{\mathrm{HF}}[\delta P(\vk)],
\end{equation}
where $\delta P(\vk)=P(\vk)-P^{0}(\vk)$.

Following \cref{eqn:1rdm}, $P^0(\vk)$ should be interpreted as the coefficients of the reference density matrix in the BM band basis.
The choice of a reference density matrix $P^0(\vk)$ is not unique and should also be viewed as part of the IBM model.
For instance, in Ref.~\cite{BernevigSongRegnaultEtAl2021}, the choice is \begin{equation}
P^0(\vk)=\frac{1}{2}I,
\end{equation}
which is called the \textit{average scheme}. 
In Ref.~\cite{XieMacDonald2020,BultinckKhalafLiuEtAl2020,PotaszXieMacDonald2021}, $P^0(\vk)$ is obtained by projecting the zero temperature limit of the density matrix corresponding to two decoupled graphene layers to the BM band basis of the TBG system; this is referred to as the \textit{decoupled scheme}.
In the computational simulations presented here, we follow the procedure used in Ref.~\cite{SoejimaParkerBultinckEtAl2020}, where terms from the frozen negative energy sea of the BM model are not included in the decoupled subtraction Hamiltonian (cf.~\cite[Eq. 2]{PotaszXieMacDonald2021}). For additional details on the different decoupled schemes used in the literature, see~\cref{sec:dec_subtraction}.


The zero temperature limit ensures that $P^0$ is uniquely defined even if some of the band energies of the two decoupled graphene layers may become degenerate. Furthermore, the choice of $P^0$ is only used to define the quadratic part of the IBM Hamiltonian and is assumed to be independent of the filling factor $\nu$. Unless otherwise specified, we adopt the decoupled scheme in all calculations.



\subsection{Coupled Cluster Theory}

Coupled cluster theory is one of the most widely used post-Hartree-Fock correlated wavefunction methods in quantum chemistry~\cite{BartlettMusial2007,Helgaker2014}. 
In this ansatz, the ground-state wavefunction takes the form
\begin{equation}
\label{eqn:ExponentialParametrization}
|\Psi \rangle 
=
e^{T(\vt)}|\Phi_0 \rangle,
\end{equation}
where 
\begin{equation}
\label{eqn:ClusterOperator}
T(\vt)=\sum_\mu t_\mu X_\mu
\end{equation}
is the cluster operator determined by the cluster amplitudes $\vt$, and $|\Phi_0 \rangle$ is a chosen reference Slater determinant (most commonly the Hartree-Fock solution). 
The operators $X_\mu$ are the excitation operators with respect to the chosen reference $|\Phi_0\rangle $, i.e., 
\begin{equation}
\label{eqn:ExcitationOperator}
X_\mu
=
X_{i_1,...,i_k}^{a_1,...,a_k}
=
\hat{a}_{a_1}^\dagger...\hat{a}_{a_k}^\dagger \hat{a}_{i_k}...\hat{a}_{i_1},
\end{equation}
where, for the sake of compactness, we have combined the occupied orbital indices $\{i_l\}$ and virtual orbital indices $\{a_l\}$ in the multi-index $\mu$. 
The ground-state energy can then be computed as
\begin{equation}
\label{eqn:CCEnergy}
\mathcal{E}(\vt)
=
\langle\Phi_0|e^{-T(\vt)} H e^{T(\vt)}| \Phi_0 \rangle.
\end{equation}
The cluster amplitudes $\vt$ are determined by the coupled cluster equations, i.e., a set of polynomial equations of at most degree four (given at most quartic terms in the Hamiltonian) with respect to $\vt$: 
\begin{equation}
\label{eqn:CCEquations}
0
=
F_\mu(\vt)
=
\langle
\Phi_0|X_\mu^\dagger e^{-T(\vt)}H e^{T(\vt)} |\Phi_0
\rangle,\qquad \forall \mu.
\end{equation}
More compactly,~\cref{eqn:CCEnergy,eqn:CCEquations} can be combined in the coupled cluster Lagrangian  
\begin{equation}
\label{eqn:CCLagrangian}
\mathcal{L}(\vt, \boldsymbol\lambda)
=
\mathcal{E}(\vt)+\langle \boldsymbol\lambda, \vF(\vt)\rangle
=
\langle \Phi_0|(I+\Lambda(\boldsymbol\lambda))e^{-T(\vt)}H e^{T(\vt)} |\Phi_0 \rangle,
\end{equation}
where 
\begin{equation}
\label{eqn:LambdaClusterOperator}
\Lambda(\boldsymbol\lambda)
=
\sum_\mu \lambda_\mu X^\dagger_\mu.
\end{equation}
The states $|e^{T(\vt)} \Phi_0 \rangle$ and $\langle \Phi_0(I+\Lambda(\lambda))e^{-T(\vt)}|$ are commonly referred to as the right and left coupled cluster solutions, respectively.
The $N$-RDM in coupled cluster theory is given by 
\begin{equation}
\label{eqn:ccNrdm}
\varrho_{\rm CC}(\vt, \boldsymbol\lambda)
=
|e^{T(\vt)} \Phi_0 \rangle\langle
\Phi_0(I+\Lambda(\boldsymbol\lambda))e^{-T(\vt)}|
\end{equation}
ensuring that ${\rm Tr}[H\varrho_{\rm CC}(\vt, \boldsymbol\lambda)]= \mathcal{E}(\vt)$.
The corresponding 1-RDM is then given by
\begin{equation}
\label{eqn:cc1rdm}
[P_{\rm CC}(\vt, \boldsymbol\lambda)]_{p, q}
=
\langle \Phi_0(I+\Lambda(\boldsymbol\lambda))e^{-T(\vt)}|
a_p^\dagger a_q
|e^{T(\vt)} \Phi_0 \rangle,
\end{equation}
see~\cite{Helgaker2014} for more details.
We emphasize that this ansatz, in its untruncated form, is equivalent to the full configuration interaction method (i.e., the exact diagonalization method)~\cite{Monkhorst1977,Schneider2009,LaestadiusFaulstich2019}, and is thus computationally infeasible for large systems. 
In the past decades, different levels of approximation have been suggested to reduce computational complexity (see e.g.,~\cite{BartlettMusial2007,ChanKallayGauss2004,Piecuch2010,McClainSunChanEtAl2017}). The variant used in the subsequent simulations (and arguably one of the most widely used approximate versions of coupled cluster theory) is the truncation of the cluster operator in~\cref{eqn:ClusterOperator} to only contain one-body and two-body excitations, also known as the coupled cluster singles and doubles (CCSD) method.
Note that due to the exponentiation of the cluster operator, the corresponding wavefunction expansion in~\cref{eqn:ExponentialParametrization} will still contain contributions from higher excited determinants.
One of the central benefits of the exponential ansatz is that it ensures that the energy is size consistent and size extensive, in particular, for (rank complete) truncations of $T$ such as in CCSD~\cite{Helgaker2014}.
As in the Hartree-Fock discussion above, we place no restrictions on the (pseudo)spin properties of the excitation operators. Thus we work with the generalized CC ansatz in this work. 

Aside from steering the accuracy of the CC approach directly through truncations of the cluster operator in~\cref{eqn:ClusterOperator}, great effort has been put into developing methods that improve the CCSD energy by means of simple, state selective, non-iterative energy corrections that, when added to the CCSD energy, 
improve the energy of the electronic states of interest~\cite{RaghavachariTrucksPopleEtAt1989,PiecuchKowalskiPimientaEtAt2002,LochLodriguitoPiecuchEtAl2006,PiecuchKowalskiPimientaEtAl2004,JankowskiPaldusPiecuch1991,KowalskiPiecuch2000,FanKowalskiPiecuch2005}.
This includes the CCSD(T)~\cite{RaghavachariTrucksPopleEtAt1989} method, which yields a perturbative non-iterative energy correction that accounts for the effect of triexcited clusters (i.e., triples) using arguments based on the many-body perturbation theory. 

\subsection{Implementation in PySCF}
\label{sec:implementation-PySCF}

We use the Python-based Simulations of Chemistry Framework (PySCF)~\cite{SunBerkelbachBluntEtAl2018, SunZhangBanerjeeEtAl2020} to perform calculations for the IBM model in \cref{eqn:IBM_Ham}, which can be defined as a ``customized Hamiltonian'' accessed through the one- and two-electron integrals referred to as \texttt{h1e} and \texttt{eri}.
These integrals are complex-valued, therefore, minor adjustments to PySCF need to be made to enable calculations using these customized Hamiltonians.
We also use the ``molecular'' formulation in PySCF, i.e., the \texttt{h1e} and \texttt{eri} are stored without taking advantage of the $\vk$-point symmetry~\cite{McClainSunChanEtAl2017}. 
This can increase the storage cost by a factor of $N_\vk$, and the computational cost by a polynomial of $N_\vk$. 
Interfacing the $\vk$-point symmetry (periodic boundary condition or ``pbc'') modules of PySCF is possible and is left here for future work.


Once \texttt{h1e} and \texttt{eri} are constructed, the PySCF software package allows us to perform GHF and GCCSD calculations on the same footing with a simple code structure. Here is an example:

\begin{lstlisting}[language=Python, caption=Example code running GHF and GCCSD in PySCF from precomputed integrals.]
from PySCF import gto, scf, cc 

def get_veff(mol, dm, *args):       
    vj, vk= scf.hf.dot_eri_dm(eri, dm)
    return vj- vk                  
                                    
mol= gto.M()                       
mol.incore_anyway= True            
mol.nelectron= nelec 
                                    
ghf_mf= scf.GHF(mol)               
ghf_mf.get_hcore= lambda *args: h1e
ghf_mf.get_ovlp= lambda *args: ovlp
ghf_mf._eri= eri
ghf_mf.get_veff= get_veff          
                                    
# Running GHF
ghf_mf.kernel()

# Running GCCSD
gcc = cc.GCCSD(ghf_mf)  
gcc.kernel()            
\end{lstlisting}

After the calculations, PySCF also provides compact instructions to evaluate the 1-RDMs so that we can evaluate the observables to detect the symmetry breaking in  Section~\ref{sec:symm}. 

Similarly, the \texttt{h1e} and \texttt{eri} objects may be saved and used to define the Hamiltonian for the Block2 program for a QC-DMRG calculation (DMRG calculations can be performed directly through a PySCF interface). Thus DMRG calculations can be used to assess the same ground state as targeted by the HF and CC calculations, and we will use such results for benchmarking in this work. Further details of the DMRG calculations are provided in Appendix~\ref{sec:dmrg}.

\section{Symmetries} \label{sec:symm}
Both the BM and IBM models satisfy a number of symmetries which have been used extensively to analyze the properties of both models, particularly in the chiral limit~\cite{BultinckKhalafLiuEtAl2020,BernevigSongRegnaultEtAl2021}.
For TBG, the symmetries of interest are point-group symmetries, time-reversal symmetry, and their compositions.
Point-group symmetries are unitary and time-reversal symmetry is antiunitary.  
Some relevant symmetries in the valley and spin-polarized BM and IBM models are summarized in~\cref{tab:symmetry}.

\begin{table}[ht]
\begin{center}
\renewcommand{\arraystretch}{1.2} 
\begin{tabular}{l|l|l|l}
\hline Symmetry & Real space & Momentum space &Type \\
\hline$C_{2 z}$ & swap sublattice & swaps valleys; \\
& & $\mathbf{k} \rightarrow-\mathbf{k}$ & Unitary\\
\hline$C_{3z}$ & rotate by $120^{\circ}$ & $\mathbf{k} \rightarrow C_{3z}\mathbf{k}$ & Unitary\\
\hline $\mathcal{T}$ & & swaps valleys; \\
& & $\mathbf{k} \rightarrow-\mathbf{k}$ & Antiunitary\\
\hline $C_{2 z} \mc{T}$ & swap sublattice & $\vk\to\vk$ & Antiunitary\\
\hline
\end{tabular}
\end{center}
\caption{Some relevant symmetry operations for the spinless, valleyless IBM model. }
\label{tab:symmetry}
\end{table}

In this section, we propose a set of gauge-invariant order parameters which can be used to detect spontaneous symmetry breaking in the 1-RDM $P(\vk)$.
Our final results are summarized in~\cref{tab:symm-breaking}.
We defer proofs of the claims given in this section to \cref{sec:sewing_matrix,sec:symm_order_param}.

\subsection{Detecting Symmetry Breaking: Unitary Case} \label{sec:detect_symm_unitary}
We begin by considering the simpler case of unitary symmetries. For a point-group symmetry $g$, due to the properties of the Bloch transform, there exists a unitary $D(g)$, called the \textit{representation matrix}, so that the creation operators, $c_{\vk}^\dagger$, transform via the rule
\begin{equation}
\begin{split}
    (g \hat{c}_{\vk}^\dag g^{-1})(\alpha) & = \sum_{\alpha'} \hat{c}_{g\vk}^\dag(\alpha') [D(g)]_{\alpha',\alpha}, \\
    (g \hat{c}_{\vk} g^{-1})(\alpha) & = \sum_{\alpha'} \hat{c}_{g\vk}(\alpha') [D(g)]_{\alpha',\alpha}^{*}. 
  \end{split}
\end{equation}
For instance, $C_{3z}$ is a unitary symmetry, and it maps $\vk$ to $C_{3z}\vk$. Its representation matrix in the primitive basis can be written as
\begin{equation}
[D(C_{3z})]_{\vG'\sigma' l',\vG\sigma l}=\delta_{\vG',C_{3z}\vG} (e^{\I \frac{2\pi}{3} \sigma_z})_{\sigma',\sigma}\delta_{l',l}.
\end{equation}

Since the IBM model is defined in terms of the band creation operators, $\{f_{n\vk}^\dagger\}$, we need to determine how the symmetry $g$ acts on the band creation operators.
The object which encodes this symmetry action is known as the \textit{sewing matrix}~\cite{FangGilbertBernevig2012,BernevigSongRegnaultEtAl2021}.
Given a set of bands $\{ u_{n\vk} \}$ and a unitary symmetry operation $g$, the sewing matrix $[B(g)]_{\vk}$ is defined as:
\begin{equation}
  \label{eqn:unitary_sewing}
  [B(g)]_{\vk,mn} := \braket{u_{m,g\vk} | D(g) | u_{n\vk}}.
\end{equation}
Assuming $[B(g)]_{\vk}$ is unitary, the band creation operators transform under $g$ by the rule (see~\cref{sec:sewing_unitary_case}):
\begin{equation}
\label{eqn:sewing_transformation}
  \begin{split}
    g \hat{f}^{\dag}_{n\vk} g^{-1} & = \sum_{m} \hat{f}^{\dag}_{m,g\vk}  [B(g)]_{\vk,mn}, \\
    g \hat{f}_{n\vk} g^{-1} & = \sum_{m} \hat{f}_{m,g\vk}  [B(g)^\dagger]_{\vk,mn}.
  \end{split}
\end{equation}
The unitarity of $[B(g)]_{\vk}$ is satisfied when the energy bands $\{ u_{n\vk} \}$ are isolated, i.e., there is an energy gap between the chosen bands and the rest of the energy bands (\cref{sec:sewing_matrix_unitary}).

Using this transformation rule and recalling that the 1-RDM for a state $\ket{\Psi}$ is defined by $[P(\vk)]_{mn} = \braket{\Psi | \hat f_{n\vk}^\dagger \hat f_{m\vk} | \Psi}$, we can conclude that if the following commutator-like quantity
\begin{equation}
  \label{eqn:order_parameter_unitary}
  \begin{split}
    \mc{C}_{\vk}(g) & = \norm{[B(g)]_{\vk}^{\dag}P(g\vk)[B(g)]_{\vk}-P(\vk)} \\ 
    &= \norm{P(g\vk)[B(g)]_{\vk}-[B(g)]_{\vk}P(\vk)}
   \end{split}
\end{equation}
does not vanish, then the 1-RDM breaks the symmetry $g$.
Here $\norm{\cdot}$ can be any unitarily invariant norm. 
Additionally, it can be shown that $\mc{C}_{\vk}(g)$ is invariant under gauge transformations of the band creation operators (see~\cref{sec:symm_order_param_unitary}).

\subsection{Detecting Symmetry Breaking: Antiunitary Case}
We now turn to consider the case of antiunitary symmetries.
Any antiunitary symmetry $\wt{g}$ can be written as $\wt{g} = g \mc{K}$. Here $g$ is a unitary symmetry and $\mc{K}$ is complex conjugation satisfying $\mc{K}(a\ket{\vG,\sigma,l})=a^*\ket{\vG,\sigma,l}$ for any $a\in\CC$.
For an antiunitary symmetry $g \mc{K}$, we define the representation matrix as $D(g \mc{K}):=D(g)$. For instance, $C_{2z}\mc{T}$ is an antiunitary symmetry.
It satisfies $(C_{2z}\mc{T})\vk=\vk$, and its representation matrix in the primitive basis can be written as
\begin{equation}
[D(C_{2z}\mc{T})]_{\vG'\sigma' l',\vG\sigma l}=\delta_{\vG',\vG} (\sigma_x)_{\sigma',\sigma}\delta_{l',l}.
\label{eqn:c2t_D}
\end{equation}

Given a set of bands $\{ u_{n\vk} \}$ and an antiunitary symmetry operation $g \mc{K}$, the corresponding sewing matrix $[B(g \mc{K})]_{\vk}$ is defined by the formula:
\begin{equation}
  \label{eqn:antiunitary_sewing}
  [B(g \mc{K})]_{\vk,mn} := \braket{u_{m,g\vk} | D(g) | u^*_{n\vk}} .
\end{equation}
As before, when these bands are isolated, $[B(g \mc{K})]_{\vk}$ is unitary (\cref{sec:sewing_matrix_unitary}) and the band creation operators transform under $g \mc{K}$ by the same rule as in \cref{eqn:sewing_transformation}.
Similar to calculations to the unitary case, if the following commutator-like quantity
\begin{equation}
  \label{eqn:order_parameter_antiunitary}
  \begin{split}
    \mc{C}_{\vk}(g \mc{K}) & = \norm{[B(g \mc{K})]_{\vk}^{\top}P(g\vk)^*[B(g \mc{K})]_{\vk}^* - P(\vk)} \\
     & = \norm{P(g\vk)[B(g \mc{K})]_{\vk}-[B(g \mc{K})]_{\vk}P(\vk)^*}
  \end{split}
\end{equation}
does not vanish, then the 1-RDM breaks the antiunitary symmetry $g\mc{K}$.
Furthermore,  $\mc{C}_{\vk}(g \mc{K})$ is invariant under gauge transformations of the band creation operators (see~\cref{sec:symm_order_param_antiunitary}).

\begin{center}
\begin{table}[ht]
\renewcommand{\arraystretch}{1.5}
\begin{tabular}{l|lcl}
     & Sewing matrix && Order parameter \\
     \hline 
    Unitary ($g$) & $\braket{u_{m,g\vk} | D(g) | u_{n\vk}}$ && $\norm{P(g\vk)[B(g)]_{\vk}-[B(g)]_{\vk}P(\vk)}$ \\
    Antiunitary ($g \mc{K}$) & $\braket{u_{m,g\vk} | D(g) | u^*_{n\vk}}$ && $\norm{P(g\vk)[B(g \mc{K})]_{\vk}-[B(g \mc{K})]_{\vk}P(\vk)^*}$
\end{tabular}
\caption{The definitions of the sewing matrix and symmetry order parameter for a unitary symmetry $g$ and an antiunitary symmetry $g \mc{K}$.}
\label{tab:symm-breaking}
\end{table}
\end{center}

\subsection{Connection with the $C_{2z}\mathcal{T}$ order parameter in the Chern band basis}

Let us also show the connection between $\mc{C}_{\vk}(\mc{C}_{2z}\mc{T})$ and the order parameter used in \cite{SoejimaParkerBultinckEtAl2020} using a particular gauge fixing called the Chern band basis. 
According to the gauge choice of the Chern band basis, the sewing matrix takes the form
\[
[B(C_{2z}\mc{T})]_{\vk}=\sigma_x  e^{\I \theta(\vk)}.
\]
The sewing matrix in this basis resembles the representation matrix in the primitive basis in \cref{eqn:c2t_D}, except that $\theta(\vk)$ is a $\vk$-dependent phase factor.
In this basis, the $C_{2z}\mc{T}$ symmetry breaking can be detected by computing
\begin{equation}
\gamma_z(\vk)=\Tr[P(\vk)\sigma_z]=P_{11}(\vk)-P_{22}(\vk).
\label{eqn:chernband_c2t_order}
\end{equation}

Note that the commutator for the $C_{2z}\mc{T}$ symmetry satisfies 
\[
P(\vk)[B(C_{2z}\mc{T})]_{\vk}-[B(C_{2z}\mc{T})]_{\vk}P(\vk)^*=e^{\I \theta(\vk)}
\begin{pmatrix}
P_{21}(\vk)-P_{12}(\vk)^* & P_{22}(\vk)-P_{11}(\vk)\\
P_{11}(\vk)-P_{22}(\vk) & P_{12}(\vk)-P_{21}(\vk)^*
\end{pmatrix},
\]
where we have used the fact that $P_{11}(\vk)$ and $P_{22}(\vk)$ are real.
Therefore $\gamma_z(\vk)$ can be interpreted as checking the magnitude of the off-diagonal element of the commutator in the Chern band basis. However, the order parameter $\gamma_z(\vk)$ is designed specifically for the Chern band basis and $C_{2z}\mathcal{T}$ symmetry, 
and does not generalize to other
band bases and other symmetries. On the other hand, the commutator can be used with any symmetry of interest and works for any band basis.

\section{Numerical results}
\label{sec:numerical}
%

\begin{figure}
    \centering
    \includegraphics[width=\textwidth]{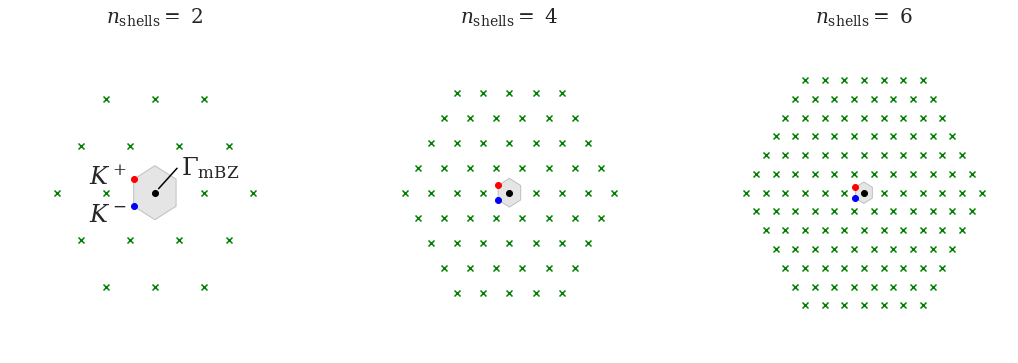}
    \caption{A plot of the moir{\' e} reciprocal lattice points included for $n_{\rm shells} = 2, 4, 6$ with the two valleys ($K^+, K^-$) and the Gamma point ($\Gamma_{\rm mBZ}$) marked. Note that the points are closed under $C_{3x}$ rotation and can be given by the formula $\{ \vG_{\rm moir\acute{e}} [m, n]^\top : | m + n | \leq n_{\rm shells}, (m,n) \in \mathbb{Z}^2 \}$ where $\vG_{\rm moir\acute{e}}$ is the generating matrix for the moir\'e lattice.}
    \label{fig:shells_diagram}
\end{figure}

Throughout our tests, we will use  $\vk$-mesh of size $(n_{k_x}, n_{k_y})$, and we always fix $n_{k_x} = 2 n_{k_y}$. The number of $\vG$ vectors is controlled by the number of \textit{shells} $n_{\mathrm{shell}}$, which specifies a number moir{\' e} reciprocal lattice vectors used in the interlayer coupling term in the BM model (see~\cref{fig:shells_diagram}). The number of included moir{\' e} reciprocal lattice vectors is bounded by $3 (n_{\mathrm{shell}} + 1)^2$.
The inverse temperature used in the decoupled subtraction scheme (\cref{eq:p0_dec}) is $\beta = 1000~\text{eV}^{-1}$.
We express $\mc{C}_{\vk}(C_{2z}\mc{T})$ in~\cref{eqn:order_parameter_antiunitary} in the spectral norm, and report the order parameter averaged over the number of $\vk$-points. 
We begin by studying the convergence of the IBM model with respect to discretization parameters in~\cref{sec:convergence-hf}. Then we report the results of HF and post-HF calculations in the integer filling regime in~\cref{sec:CCSD_integer_filling} and compare the effects of different subtraction schemes in~\cref{sec:model_discr}. Finally, we report the effects that initialization has on HF and post-HF calculations
in the non-integer filling regime in~\cref{sec:CCSD_noninteger_filling}.

\subsection{Convergence of parameters at the Hartree-Fock level} \label{sec:convergence-hf}

As mentioned in~\cref{sec:implementation-PySCF}, we do not exploit $\vk$-point symmetry in our current implementation using PySCF. As such, for larger $\vk$-meshes we incur significantly higher memory costs as compared to code which does exploit this symmetry. For our convergence tests, we test system sizes $n_{k_x} = 4, 8, 12$ exclusively using PySCF, and system sizes $n_{k_x} = 16, 20$ are tested using a separate code used in Ref.~\cite{SoejimaParkerBultinckEtAl2020}.

\begin{figure*}[h!]
\centering
\begin{subfigure}[b]{.48\textwidth}
    \centering
    \includegraphics[width=\linewidth]{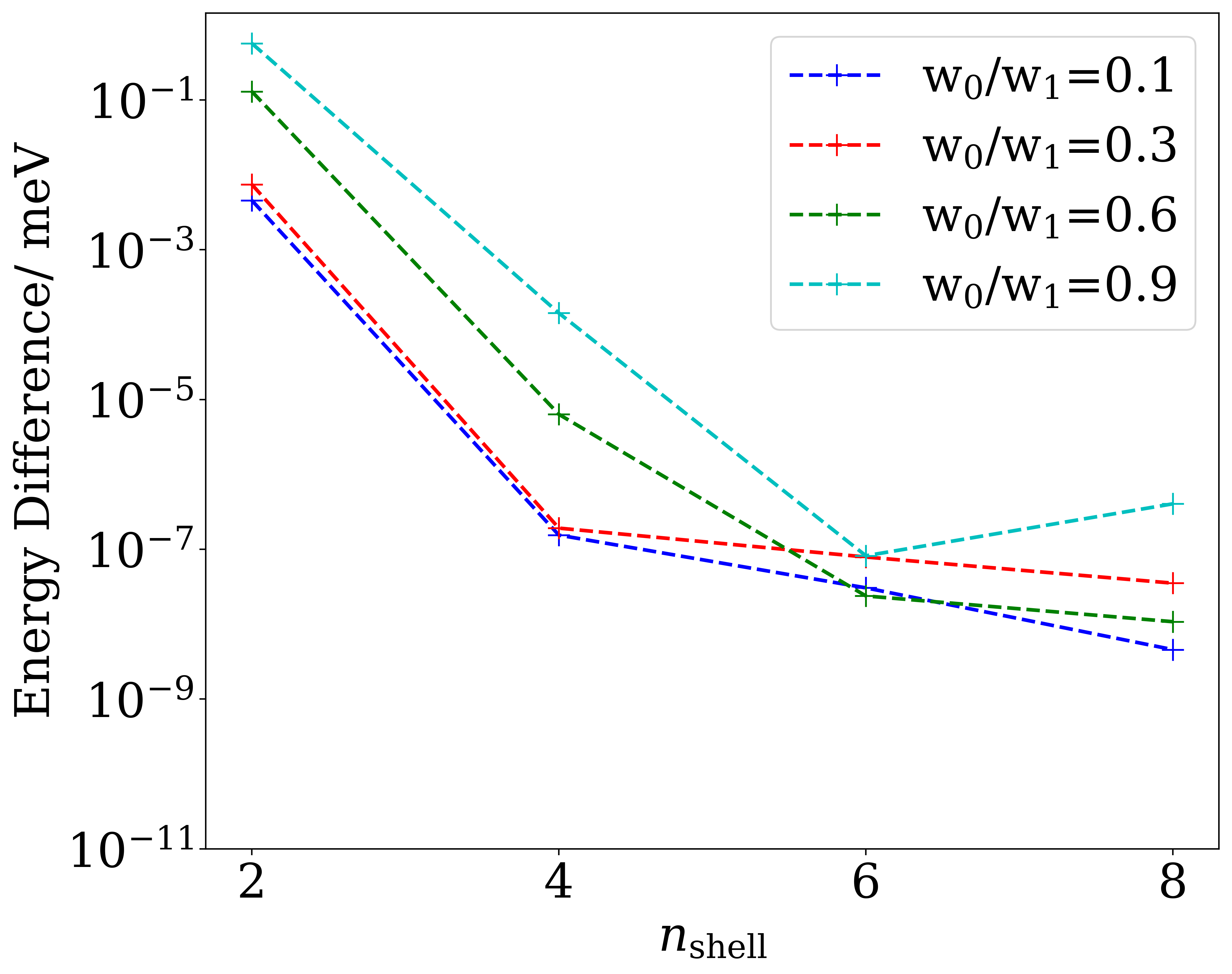} 
    \caption{}
    \label{fig:shell_convergence_test}
\end{subfigure}
\hfill
\begin{subfigure}[b]{.48\textwidth}
    \centering
    \includegraphics[width=\linewidth]{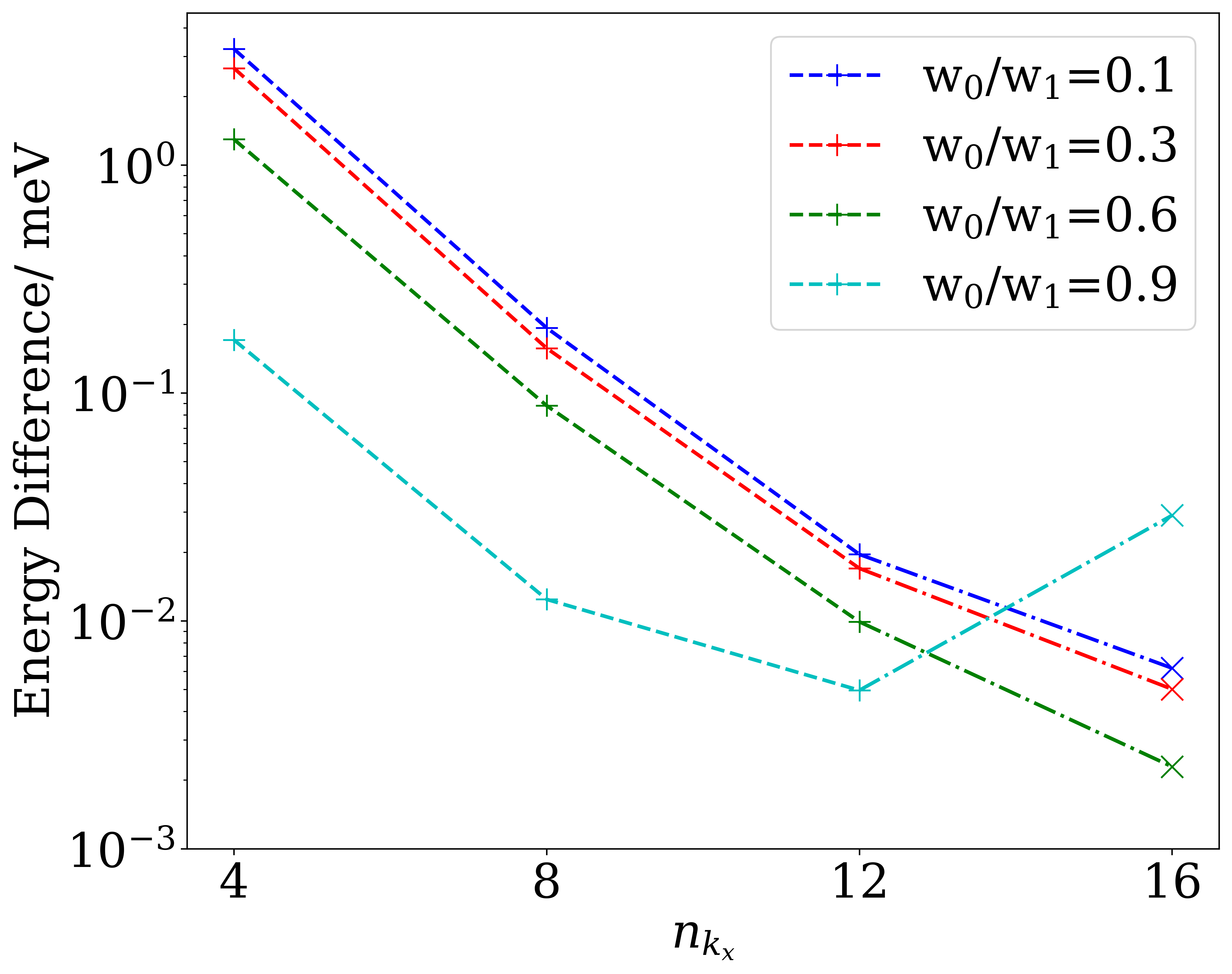}
    \caption{}
    \label{fig:kpts_convergence_test}
\end{subfigure}
\caption{\label{fig:energy_hf_dec}(\raggedright\subref{fig:shell_convergence_test}) The convergence test of Hartree-Fock with respect to the number of shells $n_{\mathrm{shell}}$ at ratio $w_0/w_1=0.0,0.3,0.6, 0.9$. $n_{k_x} = 2 n_{k_y} = 8$ is fixed.
(\subref{fig:kpts_convergence_test}) The convergence test of Hartree-Fock with respect to the number of $\vk$-points $n_{k_x} = 2 n_{k_y}$ at ratio $w_0/w_1=0.0,0.3,0.6, 0.9$. $n_{\mathrm{shell}}=6$ is fixed. Calculations with $n_{k_x} = 4, 8, 12$ are computed using the molecular structure module provided by PySCF, while $n_{k_x} = 16, 20$ are computed using a separate code exploiting  $\vk$-point symmetry. 
}
\end{figure*}

In~\cref{fig:shell_convergence_test}, we show the results of testing the convergence of Hartree-Fock energy with respect to the number of shells $n_{\mathrm{shell}} = 2, 4, 6, 10$ at ratios $w_0/w_1=0,0.3,0.6, 0.9$ with the $\vk$-mesh fixed to $n_{k_x} = 12$.
In~\cref{fig:kpts_convergence_test}, we show the results of the convergence test of Hartree-Fock energy per electron with respect to the number of $\vk$-points $n_{k_x} = 4, 8, 12, 16, 20$ at ratios $w_0/w_1=0.0,0.3,0.6, 0.9$ with the number of shells fixed to $n_{\mathrm{shell}}=6$.
The energy differences reported in~\cref{fig:shell_convergence_test,fig:kpts_convergence_test} are the differences between consecutive energies of $n_{\rm shell}$ and $n_{k_x}$, respectively.
From these experiments, we find that the choice $n_{k_x} = 8$ and $n_{\mathrm{shell}} = 8$ provides a good compromise between accuracy and required computation time.

\begin{figure*}[h!]
\centering
\begin{subfigure}[b]{.48\textwidth}
    \centering
    \includegraphics[width=\linewidth]{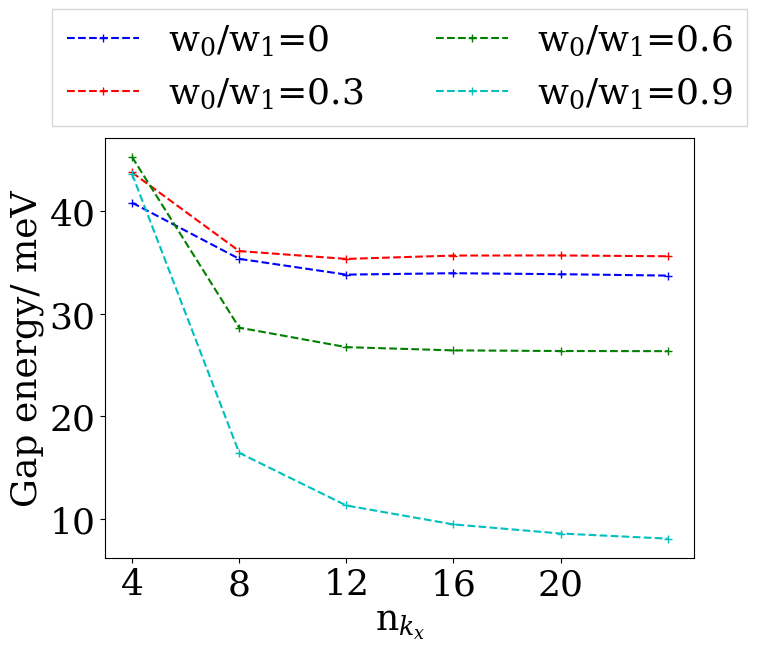} 
    \caption{}
    \label{fig:kpts_homo_lumo}
\end{subfigure}
\hfill
\begin{subfigure}[b]{.48\textwidth}
    \centering
    \includegraphics[width=\linewidth]{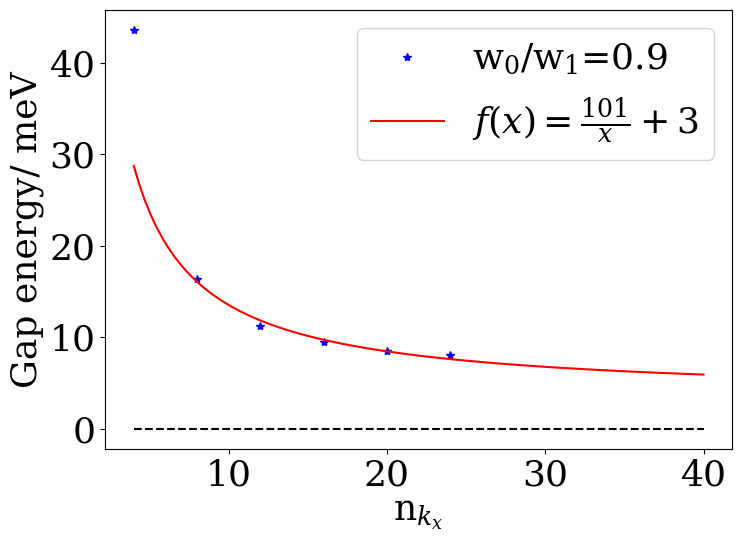}
    \caption{}
    \label{fig:hoomo_lumo_ext}
\end{subfigure}
\caption{\label{fig:kpts_conv_homo_lumo}\raggedright(\subref{fig:kpts_homo_lumo}) 
The convergence test of the HOMO-LUMO gap with respect to the number of $\vk$-points $n_{k_x} = 2 n_{k_y}$ at ratio $w_0/w_1=0.0,0.3,0.6, 0.9$. $n_{\mathrm{shell}}=6$ is fixed. Calculations with $n_{k_x} = 4, 8, 12$ are computed using the molecular structure module provided by PySCF.
(\subref{fig:hoomo_lumo_ext}) Extrapolation of the HOMO-LUMO gap as a function of $n_{k_x}$ at $w_0/w_1=0.9$. The fitted function is $f(x)=\frac{a}{x}+c$ where $a=101$, and $c=3$.
}
\end{figure*}

Aside from the Hartree-Fock energy, we investigate the convergence of the HOMO-LUMO gap with respect to the number of $\vk$-points $n_{k_x} = 4, 8, 12, 16, 20$, see~\cref{fig:kpts_homo_lumo}.
The computations suggest that $w_0/w_1=0.9$ is in the metallic region since the gap closes as the Brillouin zone sampling is being refined. 
We confirm this by fitting the function $f(x)=\frac{a}{x}+c$ to the samples, showing the inverse proportionality of the HOMO-LUMO gap to the number of $k$-points.   

\subsection{Integer filling}
\label{sec:CCSD_integer_filling}

We here present HF, CCSD, CCSD(T), and DMRG calculations for twisted bilayer graphene at integer filling, i.e., $\nu=0$ which amounts to one electron per moir\'e site. 
The subsequently presented results are obtained for a discretization of TBG using $n_{k_x} = 2 n_{k_y} = 8$, $n_{\mathrm{shell}} = 8$, and using the decoupled subtraction scheme.
The computations are performed for different ratios of the interlayer moir\'e potential parameters, i.e., $w_0/w_1\in[0,0.95]$.
The correlation energy per moir\'e site is defined to be the difference between the total energies from the correlated wavefunction
method, i.e., CCSD, CCSD(T), or DMRG, and the HF energy. 
All energies are reported per moir\'e site.

\begin{figure*}[h!]
\centering
\begin{subfigure}[b]{.48\textwidth}
    \centering
    \includegraphics[width=\linewidth]{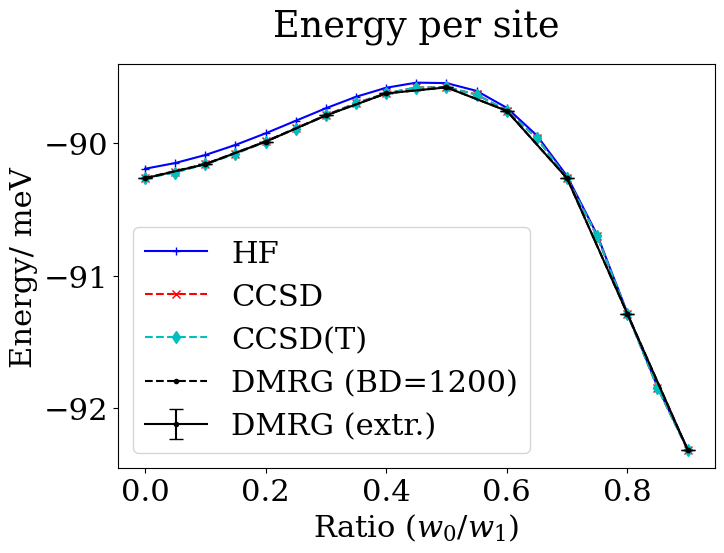} 
    \caption{}
    \label{fig:abs_energies_dec}
\end{subfigure}
\hfill
\begin{subfigure}[b]{.48\textwidth}
    \centering
    \includegraphics[width=\linewidth]{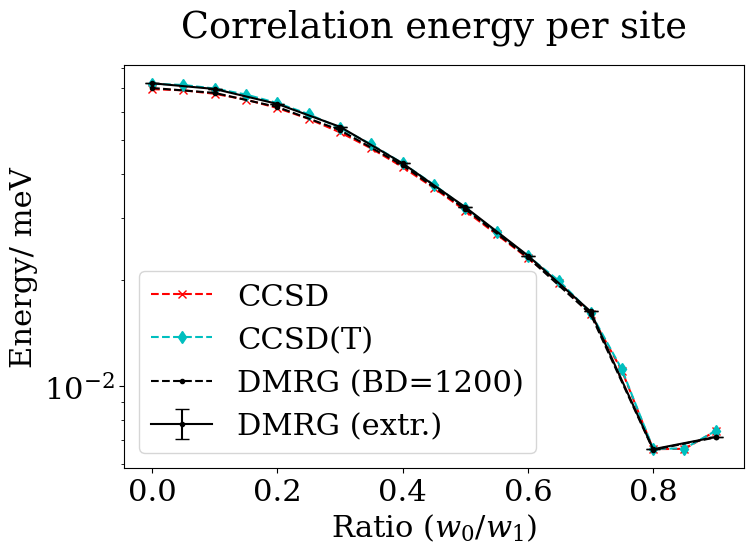}
    \caption{}
    \label{fig:corr_energies_dec}
\end{subfigure}
\caption{\label{fig:energy_CCSD_dec}\raggedright(\subref{fig:abs_energies_dec}) The HF, CCSD, CCSD(T), and DMRG (bond-dimension is $BD=1200$ and extrapolated to the infinite bond dimension limit [DMRG (extr.)]) energies per moir\'e site in meV as a function of the ratio $w_0/w_1$.  
(\subref{fig:corr_energies_dec}) The absolute value of CCSD, CCSD(T), and DMRG (bond-dimension $BD= 1200$ and extrapolated to the infinite bond dimension limit [DMRG (extr.)]) correlation energies in meV per moir\'e site as a function of the ratio $w_0/w_1$. 
}
\end{figure*}

Fig.~\ref{fig:abs_energies_dec} shows that the total energy is not monotone with respect to ratio $w_0/w_1$, and attains a maximum at around $w_0/w_1=0.5$. 
However, the correlation energy monotonically decreases with respect to the ratio until $w_0/w_1=0.8$, see~\cref{fig:corr_energies_dec}. 
The magnitude of the correlation energy per site is small, which qualitatively agrees with the theoretical prediction that the correlation energy vanishes (i.e., Hartree-Fock theory gives the exact ground state energy) at the chiral limit~\cite{BultinckKhalafLiuEtAl2020}.
However, the reason why the correlation energy does not exactly vanish at the chiral limit is due to the choice of the subtraction Hamiltonian, which we elaborate on in more detail in \cref{sec:model_discr}.
Compared to the energy evaluated at the CCSD level, the additional correlation energy obtained by CCSD(T) is negligible, see~\cref{fig:corr_energies_dec}.
Further comparison of the CCSD and CCSD(T) energies with DMRG energies extrapolated to the infinite bond-dimension limit shows that CCSD and CCSD(T) recover 95.4--100\% and 98.5--100\% of the correlation energy, respectively. 
Note that due to the high computational cost, we only compute extrapolated DMRG results for every other point in~\cref{fig:abs_energies_dec,fig:corr_energies_dec}; this suffices since there are no significant details in the intermediate range. 

We also report the Fermi-Dirac entropy per moir\'e site:
\begin{equation}
S_{\mathrm{FD}}=-\frac{1}{n_{k_x}n_{k_y}}\sum_{i}\left(p_i\ln p_i+(1-p_i)\ln(1-p_i)\right),
\label{eqn:fd_entropy}
\end{equation}
where $\{p_i\}$ are the eigenvalues of the 1-RDM. 
By construction, $S_{\mathrm{FD}}=0$ in the Hartree-Fock theory.  We find that the Fermi-Dirac entropy is between 0.009 and 0.032 from the DMRG calculations. This reveals that the solutions for all parameter ratios are close to being single Slater determinants.

\begin{figure}[h!]
    \centering
    \includegraphics[width=0.5 \textwidth]{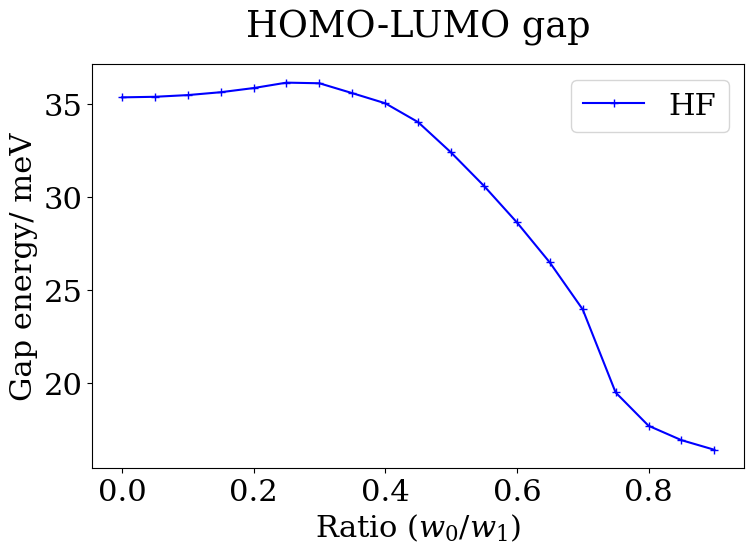}
    \caption{The HOMO-LUMO gap as a function of the ratio $w_0/w_1$.}
    \label{fig:homo-lumo-gap_dec}
\end{figure}

Investigating the HOMO-LUMO gap, we observe a gap closing as we transition from the chiral limit to $w_0/w_1=1$, see~\cref{fig:homo-lumo-gap_dec}.
The HOMO-LUMO gap closes around $w_0/w_1=0.85$, indicating a transition from an insulating to a metallic phase. 
This is in agreement with the finding in~\cref{fig:kpts_homo_lumo}.

\begin{figure*}[h!]
\centering
\begin{subfigure}[b]{.48\textwidth}
    \centering
    \includegraphics[width=\linewidth]{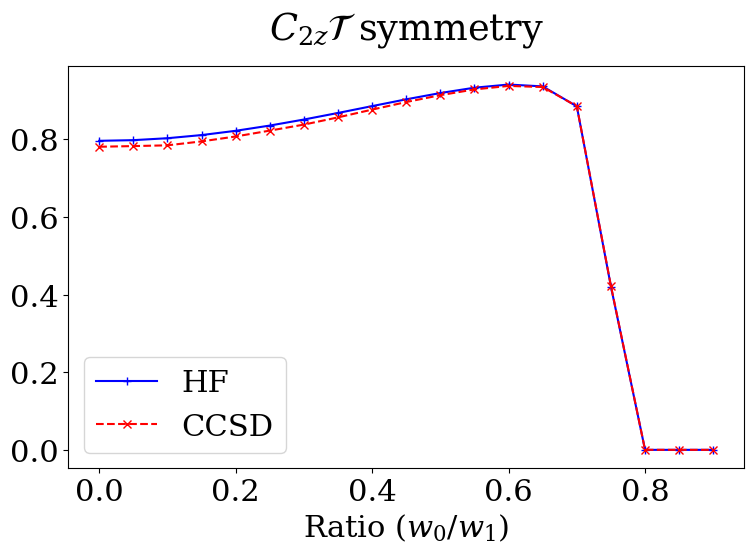} 
    \caption{}
    \label{fig:c2t_symm_dec}
\end{subfigure}
\hfill
\begin{subfigure}[b]{.48\textwidth}
    \centering
    \includegraphics[width=\linewidth]{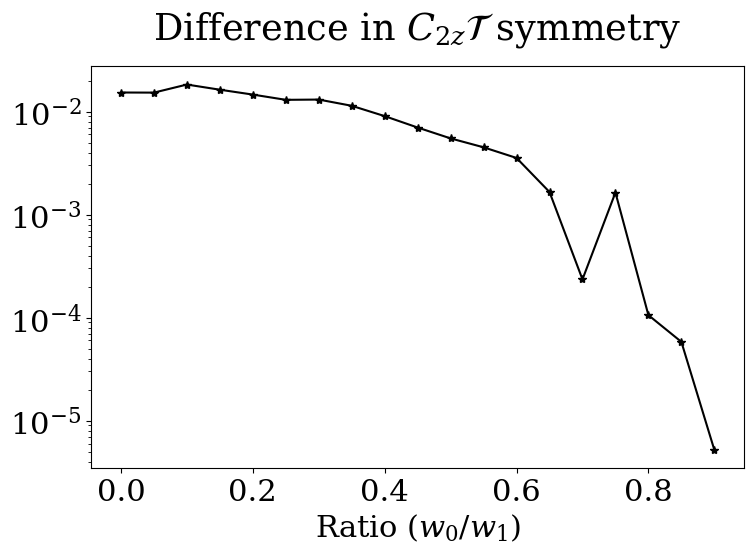}
    \caption{}
    \label{fig:c2t_symm_diff_dec}
\end{subfigure}
\caption{\label{fig:c2t_ccsd_dec}\raggedright(\subref{fig:c2t_symm_dec}) The HF and CCSD $C_{2z}\mathcal{T}$ symmetry predictions as a function of the ratio $w_0/w_1$.  
(\subref{fig:c2t_symm_diff_dec}) The absolute value of the difference of the $C_{2z}\mathcal{T}$ symmetry characteristic of CCSD and the HF as a function of the ratio $w_0/w_1$. }
\end{figure*}

Next, we investigate the effect of electronic correlations on the order parameter $\mc{C}_{\vk}(C_{2z}\mc{T})$ in \cref{eqn:order_parameter_antiunitary}. 
Fig.~\ref{fig:c2t_ccsd_dec} reports the $C_{2z}\mathcal{T}$ order parameter as a function of the ratio $w_0/w_1$, which shows a transition from the $C_{2z}\mathcal{T}$ broken phase to a $C_{2z}\mathcal{T}$ symmetric phase, and the phase transition occurs around $w_0/w_1=0.8$. 
This agrees with the result in~\cite{SoejimaParkerBultinckEtAl2020}, where the order parameter uses the expression \cref{eqn:chernband_c2t_order} in the Chern band basis.
Fig.~\ref{fig:c2t_symm_diff_dec} shows that compared to CCSD, HF slightly overestimates the symmetry breaking, and the difference between
 HF and CCSD decreases as the ratio $w_0/w_1$ increases.

\subsection{Model Discrepancies due to the subtraction Hamiltonian} 
\label{sec:model_discr}

To assess the effect of the subtraction Hamiltonian obtained from the decoupled scheme, we report the results using another subtraction Hamiltonian obtained from the average scheme (see \cref{sec:subtraction}).
We demonstrate the differences of the total energy and the $C_{2z}\mathcal{T}$ order parameter. 
Additionally, we compute and compare the effect of the subtraction Hamiltonians on the band structure, see~\cref{sec:cnp_clacs}. 

\begin{figure*}
\begin{center}
\begin{subfigure}[c]{.45\textwidth}
    \includegraphics[width=\textwidth]{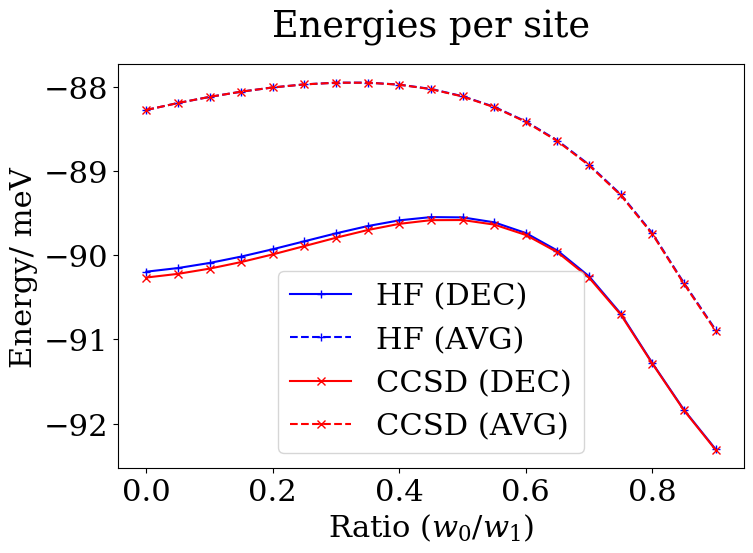}
    \caption{}
    \label{fig:energies_dec_cnp}
\end{subfigure}
\hfill
\begin{subfigure}[c]{.45\textwidth}
    \includegraphics[width=\textwidth]{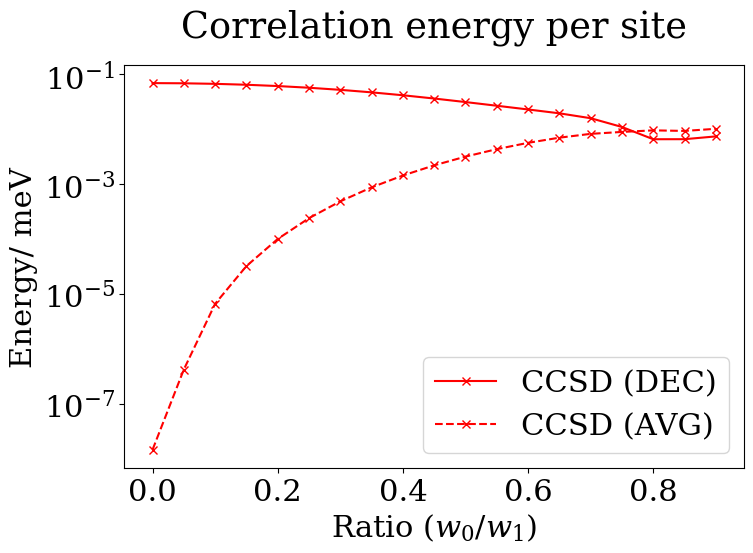} 
    \caption{}
    \label{fig:corr_energies_dec_cnp}
\end{subfigure}
\end{center}
\caption{
\raggedright(\subref{fig:energies_dec_cnp}) The HF and CCSD energies per moir\'e site in meV as a function of the ratio $w_0/w_1$ for decoupled and average scheme subtraction Hamiltonians.  
(\subref{fig:corr_energies_dec_cnp}) The absolute value of CCSD correlation energies in meV per moir\'e site as a function of the ratio $w_0/w_1$ for decoupled and average scheme subtraction Hamiltonians.
}
\end{figure*}

Comparing energies at the HF and CCSD level of theory we first note that using the decoupled scheme yields a more pronounced maximum in the energy, i.e., the curvature around the maximum is greater when employing the decoupled scheme, see Fig.~\ref{fig:energies_dec_cnp}. 
Moreover, we observe that employing the average scheme subtraction Hamiltonian yields an overall lower correlation energy, see Fig.~\ref{fig:corr_energies_dec_cnp}.
Interestingly, both subtraction Hamiltonians yield a similar amount of electronic correlation near $w_0/w_1\geq 0.8$.
Aside from the magnitude of the correlation, we find that the electronic correlation increases as a function of $w_0/w_1$ when using the average scheme subtraction Hamiltonian whereas the electronic correlation decreases as a function of $w_0/w_1$ when using the decoupled scheme subtraction Hamiltonian.

\begin{figure*}
\begin{center}
\begin{subfigure}[c]{.45\textwidth}
    \includegraphics[width=\textwidth]{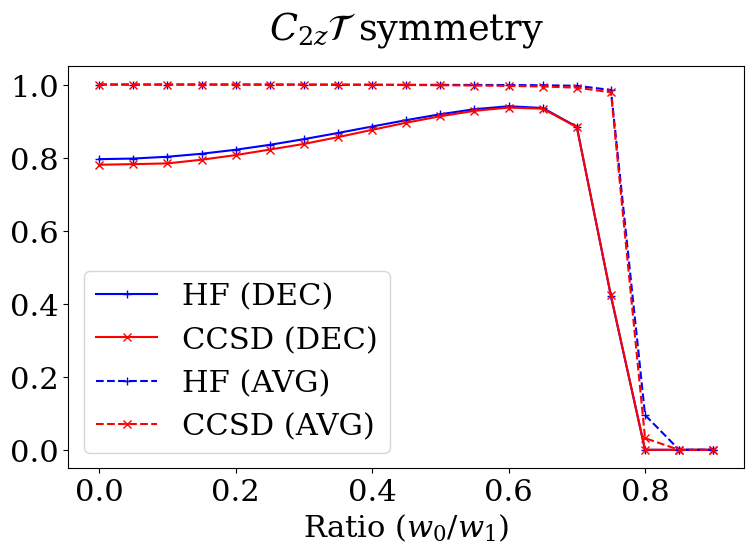}
    \caption{}
    \label{fig:c2t_dec_cnp}
\end{subfigure}
\hfill
\begin{subfigure}[c]{.45\textwidth}
    \includegraphics[width=\textwidth]{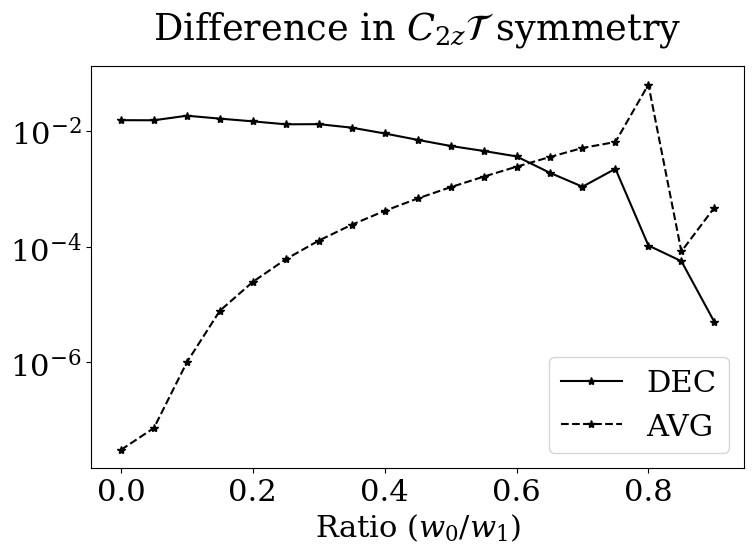} 
    \caption{}
    \label{fig:c2t_diff_dec_cnp}
\end{subfigure}
\end{center}
\caption{\label{fig:c2t_syms_dec_cnp}
\raggedright(\subref{fig:c2t_dec_cnp}) The HF and CCSD $C_{2z}\mathcal{T}$ symmetry predictions as a function of the ratio $w_0/w_1$ for decoupled and average scheme subtraction Hamiltonian.  
(\subref{fig:c2t_diff_dec_cnp}) The absolute value of the difference of the $C_{2z}\mathcal{T}$ symmetry characteristic of CCSD and the HF as a function of the ratio $w_0/w_1$ for decoupled and average scheme subtraction Hamiltonian.  
}
\end{figure*}

The different subtraction Hamiltonians also affect the order parameter, see Fig.~\ref{fig:c2t_syms_dec_cnp}.
We observe a very clean first-order phase transition when employing the average scheme subtraction Hamiltonian whereas the decoupled scheme subtraction Hamiltonian yields a more continuous transition.
This agrees with earlier numerical results in \cite[Fig. 6]{KangVafek2020}. Correlation effects on the order parameter appear to be larger in the decoupled scheme near the chiral limit, see~\cref{fig:c2t_diff_dec_cnp}.


\begin{figure*}[h!]
\centering
\begin{subfigure}[b]{.48\textwidth}
    \centering
    \includegraphics[width=\linewidth]{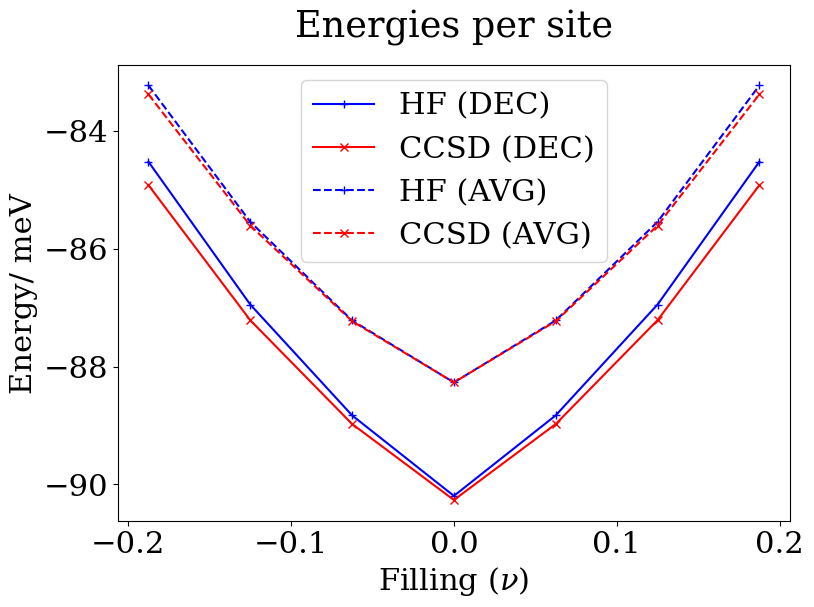} 
    \caption{}
    \label{fig:abs_energies_chiral_NIF_dec_cnp}
\end{subfigure}
\hfill
\begin{subfigure}[b]{.48\textwidth}
    \centering
    \includegraphics[width=\linewidth]{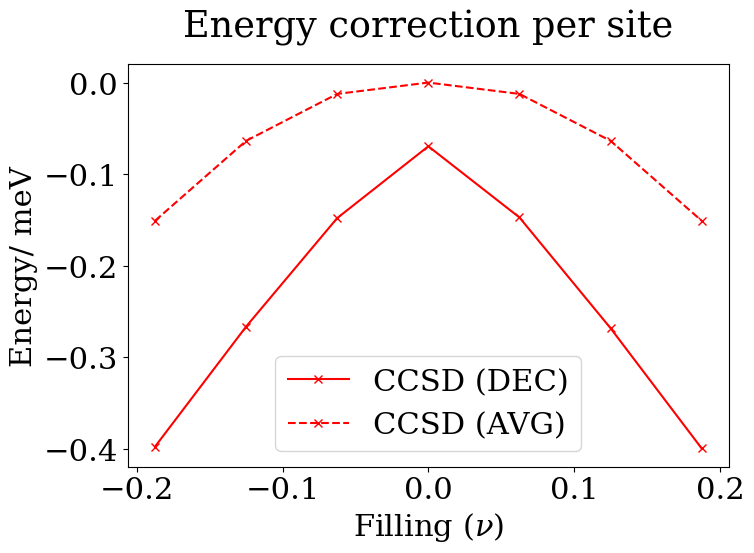}
    \caption{}
    \label{fig:corr_energies_chiral_NIF_dec_cnp}
\end{subfigure}
\raggedright\caption{\label{fig:ccsd_corr_chiral_e_NIF_dec_cnp}(\subref{fig:abs_energies_chiral_NIF_dec_cnp}) The HF and CCSD energies per moir\'e site in meV in the chiral limit as a function of the filling for decoupled and average scheme subtraction Hamiltonian. 
(\subref{fig:corr_energies_chiral_NIF_dec_cnp}) The CCSD energy correction per moir\'e site in meV in the chiral limit as a function of the filling for decoupled and average scheme subtraction Hamiltonian}
\end{figure*}

In the non-integer filling regime, we observe that the decoupled and average scheme subtraction Hamiltonian yield qualitatively similar results, see Fig.~\ref{fig:abs_energies_chiral_NIF_dec_cnp}. We here initialize the HF computations with a one-particle reduced density matrix following~\cite{SoejimaParkerBultinckEtAl2020}. 
We find that for the decoupled scheme subtraction Hamiltonian, the total energy changes more rapidly with respect to $\nu$ (i.e., a larger curvature in $\nu$), and the energy correction through post-HF methods is smaller than the energy corrections using the average scheme, i.e., using the decoupled scheme subtraction Hamiltonian yields stronger electronic correlation effects.

\subsection{Non-integer filling}
\label{sec:CCSD_noninteger_filling}


We now proceed to HF, CCSD, CCSD(T), and DMRG calculations at non-integer fillings. 
The subsequently presented results are again obtained for a discretization of TBG using $n_{k_x} = 2n_{k_y} = 8$, $n_{\mathrm{shell}} = 8$. 
The TBG is here modeled with $n_{\rm elec}\in \{26, 28, 30, 32, 34, 36, 38\}$, i.e., with a filling factor of $\nu= n_{\rm elec}/32-1$, and $\abs{\nu}<0.2$.
We moreover fix the initialization of the HF calculations following~\cite{SoejimaParkerBultinckEtAl2020} while adjusting the particle number correspondingly. 
We will investigate the effect of correlated methods first by varying the filling factor $\nu$ at the chiral limit, and then by varying both the filling factor $\nu$ and the interlayer coupling ratio $w_0 / w_1$.
In the next subsection, we will see that the ``true'' HF global minimum can be sensitive to the initial guess and difficult to reach. Hence, we will refer to the difference between post-HF energies and the HF energy only as an ``energy correction'' rather than the ``correlation energy''.

\begin{figure*}[h!]
\centering
\begin{subfigure}[b]{.42\textwidth}
    \centering
    \includegraphics[width=\linewidth]{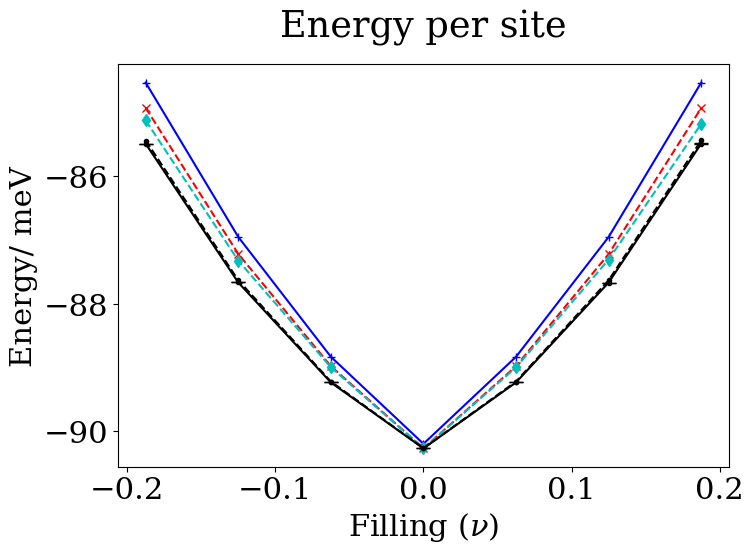} 
    \caption{$w_0/w_1=0$}
    \label{fig:abs_energies_chiral_NIF_dec}
\end{subfigure}
\hfill
\begin{subfigure}[b]{.54\textwidth}
    \centering
    \includegraphics[width=\linewidth]{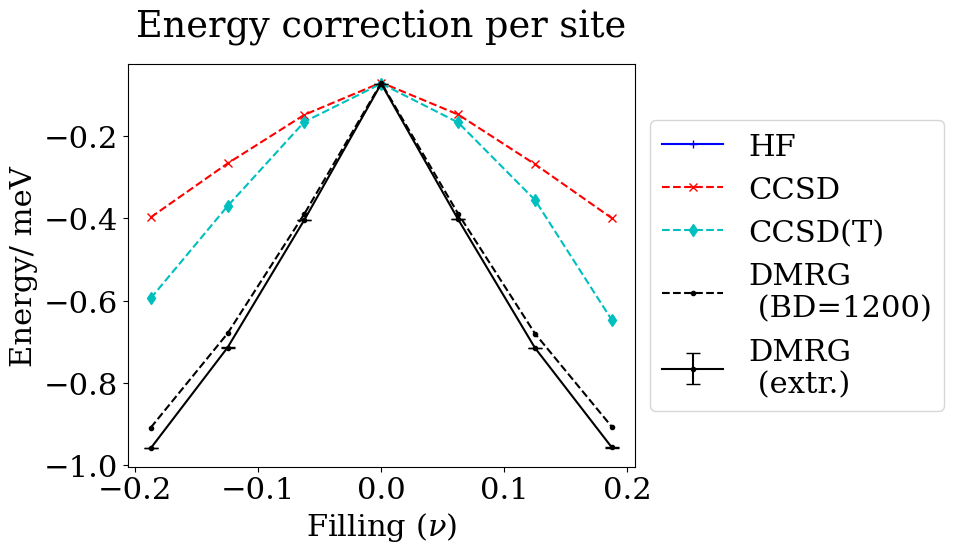}
    \caption{$w_0/w_1=0$}
    \label{fig:corr_energies_chiral_NIF_dec}
\end{subfigure}
\caption{\label{fig:ccsd_corr_chiral_e_NIF_dec}\raggedright(\subref{fig:abs_energies_chiral_NIF_dec}) The HF and CCSD, CCSD(T), and DMRG energies per moir\'e site in meV in the chiral limit ($w_0/w_1=0$) as a function of the filling.  
(\subref{fig:corr_energies_chiral_NIF_dec}) The CCSD, CCSD(T), and DMRG correlation energies per moir\'e site in meV in the chiral limit as a function of the filling.  
}
\end{figure*}


At the chiral limit, Fig.~\ref{fig:abs_energies_chiral_NIF_dec} and~\ref{fig:corr_energies_chiral_NIF_dec} show that the energy correction by means of post-HF methods increases as $\nu$ deviates from $0$. 
However, DMRG benchmark computations reveal that the Fermi-Dirac entropy in \cref{eqn:fd_entropy} is very small for all filling factors under consideration (see~\cref{tab:FD_ni_filling}). This indicates that the solution is relatively well described by a single Slater determinant, and thus by the HF theory. 
\begin{table}[h!]
    \centering
\renewcommand{\arraystretch}{1.4}
    \begin{tabular}{c|ccccccc}
         \hline
         Filling ($\nu$)     &  
         -0.188& -0.125& -0.062&  0&  0.062&  0.125&  0.188\\
         \hline
         S$_{\rm FD}$&  
         0.067& 0.051& 0.107& 0.033& 0.107& 0.052& 0.069\\
         \hline        
    \end{tabular}
    \caption{Fermi-Dirac entropy of the DMRG computations with bond dimension 1100 in the chiral limit for different fillings $\nu$. }
    \label{tab:FD_ni_filling}
\end{table}

\begin{figure*}[h!]
\centering
\begin{subfigure}[b]{.51\textwidth}
    \centering
    \includegraphics[width=\linewidth]{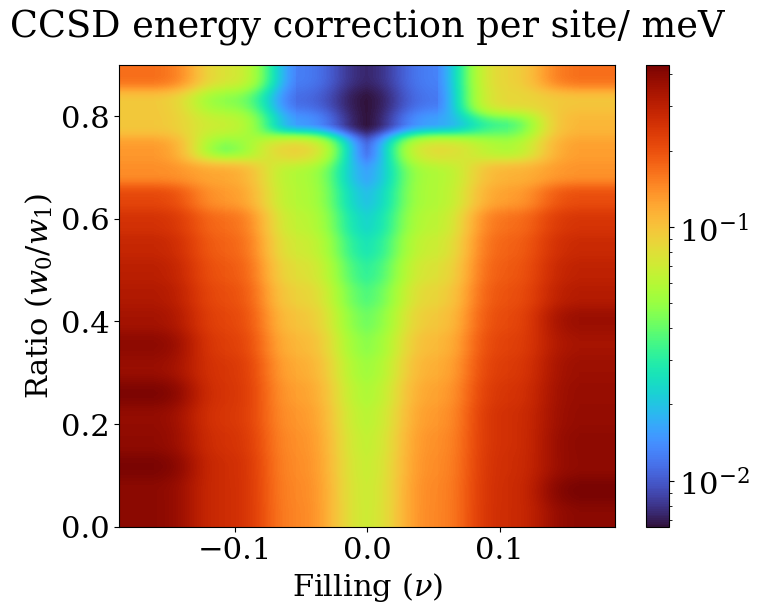}
    \caption{}
    \label{fig:ccsd_corr_energy_pd_dec}
\end{subfigure}
\hfill
\begin{subfigure}[b]{.48\textwidth}
    \centering
    \includegraphics[width=\linewidth]{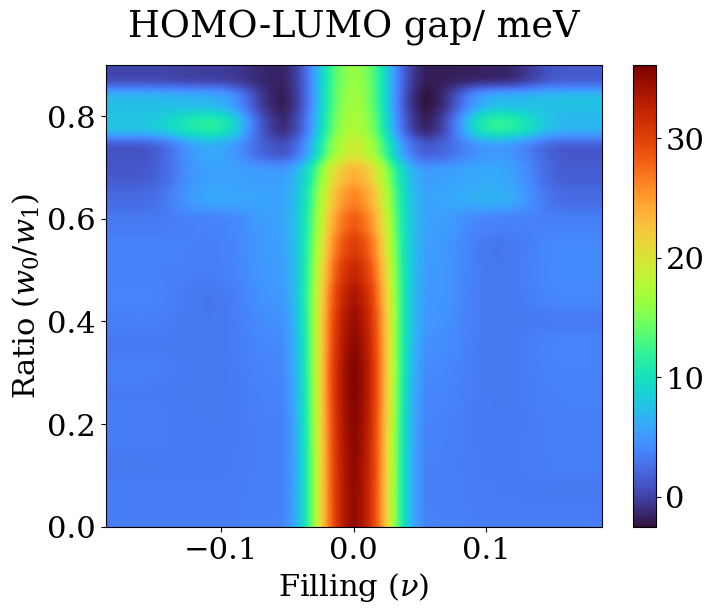}
    \caption{}
    \label{fig:homo-lumo-gap_pd_dec}
\end{subfigure}
\caption{
\raggedright(\subref{fig:ccsd_corr_energy_pd_dec}) Energy surface of the CCSD energy correction per moir\'e site in meV with respect to the filling and the ratio $w_0/w_1$.
(\subref{fig:homo-lumo-gap_pd_dec}) Phase diagram of the HOMO-LUMO gap in meV with respect to the filling and the ratio $w_0/w_1$.
}
\end{figure*}

We find that at this point, DMRG calculations are too expensive to be applied to evaluate the entire 2D phase diagram. Hence we investigate the landscape of the energy correction provided by CCSD with respect to the filling and the ratio $w_0/w_1$, we observe that the magnitude of the energy correction increases with respect to $\abs{\nu}$ (see Fig.~\ref{fig:ccsd_corr_energy_pd_dec}). 
Note that Fig.~\ref{fig:ccsd_corr_energy_pd_dec} is on a logarithmic scale, that is, we here depict the absolute values of the obtained energy corrections. The computed energy corrections are consistently negative.
We also investigate the HOMO-LUMO gap landscape with respect to the filling and the ratio $w_0/w_1$ in Fig.~\ref{fig:homo-lumo-gap_pd_dec}.
We observe that the HOMO-LUMO gap reaches its maximum at the chiral limit at $\nu=0$. 
When transitioning into the fractional filling regime (i.e., at $\abs{\nu}>0.0625$), the HOMO-LUMO gap decreases by one order of magnitude, indicating a metallic phase. 


\begin{figure*}[h!]
\centering
\begin{subfigure}[b]{.48\textwidth}
    \centering
    \includegraphics[width=\linewidth]{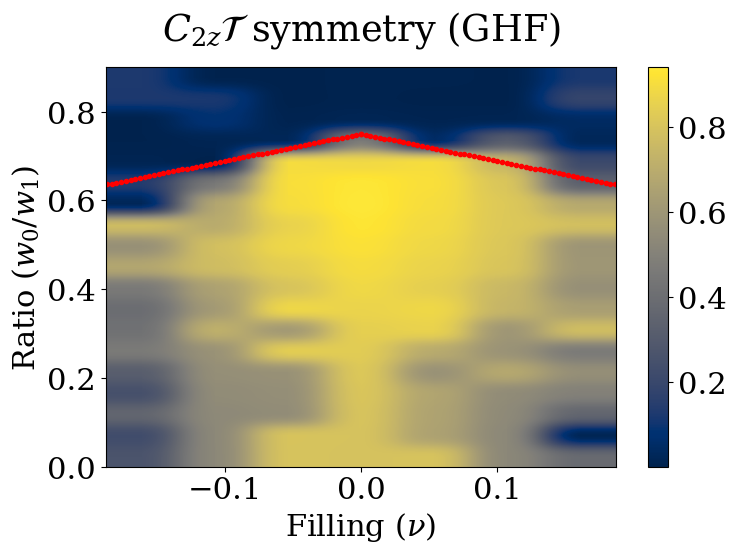} 
    \caption{}
    \label{fig:ccsd_c2t_pd_dec}
\end{subfigure}
\hfill
\begin{subfigure}[b]{.48\textwidth}
    \centering
    \includegraphics[width=\linewidth]{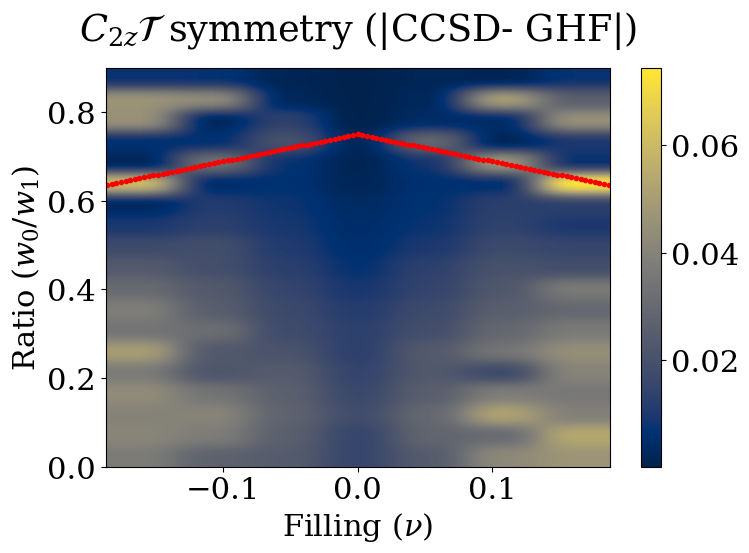}
    \caption{}
    \label{fig:ccsd_c2t_diff_pd_dec}
\end{subfigure}
\caption{
\raggedright(\subref{fig:ccsd_c2t_pd_dec}) 
Phase diagram of $C_{2z}\mathcal{T}$ symmetry predictions at the HF level of theory.
The dotted red line indicates the phase transition as a function of the filling $\nu$.
(\subref{fig:ccsd_c2t_diff_pd_dec}) Phase diagram of the difference of $C_{2z}\mathcal{T}$ symmetry predictions comparing CCSD and HF. The dotted red line indicates the phase transition as a function of the filling $\nu$.}
\end{figure*}

In~\cref{fig:ccsd_c2t_pd_dec} we report the phase diagram of the order parameter for the $C_{2z}\mathcal{T}$ symmetry with respect to the filling $\nu$ and the ratio $w_0/w_1$. 
We find that the difference between the order parameters obtained by HF and CCSD also increases as $\abs{\nu}$ deviates from $0$, but the phase diagrams qualitatively agree with each other, see~\cref{fig:ccsd_c2t_diff_pd_dec}.
The phase diagram indicates that the location of the phase transition from a $C_{2z}\mathcal{T}$ broken phase to a $C_{2z}\mathcal{T}$ symmetric phase is a function of the filling $\nu$. 
We highlight this dependence with a dotted red line in~\cref{fig:ccsd_c2t_diff_pd_dec}.
Recall that at integer filling, the system is either in a $C_{2z}\mc{T}$ symmetry breaking and insulating state, or in a $C_{2z}\mc{T}$ trivial and metallic state~\cite{SoejimaParkerBultinckEtAl2020}.
However, in the non-integer filling case, we find that the system can be in a $C_{2z}\mc{T}$ symmetry breaking and metallic state.
We also find that the difference between CCSD and HF is negative except for a few points on the phase diagram, indicating that HF tends to slightly over-polarize the $C_{2z}\mathcal{T}$ order parameters.

\subsection{Impact of the Initial One-Particle Reduced Density Matrix}

\begin{figure*}[h!]
\centering
\begin{subfigure}[b]{.48\textwidth}
    \centering
    \includegraphics[width=\linewidth]{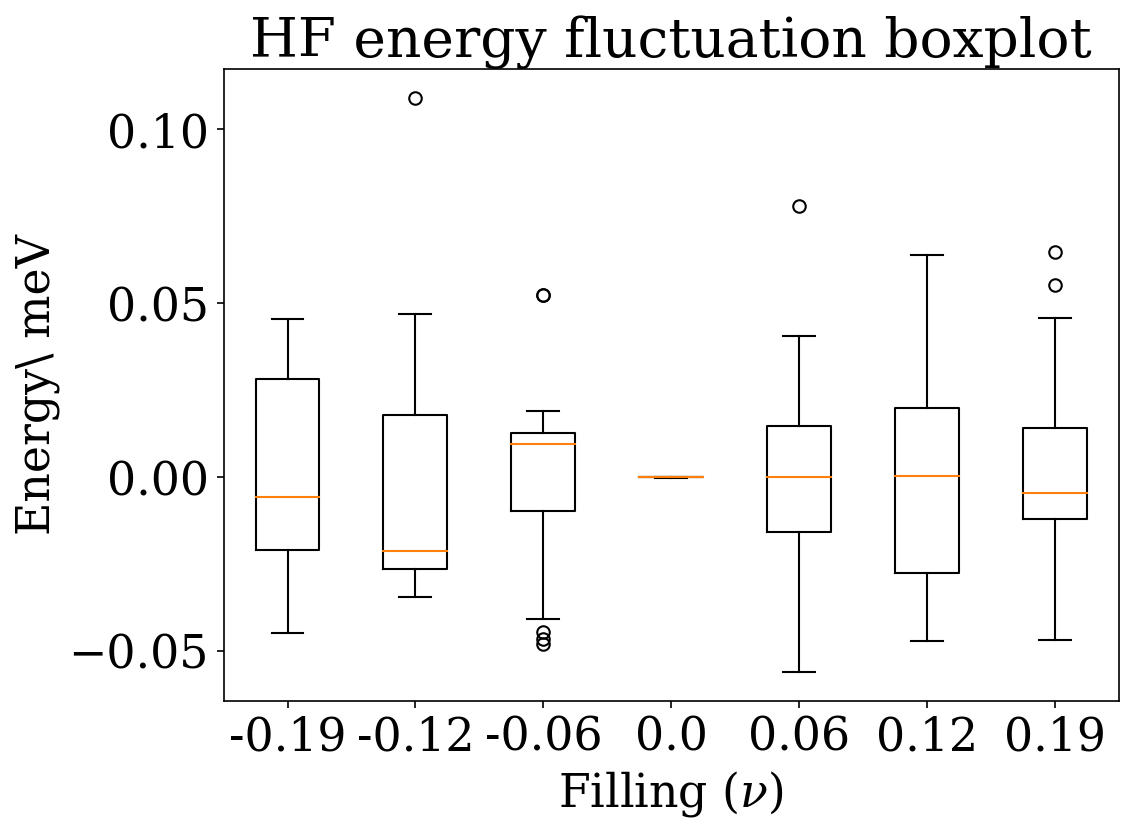} 
    \caption{}
    \label{fig:hist_ghf_dec}
\end{subfigure}
\hfill
\begin{subfigure}[b]{.48\textwidth}
    \centering
    \includegraphics[width=\linewidth]{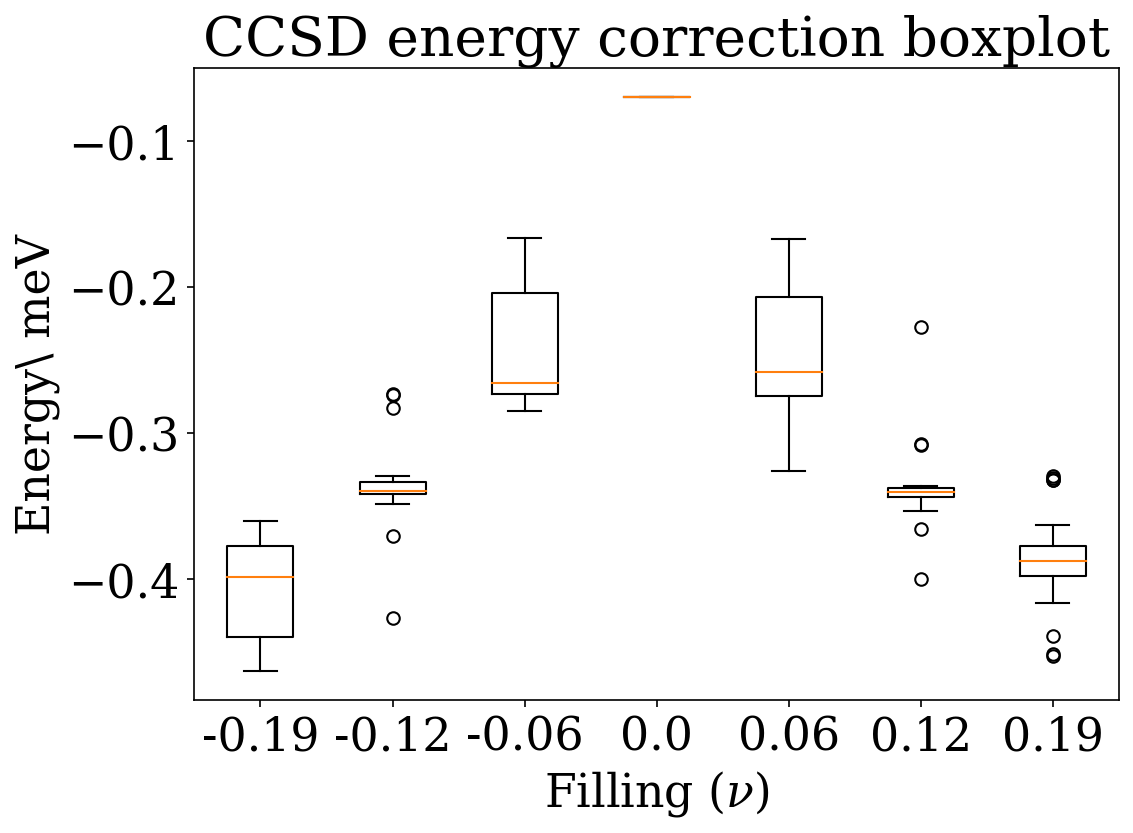}
    \caption{}
    \label{fig:hist_ghf_c2t_dec}
\end{subfigure}
\caption{\label{Fig:box_energies} Boxplot showing the initial 1-RDM dependence of energy calculations at the HF (\subref{fig:hist_ghf_dec}) and CCSD (\subref{fig:hist_ghf_c2t_dec}) level of theory for 20 random initializations.}
\end{figure*}

In the previous section, we employed a particular initial 1-RDM for the HF calculations. We now investigate the effect of the initial guess, by drawing initial 1-RDMs from a random distribution (while satisfying the particle number constraint), and perform computations at the HF and CCSD level of theory for different fillings in the chiral limit. 
Even when using a random initial guess,  HF and CCSD at half-filling can robustly converge to the global minimum.
Away from half-filling, 
even after employing various techniques in quantum chemistry calculations (e.g., level-shifting, second-order optimizers, and temperature annealing),
the HF result can still depend on the initial random guess, indicating the existence of multiple local minima. The energy differences of these local minima are small, but the magnitude of these differences can be comparable to that of the CCSD energy correction (see~\cref{Fig:box_energies}).

\begin{figure*}[h!]
\centering
\begin{subfigure}[b]{.48\textwidth}
    \centering
    \includegraphics[width=\linewidth]{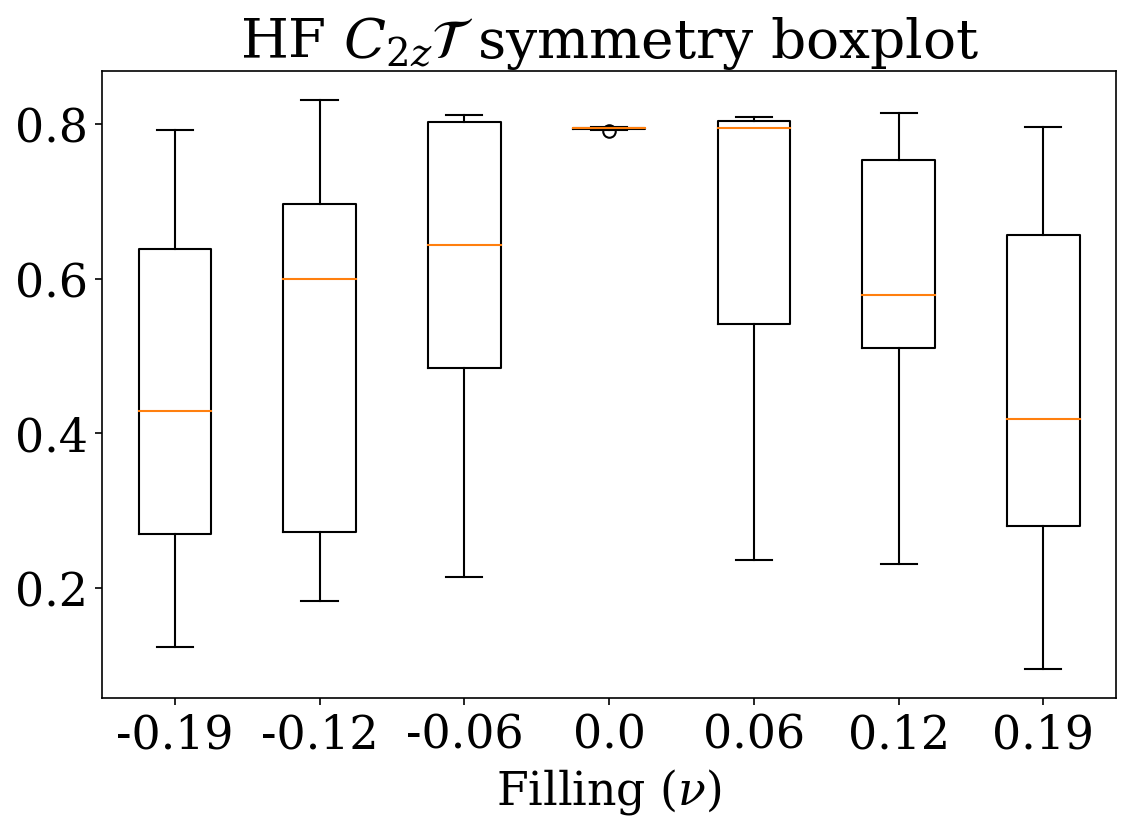} 
    \caption{}
    \label{fig:hist_ghf_dec}
\end{subfigure}
\hfill
\begin{subfigure}[b]{.48\textwidth}
    \centering
    \includegraphics[width=\linewidth]{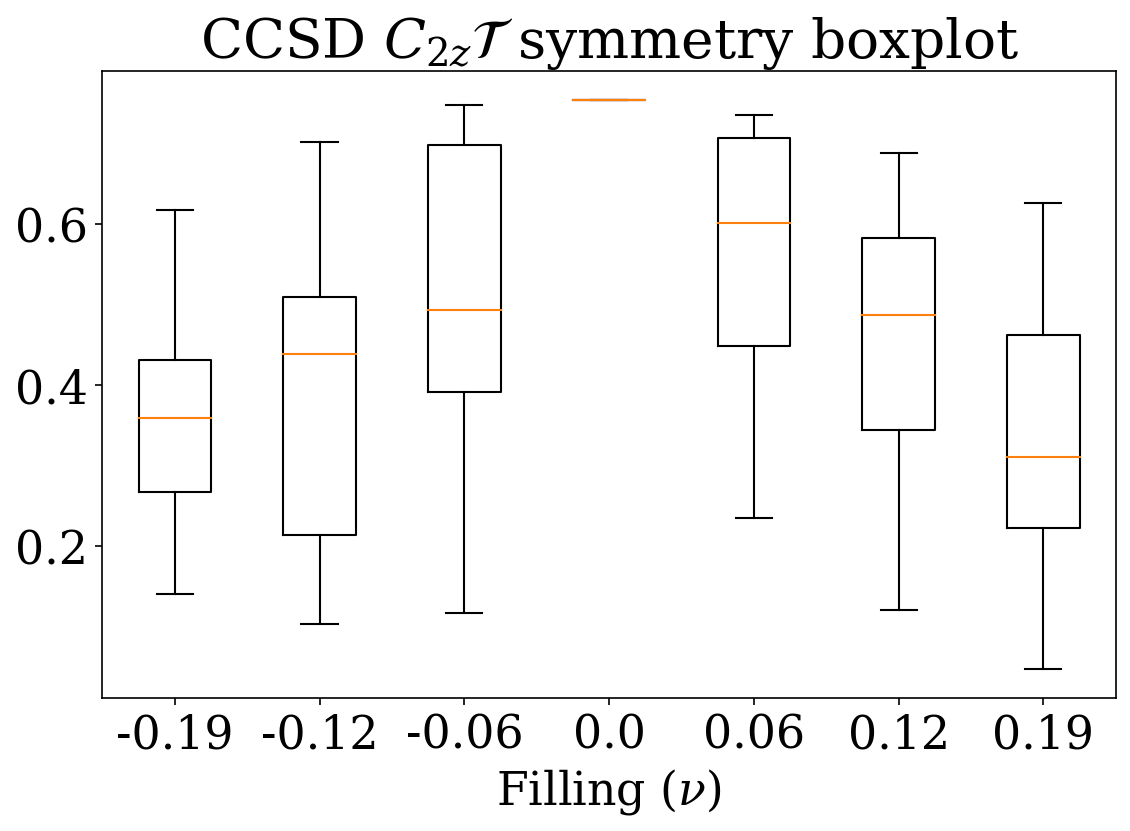}
    \caption{}
    \label{fig:hist_ghf_c2t_dec}
\end{subfigure}
\caption{\label{Fig:box_c2t} Boxplot showing the initial 1-RDM dependence of $C_{2z}\mathcal{T}$ calculations at the HF (\subref{fig:hist_ghf_dec}) and CCSD (\subref{fig:hist_ghf_c2t_dec}) level of theory for 20 random initializations.}
\end{figure*}

\begin{figure}
    \centering
    \includegraphics[width=0.5\textwidth]{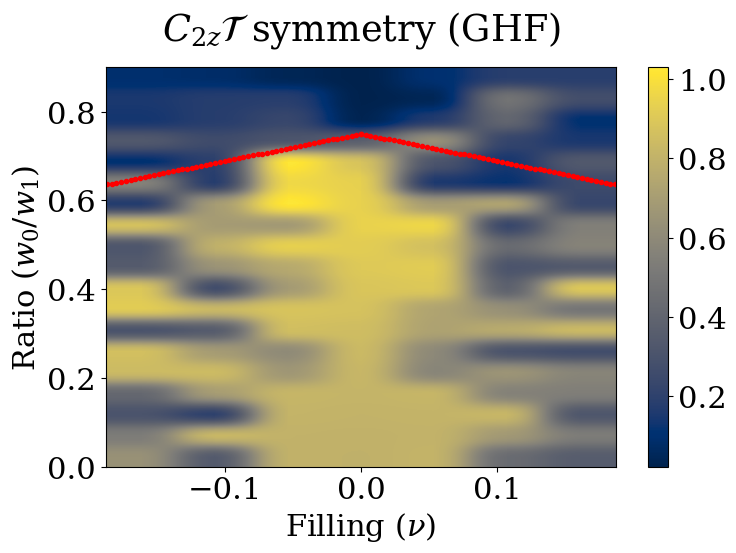}
    \caption{Phase diagram of $C_{2z}\mathcal{T}$ symmetry predictions at the HF level of theory. The
dotted red line indicates the phase transition as a function of the filling $\nu$.}
    \label{fig:c2t_pd_rand}
\end{figure}

While there are many local minima that are energetically close to the ground state, the amount of variation in the gauge-invariant $C_{2z}\mathcal{T}$ order parameter can be significantly larger, see~\cref{Fig:box_c2t}. 
This is the case both for HF and CCSD calculations. 
\cref{fig:c2t_pd_rand} shows the revised 2D HF phase diagram obtained by performing 15 independent calculations and evaluating the $C_{2z}\mathcal{T}$ order parameter from the lowest energy state. The overall shape of the phase diagram resembles that of \cref{fig:ccsd_c2t_pd_dec}. This indicates that despite the numerical fluctuation of the $C_{2z}\mathcal{T}$ order parameters at each point in the phase diagram due to the existence of many local minima, the qualitative features of the phase diagram may still be preserved.


To further study the behavior of the local minima and robustness of the numerical methods, we extract two 1-RDM initializations from the above performed experiment at filling $\nu=-0.125$ that yield different $C_{2z}\mathcal{T}$ order parameters, and we perform DMRG calculations with bond dimension $1800$. We find that the result of DMRG is close to that of HF and CCSD, in terms of the energy and the value of the $C_{2z}\mathcal{T}$ order parameter. In particular, the DMRG results are also sensitive to the choice of the initial guess, see \cref{tab:peak_bott}. Both CCSD and DMRG calculations show that the Fermi-Dirac entropy of these local minima is consistently low, suggesting that the solution is again close to being a single Slater determinant.

\begin{table}[h!]
    \centering
    \renewcommand{\arraystretch}{1.4}
    \begin{tabular}{c||c|c|c||c|c|c|c||c|c|c|c}
         \hline
         Initialization & 
         E$_{\rm HF}$ & $C_{2z}\mathcal{T}$ &  S$_{\mathrm{FD}}$ &
         E$_{\rm CCSD}$ & e$_{\rm corr}$ & $C_{2z}\mathcal{T}$ & S$_{\rm FD}$ & 
         E$_{\rm DMRG}$ & e$_{\rm corr}$ & $C_{2z}\mathcal{T}$ & S$_{\rm FD}$\\
         \hline
         Sample 1& 
         -87.522 & 0.81 & 0.00 &
         -87.865 & -0.343 & 0.64 & 0.14 & 
         -87.823 & -0.301 & 0.78 & 0.10\\
         Sample 2& 
         -87.426 & 0.33 & 0.00 &
         -87.832 & -0.404 & 0.22 & 0.14 &
         -87.747 & -0.321 & 0.31 & 0.09\\ 
         \hline
    \end{tabular}
    \caption{Energy, $C_{2z} \mc{T}$ order parameter, and Fermi-Dirac entropy using two initializations and different methods at filling $\nu = -0.125$. 
    }
    \label{tab:peak_bott}
\end{table}







\vspace{1em}
\section{Discussion}
\label{sec:conclusion}

In this paper, we demonstrated that the techniques of correlated quantum chemistry can be fruitfully applied to study interacting models of the magic angle TBG system. 
We compared Hartree-Fock, coupled cluster, and DMRG calculations for ground state properties at both integer and non-integer fillings for a spinless, valleyless IBM model. Full-flavored IBM models, excited state properties, and other quantum chemistry methods 
are also within reach and will be studied in the future. 

We find that model discrepancies can be an important source of uncertainty. An ensemble of interacting models may be needed to cross-validate the results. To some extent, this model discrepancy is baked into the design of the IBM model: we start from a non-interacting continuum BM model, and the electron-electron interaction is added as an afterthought. A more reductionist approach could be to start from an interacting electron model at the continuum level, tune the parameters at a simplified level of theory (such as Hartree-Fock), and study the electron-correlation effects by projecting the model onto a smaller number of degrees of freedom. Such an approach would be at least self-consistent, and all errors and discrepancies could eventually be attributed to the errors in the continuum model.
The gauge-invariant order parameters, which are applicable to both unitary and antiunitary symmetries, could also be convenient in this setting since their implementation does not depend on the choice of the basis.
Methods based on quantum embedding theories~\cite{GeorgesKotliar1992,KniziaChan2012,SunChan2016} may also become useful in mitigating the modeling errors and in studying electron correlation effects in this process. 

Our current implementation treats all degrees of freedom equally. This includes the BM band index (or the sublattice index in the Chern band basis) and the $\vk$-point index in the current model, but can also include other flavor indices such as spin and valley degrees of freedom. This supercell treatment of the IBM model significantly reduces implementation efforts. Proper consideration of the crystal momentum conservation can reduce the scaling of both the computational and the storage cost with respect to $N_{\vk}$ (see e.g.,~\cite{HirataPodeszwaTobitaEtAl2004,McClainSunChanEtAl2017}) and will be considered in the future. 
Quantum chemistry packages are often designed to treat one particular flavor (spin). Therefore some further modifications may be needed if we would like to perform flavor-restricted/unrestricted calculations (which generalizes the spin restricted/unrestricted calculations in standard quantum chemistry methods).

Our numerical results indicate that even in the near integer filling regime ($\abs{\nu}<0.2$), it can be very challenging to converge to the global minima. This is not only for mean-field theories such as HF, but also DMRG calculations which are often considered to be more robust and less sensitive to the initial guess. We find that in the near integer filling regime, the system can be in a $C_{2z}\mc{T}$ symmetry breaking and metallic phase. Nonetheless, the entropy of these states is observed to be small, and can thus be relatively well described by a single Slater determinant. 
It seems reasonable to expect that the nature of the states can become qualitatively different as $\abs{\nu}$ increases, as recent results indicate that at $\nu=-2/3$ (or $1/3$ filling), the state of the system can be related to a fractional quantum Hall state (FQHE) which is distinct from a Slater determinant~\cite{ParkerLedwithKhalafEtAl2021}. Our preliminary results indicate that convergence in the more heavily doped regime ($\abs{\nu}>0.2$) can be more challenging and the entropy indeed increases as $\abs{\nu}$ increases. The results will be reported in a future publication.

\vspace{1em}
\textbf{Acknowledgments}

This work was partially supported by the Air Force Office of Scientific Research under award number FA9550-18-1-0095 (F.M.F., R.K., G.K.C.), by the Simons Targeted Grants in Mathematics and Physical Sciences on Moir\'e Materials Magic (Q.Z., K.D.S.). T.S. is supported by the Masason Foundation. This material is based upon work supported by the U.S. Department of Energy, Office of Science, National Quantum Information Science Research Centers, Quantum Systems Accelerator (M.Z.), and by the U.S. Department of Energy, Office of Science, Office of Advanced Scientific Computing Research and Office of Basic Energy Sciences, Scientific Discovery through Advanced Computing (SciDAC) program under Award Number DE-SC0022198 (L.L.).
 G.K.C.  and L.L. are Simons Investigators.


\bibliographystyle{abbrvnat}
\bibliography{lin_ref_1,reference}

\widetext
\clearpage
\appendix
\renewcommand{\thesubsection}{\Alph{section}.\arabic{subsection}}

\section{The Decoupled Subtraction term}
\label{sec:dec_subtraction}
To fix notation, suppose that we are constructing a decoupled subtraction Hamiltonian for the BM Hamiltonian $\hat{H}_{BM}(\vk)$ with interlayer couplings $w_0, w_1$. For each $\vk$, we can diagonalize this Hamiltonian to get a family of periodic BM orbitals $\big\{ u_{n\vk} : n \in \{ \pm 1, \cdots, \pm N/2 \} \big\}.$ Furthermore, recall that we identify the two ``flat'' bands near zero energy with the indices $n \in \{ -1, 1\}$.

For any such BM Hamiltonian, we define a corresponding decoupled TBG Hamiltonian $\hat{H}_{BM, {\rm dec}}(\vk)$, by setting the interlayer coupling equal to zero (i.e., $w_0 = w_1 = 0$). The decoupled subtraction is obtained first by projecting the zero temperature limit of the density matrix corresponding to two decoupled graphene layers to the BM band basis of the TBG system, i.e.,
\begin{equation}
\label{eq:p0_dec}
[P^0(\vk)]_{mn} := \lim_{\beta \rightarrow \infty} \braket{u_{m\vk} | \left(\exp(-\beta \hat{H}_{BM,\text{dec}}(\vk)) + 1\right)^{-1} | u_{n\vk} },
\end{equation}
where $m,n \in \{ -1, 1 \}$ which we recall correspond to the two flat bands. 
After computing $P^0(\vk)$ using~\cref{eq:p0_dec} we define $\hat{H}_{\rm sub} := v_{\mathrm{hf}}[P^0(\vk)]$ where $v_{\mathrm{hf}}[\cdot]$ is defined in~\cref{eqn:hf-rho}.
The computational procedure to construct the decoupled subtraction Hamiltonian is summarized as pseudo code in Alg.~\ref{alg:dec_ham}.

\begin{algorithm}
\caption{\label{alg:dec_ham} Construction of decoupled subtraction Hamiltonian}
\makebox[15mm]{\hfill $\mathrm{B}_{\mathrm{FB}}$} $\leftarrow$ Compute BM band structure (with interlayer tunneling) restricted to the flat bands\;
\makebox[15mm]{\hfill $\mathrm{B}_{\mathrm{dec}}, \mathrm{E}_{\mathrm{dec}}$} $\leftarrow$ Compute BM band structure without interlayer tunneling\;
\makebox[15mm]{\hfill f$_E$} $\leftarrow$ $\big(\exp(\beta ( \mathrm{E}_{\mathrm{dec}} - \mu ) ) + 1\big)^{-1}$ (Compute Fermi-Dirac statistic)\;
\makebox[15mm]{\hfill $P^0$} $\leftarrow$ $ \mathrm{B}^\dagger_{\mathrm{FB}} \mathrm{B}_{\mathrm{dec}} \mathrm{f}_E \mathrm{B}_{\mathrm{dec}}^\dagger \mathrm{B}_{\mathrm{FB}}$\;
\makebox[15mm]{\hfill H$_{\rm sub}$} $\leftarrow$ $\hat{\rm V}_{\rm hf}[P^0]$, see~\cref{eqn:subtracted_HF}\;
\end{algorithm}
We remark that the procedure outlined in Alg.~\ref{alg:dec_ham} has been used in previous articles \cite{SoejimaParkerBultinckEtAl2020,XieMacDonald2020}. A more recent work \cite{PotaszXieMacDonald2021} has suggested that including contributions from non-flat BM bands (i.e. $\{ u_{n\vk} : |n| > 1 \}$) is important for correctly modeling phenomenological properties of TBG physics. We leave investigations of these effects to future work.


\section{Interaction term in the IBM Hamiltonian}
\label{sec:ham_i}

In the Coulomb interaction term, the summation is only over $\vk,\vk',\vk''\in\Omega^*$. This is due to the crystal momentum conservation, i.e., $\vk+\vk'-\vk''-\vk'''=\bm{0}$.
Here we follow the convention that the fourth momentum vector $\vk'''=\vk+\vk'-\vk''$ may be outside $\Omega^*$.
Many practical implementations restrict $\vk'''$ to be within the mBZ as well, and hence the crystal momentum conservation condition becomes $\vk+\vk'-\vk''-\vk'''=\vG$ for some $\vG\in\mathbb{L^*}$. We do not use this convention in this paper.
Another way to express the crystal momentum conservation condition is to write $\vk''=\vk+\vq$, $\vk'''=\vk'-\vq$. 
Note that in this convention, both $\vq$ and $\vk'''$ may be outside the mBZ.

In TBG, the flat bands are not only affected through the screened interlayer Coulomb interactions (the screening is modeled by a dielectric constant $\epsilon$) but also through the Coulomb screening from the top and bottom metallic gates.
Assume that the TBG sample is placed in a $x$\nobreakdash-$y$~plane that is of distance $d/2$ to the top and bottom metallic gate plates along the $z$ direction. 
Since the moir\'e length scale is much larger than the distance between the sublattice sites and the layers, we may assume that the Coulomb interaction depends only on the difference $\delta r= \vr-\vr'$, and does not depend on the sublattice and layer indices. 
The screened Coulomb interaction in real space (called a double gate-screened Coulomb potential) can be written as
\begin{equation}
V(\vr)=U_{d} \sum_{n=-\infty}^{\infty} \frac{(-1)^{n}}{\sqrt{\abs{\vr}^2+(nd)^{2}}}.
\end{equation}
Here, $U_d=\epsilon^{-1}$ parametrizes the strength of the screened Coulomb interaction. In Fourier space, the Coulomb interaction takes the form
\begin{equation}
V(\vq)=\int V(\vr) e^{-\I \vq \cdot \vr}\ud \vr=2\pi U_d\frac{\tanh(\abs{\vq}d/2)}{\abs{\vq}},
\label{eqn:doublegate_q}
\end{equation}
see \cite[Appendix C]{BernevigSongRegnaultEtAl2021} for more details.
Note that with a slight abuse of notation we here use $V(\vq)$ to denote the Fourier transform of $V(\vr)$.

The ERIs take the form
\begin{equation}
\begin{split}
&\braket{m\vk, m'\vk'|n\vk'', n' \vk'''}\\
\qquad&=\sum_{\sigma,l,\sigma',l'}\int V(\vr-\vr')\psi^{*}_{m\vk}(\vr,\sigma,l)
\psi^{*}_{m'\vk'}(\vr',\sigma',l')\psi_{n\vk''}(\vr,\sigma,l)
\psi_{n'\vk'''}(\vr',\sigma',l') \ud \vr \ud \vr'\\
\qquad&=\frac{1}{N_{\vk}^2}
\,\int V(\vr-\vr') e^{\I \vq\cdot (\vr-\vr')}\varrho_{m\vk,n(\vk+\vq)}(\vr) \varrho_{m'\vk',n'(\vk'-\vq)}(\vr')\ud \vr \ud \vr'.
\end{split}
\label{eqn:eri_realspace}
\end{equation}
Note that the integration is performed in the moir\'e supercell.
The pair product of the periodic components of the BM bands in the real space (summed over the sublattice and layer indices) is defined as
\begin{equation}
\varrho_{m\vk,n\vk''}(\vr)
=\sum_{\sigma,l}u^*_{m\vk}(\vr,\sigma,l)  u_{n\vk''}(\vr,\sigma,l).
\end{equation}
Expanding the periodic Bloch functions in terms of their Fourier coefficients, the pair product can be written as
\begin{equation}
\label{eqn:pair-product}
    \varrho_{m\vk,n\vk''}(\vr) = \frac{1}{|\Omega|^2} \sum_{\vG, \vG' \in \mathbb{L}^*} \sum_{\sigma,l}u^*_{m\vk}(\vG,\sigma,l)  u_{n\vk''}(\vG',\sigma,l) e^{i (\vG' - \vG) \cdot r}.
\end{equation}
Recall that by definition of the periodic Bloch functions $\psi_{n\vk''}(\vr) = \psi_{n,(\vk'' + \vG)}(\vr)$ for all $\vG \in \mathbb{L}^*$.
Therefore, $u_{n\vk''}(\vr) = e^{i \vG \cdot \vr} u_{n,(\vk''+\vG)}(\vr)$ and consequently, 
\begin{equation}
\begin{aligned}
    & &u_{n\vk''}(\vr) &= e^{i \vG \cdot \vr} u_{n,(\vk''+\vG)}(\vr) \\[1ex]
    & \Rightarrow& \sum_{\vG'} u_{n\vk''}(\vG') e^{i \vG' \cdot \vr} &=  \sum_{\vG'} u_{n,(\vk''+\vG)}(\vG') e^{i (\vG' + \vG) \cdot \vr} \\[1ex]
    & \Rightarrow& u_{n\vk''}(\vG') &= u_{n,(\vk''+\vG)}(\vG' - \vG).
\end{aligned}
\end{equation}
Hence, 
\begin{equation}
    \varrho_{m\vk,n\vk''}(\vr) = \frac{1}{|\Omega|^2} \sum_{\vG, \vG' \in \mathbb{L}^*} \sum_{\sigma,l}u^*_{m\vk}(\vG,\sigma,l)  u_{n,\vk''+\vG}(\vG' - \vG,\sigma,l) e^{i (\vG' - \vG) \cdot r}.
\end{equation}
The Fourier coefficients of the pair product then satisfy
\begin{equation}
\varrho_{m \vk,n(\vk+\vq)}(\vG)=\frac{1}{\abs{\Omega}}
\sum_{\vG'\in\mathbb{L}^*} \sum_{\sigma,l}u^*_{m\vk}(\vG',\sigma,l) u_{n(\vk+\vq+\vG)}(\vG',\sigma,l)=[\Lambda_{\vk}(\vq+\vG)]_{mn},
\label{eqn:form_factor}
\end{equation}
where the matrix $\Lambda_{\vk}(\vq+\vG)$ is called the \textit{form factor}.
Note that due to a shift in $\vG$, the Fourier coefficients of the pair product can be written as an inner product opposed to the convolution that arises directly from~\cref{eqn:pair-product}.

From \cref{eqn:form_factor} we can verify that the form factor satisfies the symmetry condition
\begin{equation}
[\Lambda_{\vk}(\vq+\vG)]^*_{mn}=[\Lambda_{\vk+\vq}(-\vq-\vG)]_{nm}.
\label{eqn:form_factor_symmetry}
\end{equation}

Using the definitions of the Coulomb interaction $V(\vq)$ in~\cref{eqn:doublegate_q} and the form factor in~\cref{eqn:form_factor}, the ERI can be expressed in the Fourier space as
\begin{equation}
\begin{split}
\braket{m\vk, m'\vk'|n\vk'', n' \vk'''}=&\frac{1}{|\Omega| N_{\vk}} \sum_{\vG\in \mathbb{L}^*} V(\vq+\vG) \varrho_{m\vk,n(\vk+\vq)}(\vG)
\varrho_{m'\vk',n'(\vk'-\vq)}(-\vG)\\
=&\frac{1}{|\Omega| N_{\vk}} \sum_{\vG\in \mathbb{L}^*} V(\vq+\vG)  [\Lambda_{\vk}(\vq+\vG)]_{mn} [\Lambda_{\vk'}(-\vq-\vG)]_{m'n'}.
\end{split}
\label{eqn:IBM_ERI}
\end{equation}
Compared to \cref{eqn:eri_realspace}, we gained an extra factor $N_\vk$ due to the integration in the moir\'e supercell.
Hence, the interaction Hamiltonian is fully determined by the form factor and the screened Coulomb potential as
\begin{equation}
\begin{split}
\hat{H}_I=&\frac{1}{2} 
\sum_{\vk,\vk',\vk''\in \Omega^*}\sum_{mm'nn'}\braket{m\vk, m'\vk'|n\vk'', n' \vk'''}\hat f_{m\vk}^{\dagger} \hat f_{m'\vk'}^{\dagger} \hat f_{n'\vk'''} \hat f_{n\vk''}\\
=& \frac{1}{2 |\Omega| N_{\vk}} \sum_{\vk,\vk',\vk+\vq=\vk''\in  \Omega^*}\sum_{mm'nn'}
\sum_{\vG\in \mathbb{L}^*} V(\vq+\vG) \braket{u_{m\vk}|u_{n(\vk+\vq+\vG)}}\\
&
\times\braket{u_{m'\vk'}|u_{n'(\vk'-\vq-\vG)}} \hat f_{m\vk}^{\dagger} \hat f_{m'\vk'}^{\dagger} \hat f_{n'(\vk'-\vq)} \hat f_{n(\vk+\vq)}\\
=&\frac{1}{2 |\Omega| N_{\vk}} \sum_{\vk,\vk',\vk+\vq=\vk''\in \Omega^*}\sum_{mm'nn'}
\sum_{\vG\in \mathbb{L}^*} V(\vq+\vG) [\Lambda_{\vk}(\vq+\vG)]_{mn}
\\
&
\times[\Lambda_{\vk'}(-\vq-\vG)]_{m'n'} \hat f_{m\vk}^{\dagger} \hat f_{m'\vk'}^{\dagger} \hat f_{n'(\vk'-\vq)} \hat f_{n(\vk+\vq)}.
\end{split}
\label{eqn:interaction_comput}
\end{equation}
This can be further simplified by treating $\vq+\vG$ as a new variable $\vq'$, which takes values in the entire reciprocal space.
Using the periodicity of the creation and annihilation operators (see~\cref{eqn:period_f}), we have
\begin{equation}
\begin{split}
\hat{H}_I=&\frac{1}{2 |\Omega| N_{\vk}} \sum_{\vk,\vk'\in \Omega^*}\sum_{mm'nn'}
\sum_{\vq'} V(\vq') [\Lambda_{\vk}(\vq')]_{mn}
\\
&
\times [\Lambda_{\vk'}(-\vq')]_{m'n'} \hat f_{m\vk}^{\dagger} \hat f_{m'\vk'}^{\dagger} \hat f_{n'(\vk'-\vq')} \hat f_{n(\vk+\vq')}.
\end{split}
\end{equation} 
Defining a pseudo density operator
\begin{equation}
\hat{\rho}_{\vq}=\sum_{\vk\in \Omega^*} \sum_{mn} \hat f_{m\vk}^{\dagger} [\Lambda_{\vk}(\vq)]_{mn} \hat f_{n(\vk+\vq)},
\label{eqn:rho_operator}
\end{equation}
and using normal ordering, i.e.,
\begin{equation}
:f_{m\vk}^{\dagger}  \hat f_{n(\vk+\vq)}f_{m'\vk'}^{\dagger}  \hat f_{n'(\vk'-\vq)}:\,
=f_{m\vk}^{\dagger}  f_{m'\vk'}^{\dagger}  \hat f_{n'(\vk'-\vq)}\hat f_{n(\vk+\vq)},
\end{equation}
we can further simplify $\hat{H}_I$ to 
\begin{equation}
\hat{H}_I=\frac{1}{2 \abs{\Omega}N_{\vk}} 
\sum_{\vq'} V(\vq'):\hat{\rho}_{\vq'}\hat{\rho}_{-\vq'}:.
\label{eqn:interaction_simplify}
\end{equation}
This form is reminiscent of the density-density Coulomb interaction in the reciprocal space. 
The expression \cref{eqn:interaction_simplify} has certain conceptual advantages, e.g., when writing down the Hartree-Fock potential. However, it does not reduce the implementation effort compared to \cref{eqn:interaction_comput}.

\section{Coulomb and exchange terms in Hartree-Fock theory}
\label{sec:hf_deriv}

For an explicit definition of the Coulomb term, we apply Wick's theorem to the interaction term~\eqref{eqn:interaction_simplify}, and use the relation in \cref{eqn:average_f}. 
This yields

\begin{equation}
\label{eqn:Coulomb_op_long}
\begin{split}
\hat{J}[P]=&\frac{1}{2 |\Omega| N_{\vk}} \sum_{\vk,\vk'\in \Omega^*}\sum_{mm'nn'}
\sum_{\vq'} V(\vq') [\Lambda_{\vk}(\vq')]_{mn} [\Lambda_{\vk'}(-\vq')]_{m'n'}
\\
&
\times \left(\braket{\Psi_S|\hat f_{m\vk}^{\dagger}\hat f_{n(\vk+\vq')}|\Psi_S} \hat f_{m'\vk'}^{\dagger} \hat f_{n'(\vk'-\vq')}+ \braket{\Psi_S|\hat f_{m'\vk'}^{\dagger} \hat f_{n'(\vk'-\vq')}|\Psi_S} \hat f_{m\vk}^{\dagger}\hat f_{n(\vk+\vq')}\right)\\
=&\frac{1}{2 |\Omega| N_{\vk}} \sum_{\vk,\vk'\in \Omega^*}\sum_{mm'nn'}
\sum_{\vq'} V(\vq') [\Lambda_{\vk}(\vq')]_{mn} [\Lambda_{\vk'}(-\vq')]_{m'n'}\\
&
\times \sum_{\vG\in\mathbb{L}^*} \left([P(\vk)]_{nm}\delta_{\vq',-\vG}\hat f_{m'\vk'}^{\dagger} \hat f_{n'\vk'}+[P(\vk')]_{n'm'}\delta_{\vq',\vG}\hat f_{m\vk}^{\dagger}\hat f_{n\vk}\right)\\
=&\frac{1}{|\Omega|} 
\sum_{\vG\in\mathbb{L}^*} V(\vG)\left(\frac{1}{N_{\vk}}\sum_{\vk'\in \Omega^*} \Tr[ \Lambda_{\vk'}(-\vG) P(\vk')]\right)\left(\sum_{\vk\in \Omega^*} \sum_{mn} [\Lambda_{\vk}(\vG)]_{mn}\hat f_{m\vk}^{\dagger}  \hat f_{n\vk}\right).
\end{split}
\end{equation}
In the last step we have used the fact that $V(\vq')=V(-\vq')$. 
To simplify the expression in~\cref{eqn:Coulomb_op_long}, we can define the electron density $\wt{\rho}$, i.e., 
\begin{equation}
\wt{\rho}[P](\vG)=\frac{1}{N_{\vk}} 
\sum_{\vk'\in \Omega^*}\Tr[ \Lambda_{\vk'}(\vG) P(\vk')],
\end{equation}
which is periodic with respect to the moir\'e unit cell.  
Using the definition of the pseudo-density operator in~\cref{eqn:rho_operator}, the expression of the Coulomb operator can be simplified into the standard density-density form 
\begin{equation}
\hat{J}[P]=\frac{1}{|\Omega|} \sum_{\vG} V(\vG) \wt{\rho}[P](-\vG)\hat{\rho}_{\vG}.
\end{equation}

We proceed similarly for the exchange operator. Using Wick's theorem yields
\begin{equation}
\begin{split}
\hat{K}[P]=&\frac{1}{2 |\Omega| N_{\vk}} \sum_{\vk,\vk'\in \Omega^*}\sum_{mm'nn'}
\sum_{\vq'} V(\vq') [\Lambda_{\vk}(\vq')]_{mn} [\Lambda_{\vk'}(-\vq')]_{m'n'}
\\
&
\times\left(\braket{\Psi_S|\hat f_{m\vk}^{\dagger}\hat f_{n'(\vk'-\vq')}|\Psi_S} \hat f_{m'\vk'}^{\dagger}\hat f_{n(\vk+\vq')} + \braket{\Psi_S|\hat f_{m'\vk'}^{\dagger}\hat f_{n(\vk+\vq')}|\Psi_S} \hat f_{m\vk}^{\dagger}\hat f_{n'(\vk'-\vq')}\right)\\
=&
\frac{1}{2 |\Omega| N_{\vk}} \sum_{\vk,\vk'\in \Omega^*}\sum_{mm'nn'}
\sum_{\vq'} V(\vq') [\Lambda_{\vk}(\vq')]_{mn} [\Lambda_{\vk'}(-\vq')]_{m'n'}\\
&
\times \sum_{\vG\in\mathbb{L}^*} \left([P(\vk)]_{n'm}\delta_{\vk,\vk'-\vq'+\vG}
\hat f_{m'\vk'}^{\dagger}\hat f_{n(\vk+\vq')}+[P(\vk')]_{nm'}\delta_{\vk',\vk+\vq'+\vG}
\hat f_{n'(\vk'-\vq')}\right)\\
=&
\frac{1}{ |\Omega| N_{\vk}} \sum_{\vq'}\sum_{\vk\in \Omega^*}
\sum_{mm'nn'} V(\vq') [\Lambda_{\vk}(\vq')]_{mn} [P(\vk+\vq')]_{nm'}[\Lambda_{\vk+\vq'}(-\vq')]_{m'n'}
\hat f_{m\vk}^{\dagger}\hat f_{n'\vk}. 
\end{split}
\end{equation}

\section{Properties of the Sewing Matrix} \label{sec:sewing_matrix}
For both unitary symmetries $g$ and antiunitary symmetries $g \mc{K}$, we can define unitary sewing matrices $[B(g)]_{\vk}$ or $[B(g \mc{K})]_{\vk}$ which describe how the band creation operators $\hat{f}^{\dag}_{m,g\vk}$ transform under symmetries. In particular
\begin{equation}
\label{eqn:band-sewing-eq}
\begin{split}
g \hat{f}^{\dag}_{n\vk} g^{-1}
=& \sum_{m} \hat{f}^{\dag}_{m,g\vk}  [B(g)]_{\vk,mn}, \\
(g \mc{K}) \hat{f}^{\dag}_{n\vk} (g \mc{K})^{-1} 
=& \sum_{m} \hat{f}^{\dag}_{m,g\vk}  [B(g \mc{K})]_{\vk,mn}.
\end{split}
\end{equation}
We will prove~\cref{eqn:band-sewing-eq} in~\cref{sec:sewing_unitary_case,sec:sewing_antiunitary_case}. We will also show that in both cases, assuming the existence of a direct band gap, that the resulting sewing matrix is unitary in~\cref{sec:sewing_matrix_unitary}.

\subsection{Sewing Matrix: Unitary Case}
\label{sec:sewing_unitary_case}
Recall that for any unitary symmetry operation, $g$, there exists a representation matrix $D(g)$ so that the creation operators, $c_{\vk}^\dagger$, transform via the rule
\begin{equation}
\begin{split}
    (g \hat{c}_{\vk}^\dag g^{-1})(\alpha) & = \sum_{\alpha'} \hat{c}_{g\vk}^\dag(\alpha') [D(g)]_{\alpha',\alpha}, \\
    (g \hat{c}_{\vk} g^{-1})(\alpha) & = \sum_{\alpha'} \hat{c}_{g\vk}(\alpha') [D(g)]_{\alpha',\alpha}^{*} .
  \end{split}
\end{equation}
Now we can calculate the action of this operation on any quadratic Hamiltonian in the primitive basis $\hat{H}=\sum_{\vk}\sum_{\alpha,\beta}[h(\vk)]_{\alpha\beta} \hat{c}^{\dag}_{\vk}(\alpha) \hat{c}_{\vk}(\beta)$. In particular
\begin{equation}
\begin{split}
g \hat{H} g^{-1}=&g \sum_{\vk} \sum_{\alpha\beta}[h(\vk)]_{\alpha\beta} \hat{c}^{\dag}_{\vk}(\alpha) \hat{c}_{\vk}(\beta)g^{-1}\\
=& \sum_{\vk} \sum_{\alpha\beta} \sum_{\alpha'\beta'}[h(\vk)]_{\alpha\beta} \hat{c}^{\dag}_{g\vk}(\alpha') [D(g)]_{\alpha',\alpha} \hat{c}_{g\vk}(\beta') [D(g)]^*_{\beta',\beta}\\
=& \sum_{\vk} [D(g)h(\vk)D^{\dag}(g)]_{\alpha'\beta'}\hat{c}^{\dag}_{g\vk}(\alpha') \hat{c}_{g\vk}(\beta').
\end{split}
\end{equation}

If $\hat{H}$ is invariant under $g$, i.e., $g \hat{H} g^{-1}=\hat{H}$, then this calculation shows that at the matrix level:
\begin{equation}
D(g) h(\vk)=h(g\vk) D(g).
\end{equation}
In other words,
\begin{equation}
h(g\vk)D(g)u_{n\vk}=D(g)h(\vk)u_{n\vk}=\varepsilon_{n\vk} D(g)u_{n\vk},
\end{equation}
or equivalently $D(g)u_{n\vk}$ is an eigenfunction of $h(g\vk)$ with the eigenvalue $\varepsilon_{n,g\vk}=\varepsilon_{n\vk}$. Hence, from this relation we can define the ``sewing matrix'', $[B(g)]_{\vk}$,
\begin{equation}
(D(g) u_{n\vk})(\alpha)=
\sum_{\alpha'} [D(g)]_{\alpha,\alpha'} u^*_{n\vk}(\alpha')=
\sum_{m}  u_{m,g\vk}(\alpha) [B(g)]_{\vk,mn}.
\end{equation}
Here
\begin{equation}
[B(g)]_{\vk,mn}=\braket{u_{m,g\vk} | D(g) | u_{n\vk}},
\end{equation}
where the bra-ket notation denotes contraction over the internal indices $\alpha,\alpha'$.
Similar to the unitary case, when the system is gapped, it is easily verified that $[B(g)]_{\vk}$ is a unitary matrix using a contour integral argument (see~\cref{sec:sewing_matrix_unitary}). Given the definition of sewing matrices, one can calculate that:
\begin{equation}
\begin{split}
g \hat{f}^{\dag}_{n\vk} g^{-1}=&\sum_{\alpha}  g\hat{c}_{\vk}^{\dagger}(\alpha) g^{-1}u^*_{n\vk}(\alpha)\\
=&\sum_{\alpha,\alpha'}  \hat{c}^{\dag}_{g\vk}(\alpha')[D(g)]_{\alpha',\alpha}
u^*_{n\vk}(\alpha)\\
=& \sum_{\alpha'} \hat{c}^{\dag}_{g\vk}(\alpha') \sum_{m} u_{m,g\vk}(\alpha') [B(g)]_{\vk,mn} \\
=& \sum_{m} \hat{f}^{\dag}_{m,g\vk}  [B(g)]_{\vk,mn}.
\end{split}
\label{eqn:sewing_transf_unitary}
\end{equation}

\subsection{Sewing Matrix: Antiunitary Case}
\label{sec:sewing_antiunitary_case}
We now consider the case of antiunitary symmetries. Recall that any antiunitary symmetry may be written as $g \mc{K}$ and hence the primitive creation operators transform as:
\begin{equation}
\begin{split}
    (g \mc{K}) \hat{c}_{\vk}^\dag (g \mc{K})^{-1})(\alpha) & = \sum_{\alpha'} \hat{c}_{g\vk}^\dag(\alpha') [D(g)]_{\alpha',\alpha}, \\
    (g \mc{K}) \hat{c}_{\vk} (g \mc{K})^{-1})(\alpha) & = \sum_{\alpha'} \hat{c}_{g\vk}(\alpha') [D(g)]_{\alpha',\alpha}^{*}.
\end{split}
\end{equation}
Now we can calculate the action of this operation on any quadratic Hamiltonian in the primitive basis $\hat{H}=\sum_{\vk}\sum_{\alpha,\beta}[h(\vk)]_{\alpha\beta} \hat{c}^{\dag}_{\vk}(\alpha) \hat{c}_{\vk}(\beta)$. In particular
\begin{equation}
\begin{split}
(g \mc{K}) \hat{H} (g \mc{K})^{-1}=&g \mc{K} \sum_{\vk} \sum_{\alpha\beta}[h(\vk)]_{\alpha\beta} \hat{c}^{\dag}_{\vk}(\alpha) \hat{c}_{\vk}(\beta)\mc{K} g^{-1}\\
=& \sum_{\vk} \sum_{\alpha\beta} \sum_{\alpha'\beta'}[h^*(\vk)]_{\alpha\beta} \hat{c}^{\dag}_{g\vk}(\alpha') [D(g \mc{K})]_{\alpha',\alpha} \hat{c}_{g\vk}(\beta') [D(g \mc{K})]^*_{\beta',\beta}\\
=& \sum_{\vk} [D(g \mc{K})h^*(\vk)D^{\dag}(g \mc{K})]_{\alpha'\beta'}\hat{c}^{\dag}_{g\vk}(\alpha') \hat{c}_{g\vk}(\beta').
\end{split}
\end{equation}

If $\hat{H}$ is invariant under $g \mc{K}$, i.e., $g \mc{K}\hat{H}(g \mc{K})^{-1}=\hat{H}$, then this calculation shows that at the matrix level:
\begin{equation}
D(g \mc{K}) h^*(\vk)=h(g\vk) D(g \mc{K}).
\end{equation}
In other words, 
\begin{equation}
h(g\vk)D(g \mc{K})u^*_{n\vk}=D(g \mc{K})h^*(\vk)u^*_{n\vk}=\varepsilon_{n\vk} D(g \mc{K})u^*_{n\vk},
\end{equation}
or equivalently $D(g \mc{K})u^*_{n\vk}$ is an eigenfunction of $h(g\vk)$ with the eigenvalue $\varepsilon_{n,g\vk}=\varepsilon_{n\vk}$. Hence, similar to the unitary case, we can define the sewing matrix, $[B(g \mc{K})]_{\vk}$,
\begin{equation}
(D(g \mc{K}) u^*_{n\vk})(\alpha)=
\sum_{\alpha'} [D(g \mc{K})]_{\alpha,\alpha'} u^*_{n\vk}(\alpha')=
\sum_{m}  u_{m,g\vk}(\alpha) [B(g \mc{K})]_{\vk,mn}.
\end{equation}
Here
\begin{equation}
[B(g \mc{K})]_{\vk,mn}=\braket{u_{m,g\vk} | D(g \mc{K}) | u^*_{n\vk}},
\end{equation}
where the bra-ket notation denotes contraction over the internal indices $\alpha,\alpha'$.
Similar to the unitary case, when the system is gapped, it is easily verified that $[B(g \mc{K})]_{\vk}$ is a unitary matrix using a contour integral argument (see~\cref{sec:sewing_matrix_unitary}). Given the definition of sewing matrices, one can calculate that:
\begin{equation}
\begin{split}
g \mc{K} \hat{f}^{\dag}_{n\vk} \mc{K} g^{-1}=&\sum_{\alpha}  g\hat{c}_{\vk}^{\dagger}(\alpha) g^{-1}u^*_{n\vk}(\alpha)\\
=&\sum_{\alpha,\alpha'}  \hat{c}^{\dag}_{g\vk}(\alpha')[D(g \mc{K})]_{\alpha',\alpha}
u^*_{n\vk}(\alpha)\\
=& \sum_{\alpha'} \hat{c}^{\dag}_{g\vk}(\alpha') \sum_{m} u_{m,g\vk}(\alpha') [B(g \mc{K})]_{\vk,mn} \\
=& \sum_{m} \hat{f}^{\dag}_{m,g\vk}  [B(g \mc{K})]_{\vk,mn},
\end{split}
\label{eqn:sewing_transf_antiunitary}
\end{equation}
which is of the same form as the unitary case in~\cref{eqn:sewing_transf_unitary}.

\subsection{Unitarity of Sewing Matrices} \label{sec:sewing_matrix_unitary}
Suppose that we are given a single body Hamiltonian $h(\vk)$ with eigenvectors/eigenvalue pairs $\{ (u_{n\vk}(\alpha), \epsilon_{n\vk}) \}$ where $\vk \in BZ$ and $\alpha$ is a multi-index running over the additional degrees of freedom (in this work, sublattice, layer, valley, spin).
Suppose further that $h(\vk)$ satisfies a unitary symmetry $g$ or an antiunitary symmetry $g \mc{K}$. Next, let us fix a set of occupied bands $\mc{I}_{occ}$ and define $N_b := \# |\mc{I}_{occ}|$. 
For such a system, we can define the sewing matrices $[B(g)]_{\vk} \in \CC^{N_b \times N_b}$ and $[B(g \mc{K})]_{\vk} \in \CC^{N_b \times N_b}$  as follows
\begin{equation}
\begin{split}
    [B(g)]_{\vk,mn} &= \braket{u_{m,g\vk} | D(g) | u_{n\vk}}, \\
    [B(g\mc{K})]_{\vk,mn} &= \braket{u_{m,g\vk} | D(g \mc{K}) | u^*_{n\vk}} ,
\end{split}   
\end{equation}
where the indices $m,n \in \mc{I}_{occ}$.
In this section, we prove if the occupied bands are separated in energy from the unoccupied bands, then sewing matrices $[B(g)]_{\vk}$ and $[B(g \mc{K})]_{\vk}$ are unitary. More formally, the sewing matricies as defined above are unitary if there exists a constant $c$ such that for all $\vk$ 
\[
\min_{n \in \mc{I}_{occ}} \min_{m \not\in \mc{I}_{occ}} | \epsilon_{n\vk} - \epsilon_{m\vk} | \geq c > 0.
\]

We first consider the case where $g$ is a unitary symmetry. In this case, it suffices to show that the set $\{ D(g) \ket{u_{n\vk}} : n \in \mc{I}_{occ} \}$ and the set $\{ \ket{u_{n,g\vk}} : n \in \mc{I}_{occ} \}$ are two different orthonormal bases for the same space. Due to the gap assumption, the occupied projector $P_{occ}(\vk)$ can be written a sum of exterior products:
\[
P_{occ}(\vk) = \sum_{n \in \mc{I}_{occ}} | u_{n\vk} \rangle \langle u_{n\vk} |.
\]
By the Riesz projection formula, we can also represent the occupied projector as a contour integral
\[
P_{occ}(\vk) = \frac{1}{2 \pi i} \int_{\mc{C}} (z - h(\vk))^{-1} dz,
\]
where $\mc{C}$ is a closed contour in the complex plane enclosing the eigenvalues $\{ \epsilon_{n\vk} : n \in \mc{I}_{occ} \}$. Conjugating both sides by the representation matrix $D(g)$ then gives that
\begin{equation}
\begin{split}
    D(g) P_{occ}(\vk) (D(g))^\dagger & = \frac{1}{2 \pi i} \int_{\mc{C}} D(g) (z - h(\vk))^{-1} (D(g))^\dagger dz \\
    & = \frac{1}{2 \pi i} \int_{\mc{C}} \Big((z - D(g) h(\vk) (D(g))^\dagger\Big)^{-1} dz \\
    & = \frac{1}{2 \pi i} \int_{\mc{C}} (z - h(g\vk) )^{-1} dz \\
    & = P_{occ}(g\vk).
\end{split}
\end{equation}
Hence
\[
P_{occ}(g\vk) =  \sum_{n \in \mc{I}_{occ}} D(g) | u_{n\vk} \rangle \langle u_{n\vk} | (D(g))^\dagger.
\]
Since $D(g)$ is unitary, it follows that $P_{occ}(\vk)$ and $P_{occ}(g\vk)$ have the same rank and $\{ D(g) \ket{u_{n\vk}} : n \in \mc{I}_{occ} \}$ is a complete orthogonal basis for the range of $P_{occ}(g\vk)$, completing the argument

We now turn to consider the sewing matrix for an antiunitary symmetry $g \mc{K}$. In this case, we instead consider $P_{occ}(\vk)^*$ which is also given by taking the complex conjugation of the Riesz projection formula:
\begin{equation}
P_{occ}(\vk)^* = \frac{-1}{2 \pi i} \int_{\mc{C}} (\overline{z} - h^{*}(\vk))^{-1} dz.
\end{equation}
We may assume without loss of generality that the contour $\mc{C}$ is symmetric about the real axis, therefore performing the change of variables $z \mapsto \overline{z}$ and reversing the orientation of the contour we conclude that
\begin{equation}
P_{occ}(\vk)^* = \frac{1}{2 \pi i} \int_{\mc{C}} (z - h^{*}(\vk))^{-1} dz.
\end{equation}
Finally, conjugating by $D(g)$ yields
\begin{equation}
\begin{split}
    D(g \mc{K}) P_{occ}(\vk)^* (D(g \mc{K}))^\dagger
    & = \frac{1}{2 \pi i} \int_{\mc{C}} (z - D(g \mc{K}) h^*(\vk)(D(g \mc{K}))^\dagger)^{-1}  dz \\
     & = \frac{1}{2 \pi i} \int_{\mc{C}} (z - h(g\vk))^{-1}  dz \\
    & = P_{occ}(g\vk).
    \end{split}
\end{equation}
Hence, $P_{occ}(\vk)^*$ and $P_{occ}(g\vk)$ have the same rank and $\{ D(g) \ket{u^*_{n\vk}} : n \in \mc{I}_{occ}\}$ is a complete orthogonal basis for the range of $P_{occ}(g\vk)$ as we wanted to show.




\section{Gauge-invariant order parameter} \label{sec:symm_order_param}
\subsection{Order parameter: Unitary Case} \label{sec:symm_order_param_unitary}

If the many-body wavefunction $\ket{\Psi}$ satisfies the symmetry $g$, then
\[
g\ket{\Psi}\bra{\Psi}g^{-1}=\ket{\Psi}\bra{\Psi}.
\]
Here $g$ acting on $\ket{\Psi}$ is implemented by $g$ acting on each individual creation operators, and the vacuum is an eigenfunction of any symmetry $g\ket{0}=\ket{0}$.
Therefore the 1-RDM should satisfy
\begin{equation}
\begin{split}
[P(\vk)]_{nm}=&\braket{\Psi|\hat{f}^{\dag}_{m\vk} \hat{f}_{n\vk}|\Psi}=\Tr[\hat{f}^{\dag}_{m\vk} \hat{f}_{n\vk}\ket{\Psi}\bra{\Psi}]\\
=&\Tr[g\hat{f}^{\dag}_{m\vk} \hat{f}_{n\vk}g^{-1}g\ket{\Psi}\bra{\Psi}g^{-1}]\\
=&\braket{\Psi|g\hat{f}^{\dag}_{m\vk} \hat{f}_{n\vk}g^{-1}|\Psi}\\
=&\sum_{pq} [B(g)]_{\vk,pm} [B(g)]^*_{\vk,qn}\braket{\Psi|\hat{f}^{\dag}_{p,g\vk} \hat{f}_{q,g\vk}|\Psi}\\
=&([B(g)]_{\vk}^{\dag}P(g\vk)[B(g)]_{\vk})_{nm}.
\end{split}
\end{equation}
Therefore, on the matrix level we can check the norm 
\begin{equation}
\label{eqn:order-parameter}
\mc{C}_{\vk}(g)=\norm{[B(g)]_{\vk}^{\dag}P(g\vk)[B(g)]_{\vk}-P(\vk)}=\norm{P(g\vk)[B(g)]_{\vk}-[B(g)]_{\vk}P(\vk)}
\end{equation}
to detect broken symmetry.

Next, we will show that $\mc{C}_{\vk}(g)$ is invariant under arbitrary unitary rotation. Towards these ends, suppose that we perform the mapping
\[
\hat{f}^{\dagger}_{n\vk} \mapsto \sum_{n'}\hat{f}^{\dagger}_{n'\vk}[U(\vk)]_{n'n},
\]
where $U(\vk)$ is unitary.

Under this unitary rotation, the 1-RDM is
\begin{equation}
\begin{split}
\label{eqn:1rdm-gauge}
[P(\vk)]_{nm}\mapsto&\braket{\Psi|\sum_{m'}\hat{f}^{\dagger}_{m'\vk}[U(\vk)]_{m'm} \sum_{n'}[U(\vk)]^*_{n'n}\hat{f}_{n'\vk}|\Psi}\\
=&\sum_{m'n'}[U(\vk)]_{m'm}[U(\vk)]^*_{n'n}\braket{\Psi|\hat{f}^{\dagger}_{m'\vk} \hat{f}_{n'\vk}|\Psi},\\
=&(U^{\dagger}(\vk)P(\vk)U(\vk))_{nm}.
\end{split}
\end{equation}
By a similar calculation, it can be checked that the sewing matrices transform as:
\begin{equation}
[B(g)]_{\vk,mn} \mapsto (U^{\dagger}(g\vk)B(g)_{\vk}U(\vk))_{mn}.
\end{equation}
Therefore, for a unitary operator, the norm is
\begin{equation}
\begin{split}
\mc{C}_{\vk}(g)\mapsto&\norm{U^{\dagger}(g\vk)P(g\vk)U(g\vk)U^{\dagger}(g\vk)[B(g)]_{\vk}U(\vk)-U^{\dagger}(g\vk)[B(g)]_{\vk}U(\vk)U^{\dagger}(\vk)P(\vk)U(\vk)}\\
=&\norm{P(g\vk)[B(g)]_{\vk}-[B(g)]_{\vk}P(\vk)}.
\end{split}
\end{equation}
Hence, $\mc{C}_{\vk}(g)$ is gauge invariant.

\subsection{Order parameter: Antiunitary Case} \label{sec:symm_order_param_antiunitary}

We now turn to verify the analog of \cref{eqn:order-parameter} for antiunitary symmetries
\begin{equation}
\begin{split}
[P(\vk)]_{nm}=&\braket{\Psi|\hat{f}^{\dag}_{m\vk} \hat{f}_{n\vk}|\Psi}=\Tr[\hat{f}^{\dag}_{m\vk} \hat{f}_{n\vk}\ket{\Psi}\bra{\Psi}]\\
=& \Tr[\mc{K} \hat{f}^{\dag}_{m\vk} \hat{f}_{n\vk}\mc{K} g^{-1} \ket{\Psi}\bra{\Psi}g]^*\\
=&\braket{\Psi|g\mc{K}\hat{f}^{\dag}_{m\vk} \hat{f}_{n\vk}\mc{K} g^{-1}|\Psi}^*\\
=&\sum_{pq} [B(g \mc{K})]^*_{\vk,pm} [B(g \mc{K})]_{\vk,qn} \braket{\hat{f}^{\dag}_{p,g\vk} \hat{f}_{q,g\vk}}^* \\
=&\sum_{pq} [B(g \mc{K})]^*_{\vk,pm} [B(g \mc{K})]_{\vk,qn} [P(g\vk)^*]_{qp}\\
&=([B(g \mc{K})]_{\vk}^{\top}P(g\vk)^*[B(g \mc{K})]_{\vk}^{*})_{nm}.
\end{split}
\end{equation}
Therefore, the corresponding order parameter for antiunitary symmetries is
\begin{equation}
\begin{split}
\mc{C}_{\vk}(g \mc{K})
& = \|  [B(g \mc{K})]_{\vk}^{\top}P(g\vk)^*[B(g \mc{K})]_{\vk}^* - P(\vk) \| \\
& = \|  [B(g \mc{K})]_{\vk}^{\dagger}P(g\vk)[B(g \mc{K})]_{\vk} - P(\vk)^* \| \\
& = \|  P(g\vk)[B(g \mc{K})]_{\vk} - [B(g \mc{K})]_{\vk} P(\vk)^* \|, \\
\end{split}
\end{equation}
where in the second to last line we have used that complex conjugation preserves the spectral norm.

To see that this quantity is gauge invariant, recall~\cref{eqn:1rdm-gauge} which shows that the 1-RDM transforms under a gauge transformation as 
\begin{equation}
    [P(\vk)]_{nm} \mapsto [U(\vk)^\dagger P(\vk) U(\vk)]_{nm}.
\end{equation}
Hence, to show gauge invariance we only need to calculate how the sewing matrices transform in this case. By definition under a gauge transformation we have
\begin{equation}
\begin{split}
[B(g \mc{K})]_{\vk,mn} & \mapsto \sum_{m'n'} [U(g\vk)]_{mm'}^* [U(g\vk)]_{m'm} [U(g\vk)]_{nn'} \braket{u_{m',g\vk} | D(g \mc{K}) | u^*_{n\vk}} \\
& =  (U(g\vk)^{\dagger}B(g \mc{K})_{\vk}U(\vk)^*)_{mn}.
\end{split}
\end{equation}
Hence,
\begin{equation}
\begin{split}
    P(g\vk) [B(g \mc{K})]_{\vk} \mapsto & U(g\vk)^\dagger P(g\vk) U(g\vk) U(g\vk)^{\dagger} [B(g \mc{K})]_{\vk} U(\vk)^*, \\
    [B(g \mc{K})]_{\vk} P(\vk)^* \mapsto & U(g\vk)^\dagger  [B(g \mc{K})]_{\vk} U(\vk)^* U(\vk)^\top P(\vk)^* U(\vk)^*,
\end{split}
\end{equation}
which implies that
\begin{equation}
\begin{split}
\mc{C}_{\vk}(g \mc{K}) & \mapsto \| U(g\vk)^\dagger P(g\vk) [B(g \mc{K})]_{\vk} U(\vk)^* - U(g\vk)^\dagger  [B(g \mc{K})]_{\vk} P(\vk)^* U(\vk)^* \| \\
& = \| P(g\vk) [B(g \mc{K})]_{\vk} -  [B(g \mc{K})]_{\vk} P(\vk)^* \|.
\end{split}
\end{equation}
This proves that $\mc{C}_{\vk}(g \mc{K})$ is gauge invariant.

\section{Details of DMRG calculations}
\label{app:dmrg}
\label{sec:dmrg}

We performed the QC-DMRG calculations using the Block2 library \cite{zhai2021low}. We used the Hartree-Fock orbitals in the energy order as basis, and performed DMRG sweeps up to bond dimension 1200. 

\begin{figure*}[h!]
\begin{center}
\begin{subfigure}[c]{.48\textwidth}
    \includegraphics[width=\textwidth]{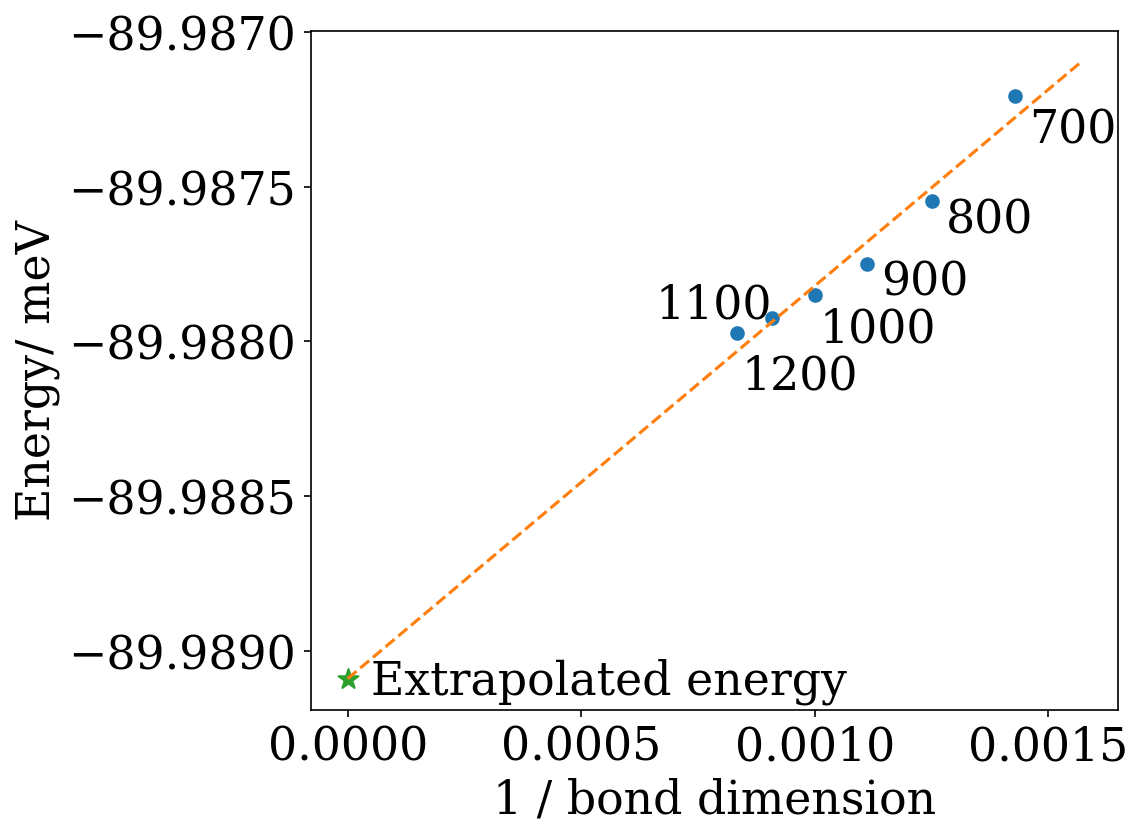}
    \caption{}
    \label{fig:dmrg_extrapolation}
\end{subfigure}
\begin{subfigure}[c]{.48\textwidth}
    \includegraphics[width=\textwidth]{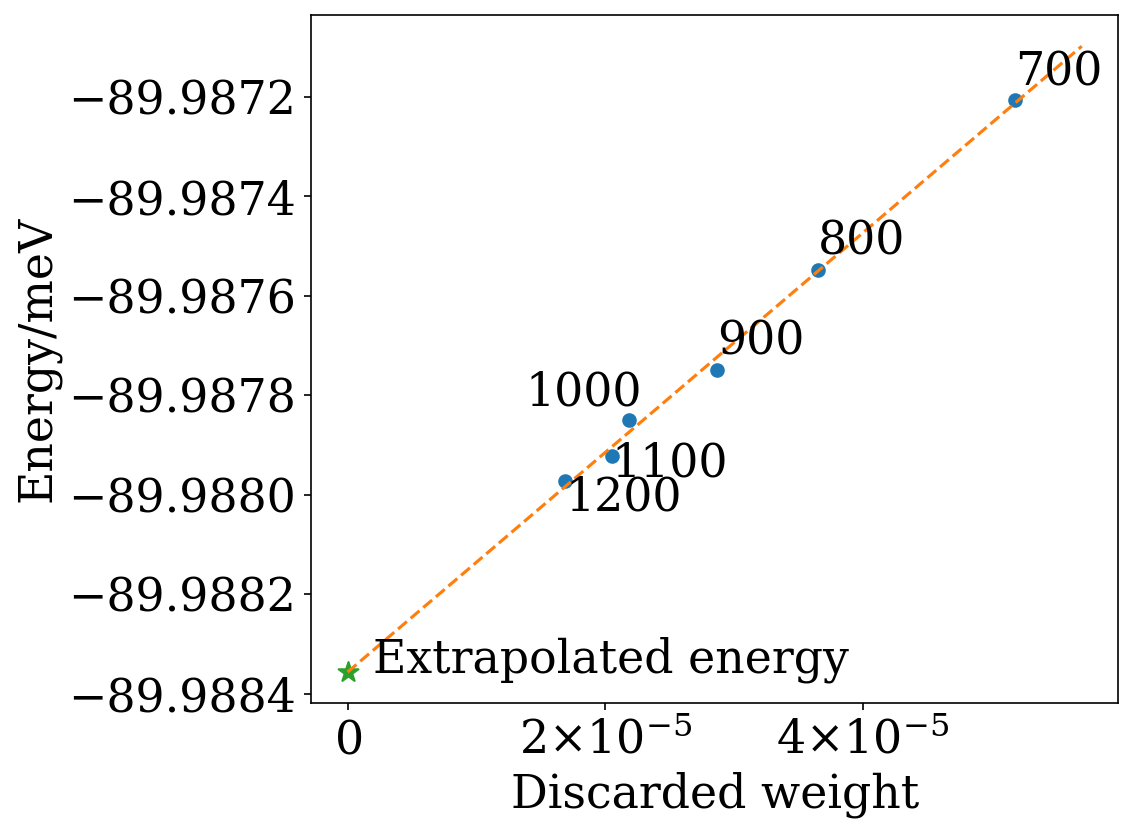}
    \caption{}
    \label{fig:dmrg_extrapolation_dw}
\end{subfigure}
\end{center}
\caption{\label{fig:dmrg_extrpl}
(\subref{fig:dmrg_extrapolation}) Extrapolation of the DMRG energy per moir\'e site in meV with respect to 1/ bond dimension
(\subref{fig:dmrg_extrapolation_dw}) Extrapolation of the DMRG energy per moir\'e site in meV with respect to the discarded weights.
}
\end{figure*}

As the bond dimension of a matrix product state (MPS) ansatz increases, the variational power of the MPS increases up to an exact state in the limit of infinite bond dimension. The DMRG algorithm becomes more expensive as one increases the bond dimension, so in practice, we extrapolate the energy from a set of ``compressed" calculations, i.e., we perform DMRG up to some bond dimension $M_{\text{max}}$, then perform additional DMRG sweeps at lower bond dimensions. The energies are nearly linear with respect to the largest discarded weight (see Fig.~\ref{fig:dmrg_extrapolation_dw}), and the infinite bond dimension limit occurs when the discarded weight is exactly $0$; the infinite bond dimension limit is thus the vertical intercept of the discarded weight-energy plot. 

For certain systems, DMRG may take a very large number of sweeps to properly converge. This makes fitting to the discarded weight difficult because one would require a very large bond dimension to satisfy both (1) sufficiently large changes in bond dimension to make the linearity clear and (2) enough points to confidently perform a linear regression. This was the case for the IBM model away from half-filling. In these cases, we considered three alternate methods of extrapolation. First, we considered the energy from the highest bond dimension calculation. Second, we used the inverse bond dimension to perform a similar procedure (i.e., $1/M=0\Leftrightarrow M\mapsto\infty$). Third, we added additional terms to the inverse bond dimension calculation (i.e., fitting to $a/M + b/M^2 + c$), see~\cref{fig:dmrg_extrapolation}. The bond dimension calculations do not have the same theoretical justification as the discarded weight fits, where the discarded weight is guaranteed to go to $0$ in the infinite bond dimension limit. However, we note that in all our calculations these three methods were in good agreement, and any differences are very small compared to the difference between DMRG and CCSD(T) (see Fig.~\ref{fig:abs_energies_chiral_NIF_dec}).

We also compute errors for the DMRG extrapolation as one-fifth of the extrapolation distance (i.e. the difference between the lowest variational energy and the extrapolated energy), an estimate sometimes used in the literature~\cite{olivares2015ab}. We used the extrapolated energies throughout the main text, and show error bars where applicable.

\section{Effect of subtraction Hamiltonians on the band structure} 
\label{sec:cnp_clacs}

\begin{figure*}[h!]
\begin{center}
\begin{subfigure}[c]{.42\textwidth}
    \includegraphics[width=\textwidth]{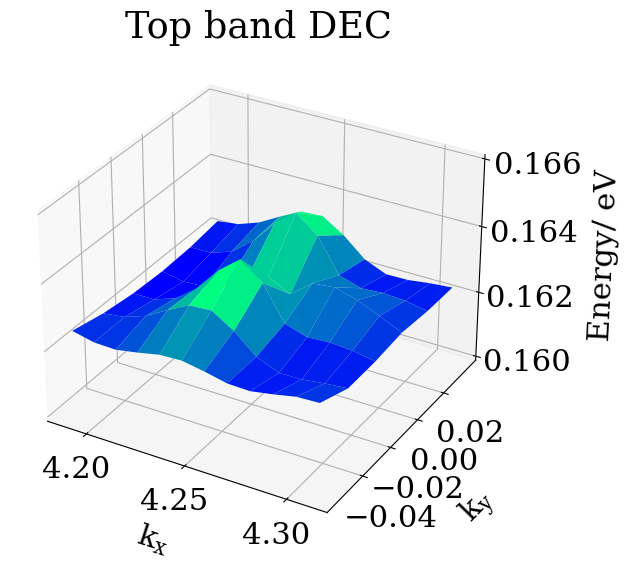}
    \caption{}
    \label{fig:band_diff_top_dec_3d}
\end{subfigure}
\begin{subfigure}[c]{.45\textwidth}
    \includegraphics[width=\textwidth]{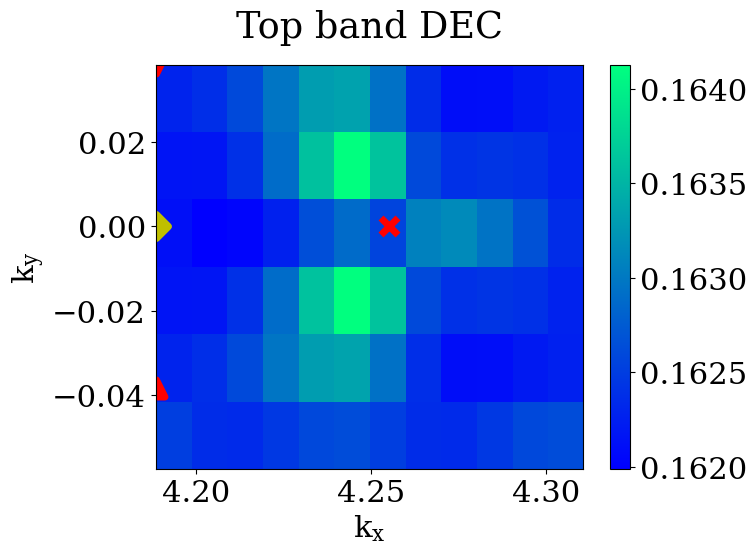} 
    \caption{}
    \label{fig:band_diff_top_dec_2d}
\end{subfigure}
\hfill
\begin{subfigure}[c]{.42\textwidth}
    \includegraphics[width=\textwidth]{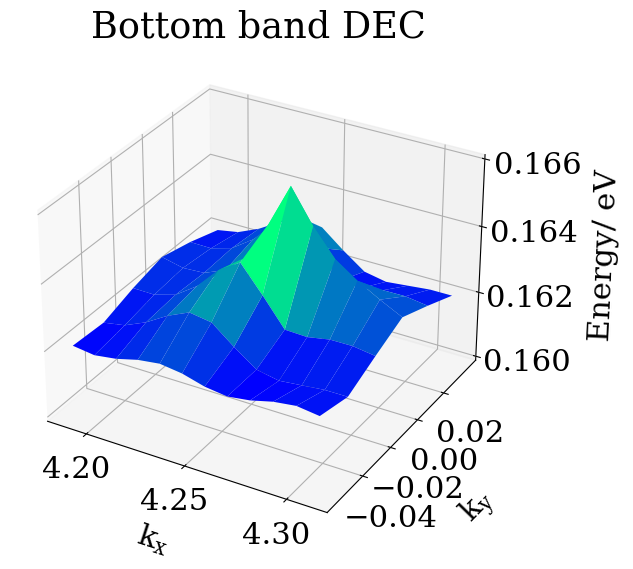}
    \caption{}
    \label{fig:band_diff_bot_dec_3d}
\end{subfigure}
\begin{subfigure}[c]{.45\textwidth}
    \includegraphics[width=\textwidth]{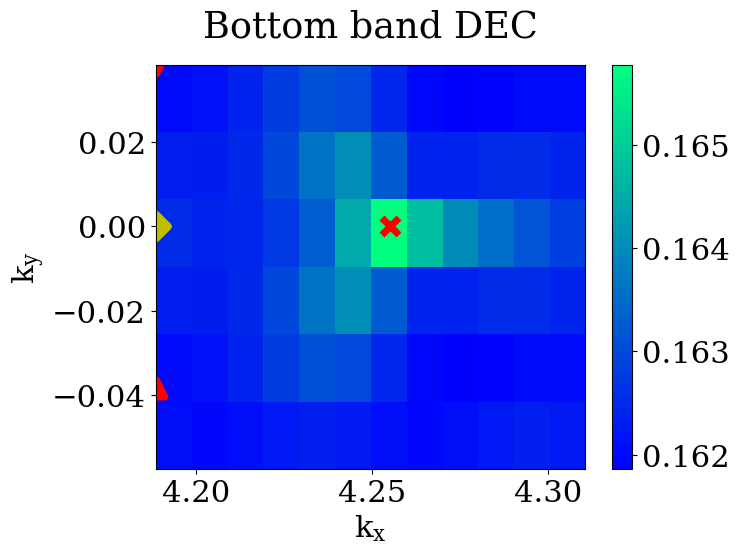}
    \caption{}
    \label{fig:band_diff_bot_dec_2d}
\end{subfigure}
\hfill
\end{center}
\caption{\label{fig:dec_diff}
(\subref{fig:band_diff_top_dec_3d}) Effect of the decoupled scheme to the top band as a function of $k$-points. (\subref{fig:band_diff_top_dec_2d}) Heat plot corresponding to (\subref{fig:band_diff_top_dec_3d}), we marked the $\Gamma_{\rm mBZ}$ point with a red cross, the K$^+$ point with a red upward triangle, the K$^-$ point with a red downward triangle, and the M point with a yellow diamond. 
(\subref{fig:band_diff_bot_dec_3d}) Effect of the decoupled scheme to the bottom band as a function of $k$-points. (\subref{fig:band_diff_bot_dec_2d}) Heat plot corresponding to (\subref{fig:band_diff_bot_dec_3d}), we marked the $\Gamma_{\rm mBZ}$ point with a red cross, the K$^+$ point with a red upward triangle, the K$^-$ point with a red downward triangle, and the M point with a yellow diamond. 
}
\end{figure*}

As pointed out in Ref.~\cite{PotaszXieMacDonald2021}, the remote band self-energy reshapes the bands principally by shifting energies near $\Gamma_{\rm mBZ}$ upward, relative to those near $K^+$, $K^-$. 
With our computations, we can confirm this inhomogeneous contribution to the band structure from the decoupled scheme numerically, see Fig.~\ref{fig:dec_diff}.
We moreover confirm that the shifting of the energies is most pronounced near $\Gamma_{\rm mBZ}$.
For the purpose of depicting this effect, we increased the $\vk$-point grid compared to the other computations. 
We here chose 12 and 6 $\vk$-points in $x$-, and $y$-direction, respectively.

\begin{figure*}[h!]
\begin{center}
\begin{subfigure}[c]{.42\textwidth}
    \includegraphics[width=\textwidth]{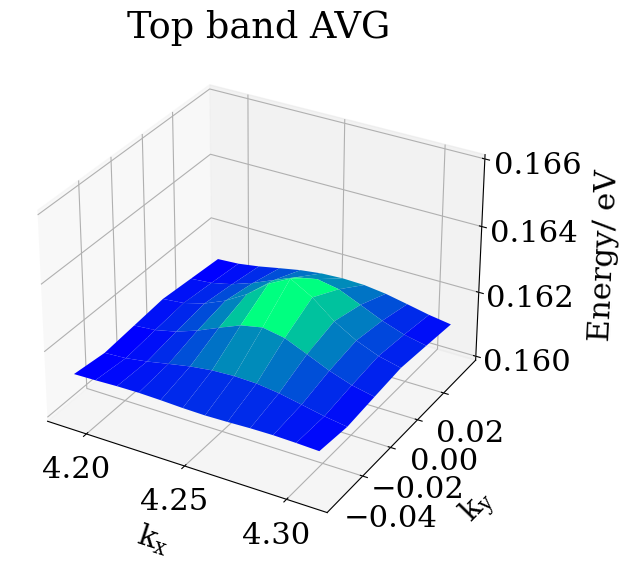}
    \caption{}
    \label{fig:band_diff_top_cnp_3d}
\end{subfigure}
\begin{subfigure}[c]{.45\textwidth}
    \includegraphics[width=\textwidth]{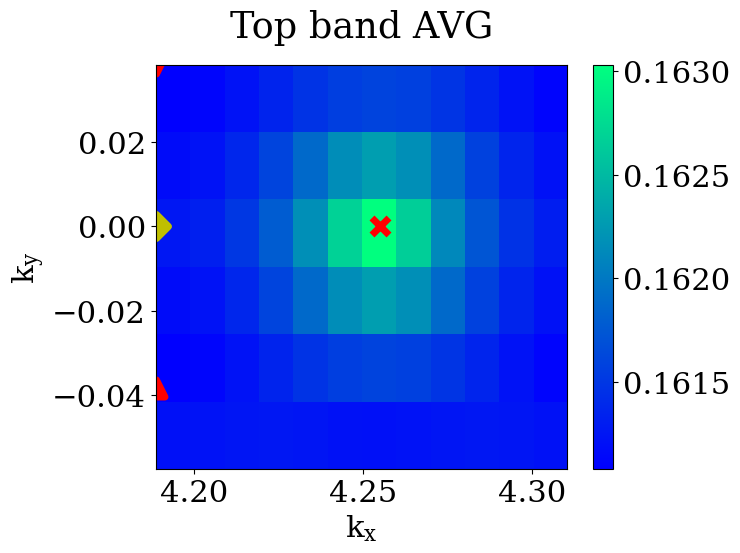} 
    \caption{}
    \label{fig:band_diff_top_cnp_2d}
\end{subfigure}
\hfill
\begin{subfigure}[c]{.42\textwidth}
    \includegraphics[width=\textwidth]{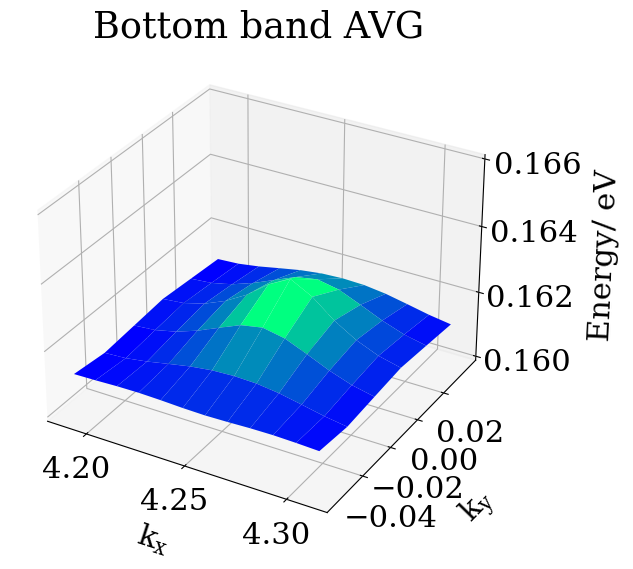}
    \caption{}
    \label{fig:band_diff_bot_cnp_3d}
\end{subfigure}
\begin{subfigure}[c]{.45\textwidth}
    \includegraphics[width=\textwidth]{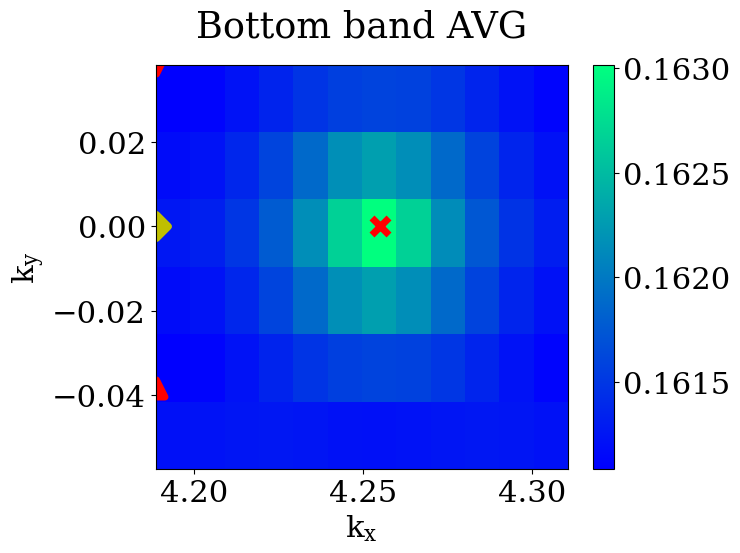}
    \caption{}
    \label{fig:band_diff_bot_cnp_2d}
\end{subfigure}
\hfill
\end{center}
\caption{\label{fig:cnp_diff}
(\subref{fig:band_diff_top_cnp_3d}) Effect of the average scheme to the top band as a function of $k$-points. (\subref{fig:band_diff_top_cnp_2d}) Heat plot corresponding to (\subref{fig:band_diff_top_cnp_3d}), we marked the $\Gamma_{\rm mBZ}$ point with a red cross, the K$^+$ point with a red upward triangle, the K$^-$ point with a red downward triangle, and the M point with a yellow diamond. 
(\subref{fig:band_diff_bot_cnp_3d}) Effect of the average scheme to the bottom band as a function of $k$-points. (\subref{fig:band_diff_bot_cnp_2d}) Heat plot corresponding to (\subref{fig:band_diff_bot_cnp_3d}), we marked the $\Gamma_{\rm mBZ}$ point with a red cross, the K$^+$ point with a red upward triangle, the K$^-$ point with a red downward triangle, and the M point with a yellow diamond. 
}
\end{figure*}

We contrast the results from the decoupled scheme with the energy shift observed for the average scheme, see Fig.~\ref{fig:cnp_diff}.
We find that the average scheme yields a significantly more homogeneous contribution to the band structure.

\end{document}